\documentclass[acmsmall]{acmart}

\usepackage{graphicx}
\usepackage{multirow,multicol}
\usepackage{subfigure}
\usepackage{color}
\usepackage{bm}
\usepackage{epsfig}
\usepackage{rotating}
\usepackage{epstopdf}

\AtBeginDocument{%
  \providecommand\BibTeX{{%
    \normalfont B\kern-0.5em{\scshape i\kern-0.25em b}\kern-0.8em\TeX}}}

\begin{document}

\title{A Comprehensive Survey on Trustworthy Recommender Systems}

\author{Wenqi Fan}
\affiliation{%
  \institution{The Hong Kong Polytechnic University}
  \country{Hong Kong}}
\email{wenqifan03@gmail.com}

\author{Xiangyu Zhao}
\authornote{Corresponding author.}
\affiliation{%
  \institution{City University of Hong Kong}
  \country{Hong Kong}}
\email{xianzhao@cityu.edu.hk}

\author{Xiao Chen}
\affiliation{%
  \institution{The Hong Kong Polytechnic University}
  \country{Hong Kong}}
\email{shawn.chen@connect.polyu.hk}

\author{Jingran Su}
\affiliation{%
  \institution{The Hong Kong Polytechnic University}
  \country{Hong Kong}}
\email{jing-ran.su@connect.polyu.hk}

\author{Jingtong Gao}
\affiliation{
  \institution{City University of Hong Kong}
  \country{Hong Kong}}
\email{jt.g@my.cityu.edu.hk}

\author{Lin Wang}
\affiliation{%
  \institution{The Hong Kong Polytechnic University}
  \country{Hong Kong}}
\email{comp-lin.wang@connect.polyu.hk}

\author{Qidong Liu}
\affiliation{%
  \institution{City University of Hong Kong}
  \country{Hong Kong}}
\email{qidongliu2-c@my.cityu.edu.hk}

\author{Yiqi Wang}
\affiliation{%
  \institution{Michigan State University}
  \city{East Lansing}
  \state{MI}
  \country{USA}}
\email{wangy206@msu.edu}

\author{Han Xu}
\affiliation{%
  \institution{Michigan State University}
  \city{East Lansing}
  \state{MI}
  \country{USA}}
\email{xuhan1@msu.edu}

\author{Lei Chen}
\affiliation{%
  \institution{The Hong Kong University of Science and Technology}
  \country{Hong Kong}}
\email{leichen@cse.ust.hk}

\author{Qing Li}
\affiliation{%
  \institution{The Hong Kong Polytechnic University}
  \country{Hong Kong}}
\email{csqli@comp.polyu.edu.hk}
\renewcommand{\shortauthors}{Fan and Zhao, et al.}


\begin{abstract}

As one of the most successful AI-powered applications, recommender systems aim to help people make appropriate decisions in an effective and efficient way, by providing personalized suggestions in many aspects of our lives, especially for various human-oriented online services such as e-commerce platforms and social media sites. 
In the past few decades, the rapid developments of recommender systems have significantly benefited human by creating economic value, saving time and effort, and promoting social good.
However, recent studies have found that data-driven recommender systems can pose serious threats to users and society, such as spreading fake news to manipulate public opinion in social media sites, amplifying unfairness toward under-represented groups or individuals in job matching services,  or inferring privacy information from recommendation results.
Therefore, systems' trustworthiness has been attracting increasing attention from various aspects for mitigating negative impacts caused by recommender systems,
so as to enhance the public's trust towards recommender systems techniques.
In this survey, we provide a comprehensive overview of \textbf{T}rustworthy \textbf{Rec}ommender systems (\textbf{TRec}) with a specific focus on six of the most important aspects; namely, Safety \& Robustness, Nondiscrimination \& Fairness, Explainability,  Privacy,  Environmental Well-being, and Accountability \& Auditability. 
For each aspect, we summarize the recent related technologies and discuss potential research directions to help achieve trustworthy recommender systems in the future.

\end{abstract}



\maketitle

\section{Introduction}

In the past few decades, the growth of information exchange through the Internet has resulted in extreme information explosion.
Thus, recommender systems have been playing an increasingly important role in people's daily lives via their successful deployments in various user-oriented online services, such as online shopping~\cite{gong2020edgerec,li2021embedding}, jobs matching~\cite{guo2020detext,geyik2019fairness},  financial product recommendations~\cite{barreau2020history}, and  medical recommendations ~\cite{zheng2022interaction}. 
As reported in \textit{MIT Technology Review 2021}\footnote{\url{https://www.technologyreview.com/2021/02/24/1014369/10-breakthrough-technologies-2021/}}, TikTok (one of the world’s fastest-growing social networks) recommendation  algorithm was awarded as one of the ``Top 10 Global Breakthrough Technologies''.
More recently, inspired by the great success of Deep Neural Networks (DNNs) in powerful representation learning abilities, DNN-based recommendation techniques have shown impressive performance across a wide range of tasks~\cite{fan2019deep,fan2022graph}. 
For example, a simple yet effective deep learning based recommender system is designed for videos recommendations in YouTube mobile app~\cite{covington2016deep}; 
A BERT based  ranking model shows great power on solving an online job search task in LinkedIn~\cite{guo2020detext}; Some works also leverage Graph Neural Networks (GNNs) to learn helpful representations of users and items in social medias~\cite{ying2018graph,fan2019graph,fan2020graph}.

Despite their great achievements in benefiting human daily lives, recent studies have shown that recommender systems could also bring negative consequences to human society. For example,  recommender systems are highly vulnerable to adversarial attacks: someone can generate and spread malicious information, thereby fooling the prediction of recommender systems to wrongly promote or demote items~\cite{fan2021attacking,chen2022knowledge}.  
In addition, recommender algorithms can also implicitly inherit and amplify the biased opinions from the data that collected from society. It will cause the models to have discriminatory biases and unfairness towards under-represented groups, such as people from various genders, races and occupations~\cite{liu2022rating}.
Moreover, recommender systems are also susceptible to the risk of leaking users' private information~\cite{yangPrivacyPreservingSocialMedia2019}. For example, someone is able to recover the private information from other users, by only exploiting the model parameters. 
Furthermore, because of the complicated architecture of DNNs, it is extremely hard to decipher and explain the prediction mechanism of recommender systems~\cite{chen2019dynamic, zhang2020explainable}. 
These vulnerabilities of recommender systems can make unreliable recommendation results and produce significant harmful effects in various real-world applications,  especially those in safety-critical areas such as finance and healthcare, resulting in severe economic, social, and security consequences.
Meanwhile, the concerns on the trustworthiness of recommender systems have significantly hindered the development and deployment of recommendation algorithms.
Therefore, how to build trustworthy recommender systems has attracted increasing attention from both academy and industry.

\begin{figure}[t]
\centering
\centering
{\includegraphics[width=0.88\linewidth]{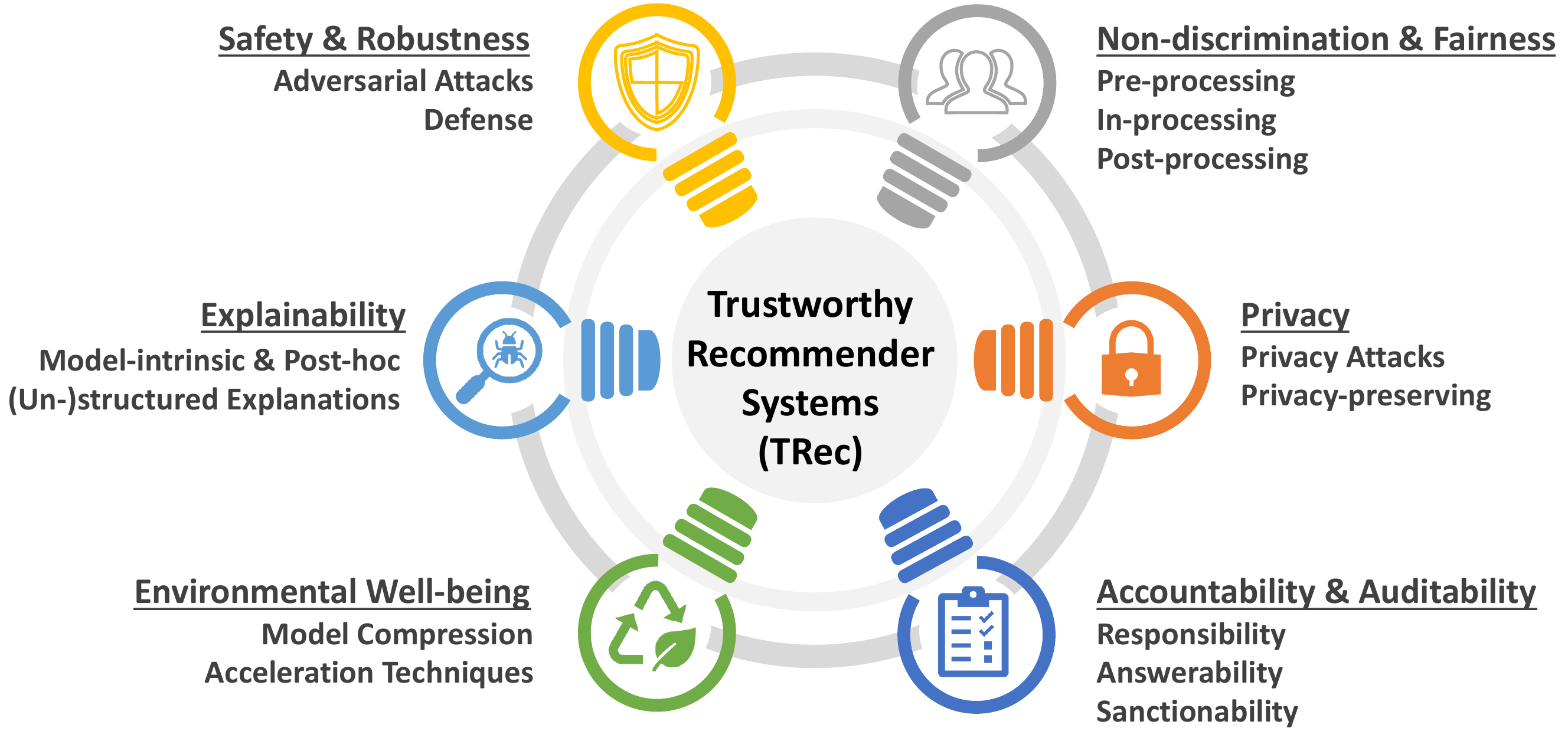}}
\caption{Six key dimensions of Trustworthy Recommender Systems (TRec).}
\label{fig:intro}
\end{figure}

More recently, the European Union (EU) has provided ethics guidelines for promoting Trustworthy Artificial Intelligence (TAI)~\cite{smuha2019eu}, in which a trustworthy AI system should obey certain ethical principles, such as \emph{Prevention of harm}, \emph{Fairness}, and \emph{Explainability}. 
These ethical principles must be unfolded in practical requirements for achieving the trustworthiness of AI systems. 
Meanwhile, building AI systems requires considerable effort from various stakeholders, including system's developers and deployers,  end-users, as well as civil society and government.
It is worth mentioning that such trustworthy principles in the context of AI systems are also suitable to characterize the trustworthiness of recommender systems, since recommender systems are one of the most successful human-centered AI applications in our daily lives.  
In this survey, we focus on \emph{SIX} of the most crucial dimensions in achieving trustworthy recommender systems: \emph{Safety \& Robustness, Non-discrimination \& Fairness, Explainability, Privacy,  Environmental Well-Being}, and \emph{Auditability \& Accountability}, as shown in Figure~\ref{fig:intro}. 

Take  recommender systems in financial applications as an example,  it plays a crucial role in various high-stakes scenarios, such as stock market, insurance products, and loan services.  
Hence, the recommender systems are expected to make particularly \emph{robust} and \emph{accurate} decisions under any potential security threats.
Meanwhile,  the demographic attributes of customers such as income, occupation, race, and genders are very \emph{private}, which requires recommender systems to avoid leakage. Thus these information require special and careful protection in recommender systems.
Furthermore, it is important that recommendation algorithms ought to mitigate \emph{discriminatory bias} or \emph{unfairness} toward certain groups or individuals for credit card and loan approval.  
Also, considering the reliability of recommender systems, it is desired to provide \emph{explanations} on how certain decisions are made for various stakeholders,  and conduct system auditing periodically from different parties.
In addition, training and fine-tuning a large-scale recommendation model typically needs huge energy and natural resources, resulting in problems of global  environmental deterioration and  resource depletion. 
Thus, it is important to consider the \emph{sustainability and environmental friendliness} of recommender systems for the benefits of our future generations.

Recent years have witnessed a growing awareness of the trustworthiness of recommender systems in both academia and industry, contributing to the emergence of a considerable body of literature that highlights various dimensions of trustworthy recommender systems~\cite{fan2021attacking, liu2022rating, zhang2020explainable}.
For example, to defend against adversarial attacks~\cite{fan2021attacking},  methods regarding robust recommendation algorithms have been proposed~\cite{fan2022graph,zhang2020gcn}.   
Debiasing technologies for building fair recommender systems have been designed for various real-world tasks such as online job matching~\cite{geyik2019fairness}.
Explainable recommendations have been proposed to improve transparency and user satisfaction in the recommendation's decision-making process~\cite{wang2019explainable}. 
Privacy-preserving techniques have been explored to reduce the risk of private data leakage~\cite{chenDifferentialPrivateKnowledge2022}.
What's more, there are several surveys regarding the trustworthiness of recommender systems focusing on specific aspects, such as Safety \& Robustness~\cite{deldjoo2021survey,si2020shilling}, Bias and Fairness~\cite{chen2020bias,li2022fairness,wang2022survey},  and Explainability~\cite{chen2022measuring, zhang2020explainable}.
In addition, recent surveys ~\cite{liu2021trustworthy,dai2022comprehensive} give a thorough review of trustworthy AI and Graph Neural Networks (GNNs). 
As one of the most successful application of human-centered AI systems, it is imperative to systematically summarize the existing achievements and challenges of trustworthy recommender systems. 
Therefore, in this survey, we provide a comprehensive overview of \textbf{T}rustworthy \textbf{Rec}ommender Systems (\textbf{TRec}) to help researchers and practitioners gain a basic understanding of trustworthy recommender systems, and then have a deeper understanding of the latest progress and facilitate the discussion of the future directions on this demanding topic.
More specifically, this survey introduces six key dimensions in realizing trustworthy recommender systems. 
For each dimension, we introduce its concepts and definitions,  as well as provide a taxonomy to review representative and state-of-the-art algorithms. 
It is worth noting that these SIX dimensions are not independent of each other for building trustworthy recommender systems. 
At last, we also provide discussions about potential interactions among different dimensions and other potential aspects to achieve the trustworthiness of recommender systems in future directions.
The remainder of this survey is organized as follows.
 
In section~\ref{sec:safety}, we describe the dimension of \textbf{Safety} \& \textbf{Robustness} from adversarial attacks and defenses aspects, in which  a recommender system is required to be robust against adversarial perturbations, so as to make reliable recommendation results. 
Recent works show that deep recommender systems can inherit vulnerability from DNNs  by generating small input perturbations.
This vulnerability has raised tremendous concerns about adopting recommender systems in safety-critical domains such as finance and healthcare.  
Therefore, it is urgent and essential to study the safety and robustness for building safe and reliable recommender systems. 

As most recommendation models are designed by our humans and trained from user behavior data, 
recommender systems can easily inherit human discrimination and unfairness toward certain groups or individuals, resulting in  trust loss from various stakeholders. 
Recently, non-discrimination \& fairness in recommender systems receives considerable attention from both academia and industry.  
In section~\ref{sec:fairness}, we detail the dimension of \textbf{Non-discrimination} \& \textbf{Fairness}, which requires a recommender systems to make  fair decisions.

In section~\ref{sec:interpret}, we introduce the dimension of \textbf{Explainability}, which expects that the working mechanism behind the predictions in recommender systems can be understandable to various stakeholders (e.g., system’s developers and end-users). 
The explainability in recommender systems is treated as an effective way to  motivate users to interact with online service,   increase   users' trust during interactions, and  assist algorithms' developers to develop and debug systems.

Since most modern recommender systems are driven by data, recent works found that users'  private data such as browsing history and credit card numbers  is likely  stored and exposed, which increases the risk of  data leakage.  
In section~\ref{sec:privacy}, we detail the dimension of \textbf{Privacy}, which requires a recommender system to prevent any private information leakage.

Modern recommender systems  heavily rely on deep learning techniques to achieve promising performance, in which the demands for large recommendation models  will constantly increase, leading to long training time, large storage space, and tremendous energy consumption.
A recent study~\cite{adnan2021accelerating} shows that  training a model on the Taobao dataset needs 621 minutes with 4 GPUs, whose average GPU power consumption is 56.39W per hour. 
In section~\ref{sec:environment}, we present the dimension of \textbf{Environmental Well-being}, which expects that a recommender system can be sustainable and environmentally-friendly.

In section ~\ref{sec:account}, we discuss the dimension of \textbf{Auditability} \& \textbf{Accountability}, which expects that the responsibility distribution can be  clearly determined  for  many different parties in the function of recommender systems.

An ideal trustworthy recommender system should  satisfy six aforementioned dimensions simultaneously, but most researches only focus on one of them and ignore their potential interactions. 
In section ~\ref{sec:relation}, we introduce the complicated interactions among  different dimensions for achieving trustworthy recommender systems.
At last, we discuss  some future directions to be explored for achieving trustworthy recommender systems in section~\ref{sec:future}.

Concurrent to our survey, Ge et al.  ~\cite{ge2022survey}   review trustworthy recommender systems from five perspectives, namely explainability, fairness, privacy, robustness, and controllability. 
Wang et al.~\cite{wang2022trustworthy} describe trustworthy recommender systems in four stages, including data preparation, data representation, recommendation generation, and performance evaluation.
In contrast, our work provides a comprehensive survey of trustworthy recommender systems from a computational perspective, discusses the interactions among different dimensions, and provides potential research directions to explore in the future.

\section{Safety \& Robustness}
\label{sec:safety}

Recommender systems play an increasingly important role in high-stake scenarios such as bank loan systems and healthcare recommendations. 
In recent years, researchers have found that recommendation systems are highly vulnerable to malicious attacks ~\cite{li2016data}, in which modifying a tiny amount of user-item interactions can manipulate recommender systems to produce incorrect results with malicious intentions~\cite{chen2022knowledge,fan2021attacking}. 
These systems cannot be fully trusted and  even be denial-of-service attacked if their vulnerabilities are exposed and exploited intentionally.
As a result, such vulnerability raises huge concerns when applied to high-stakes tasks, and hinders recommender systems' deployment.
For instance, if recommender systems are applied to financial prediction,  
there may exist some adversaries who attempt to generate fake transactions to deliberately affect the systems' predictions. In healthcare recommender systems, it is possible for an attacker to generate fake cases as a way to mislead the system's diagnosis, posing a threat to patient safety.
To prevent attackers from producing harmful effects, recommender systems are required to be robust to artificial perturbations. 
Besides, it is worth mentioning that understanding the recommendations' weaknesses can provide great opportunities to design new countermeasures against adversarial attacks.  
Therefore, it is necessary to study adversarial attacks for manipulating recommender systems and to design corresponding defense strategies, so as to improve the reliability and safety of recommender systems.

In this section, we will first introduce the concept and taxonomy of adversarial attacks and defense in recommendation systems. Then we describe how to attack recommendation systems and corresponding defense strategies in detail. All involved methods are summarized in Table ~\ref{tab:robust_summary}. Next, we present some practical applications in our daily lives where robustness is critical. Finally, we demonstrate some potential future directions for robust recommendation systems.

    \subsection{Concepts and Taxonomy}\label{concept}
    In this subsection, we introduce concepts and taxonomy related to the safety \& robustness of recommendation systems from the perspective of adversarial attack and defense.

    \subsubsection{Attackers' Goal}\label{concept:atk_goal}    
    In recommender systems, according to adversaries' goals, we can divide them into the two following categories.

    \begin{itemize}
        \item \textbf{Target Attacks}: The goal is to promote/demote a set of target items in recommendation systems, such that  target items can be recommended to as many/few users as possible. This goal is to manipulate the exposure rate of target items to achieve attackers' desires.

        \item \textbf{Untarget Attacks}: In this setting, there is no specific items to be promoted or demoted in untarget attack. Its goal is to degrade a recommendation system's overall performance, so as to reduce users' online experience and satisfaction.
    \end{itemize}

        \subsubsection{Attack Stage}\label{concept:atk_stage}
    Generally, adversarial attacks in recommendation systems can be divided into two types according to the attack stage: evasion attack and poisoning attacks, which can affect recommendation systems in the inference and training phases, respectively.

    \begin{itemize}
        \item \textbf{Evasion Attack (Inference/Test Stage)}: Evasion attack happens during the model service (test) stage. For instance, given a fixed well-trained model,  attackers can modify a target user's profile, such as historical interaction logs, so its recommendation outcome is changed.
        \item \textbf{Poisoning Attack (Training Stage)}: Poisoning attack, also called shilling attack, occurs during the data collection phase before model training. The attacker intends to inject fake users into the training data of recommendation systems, so that trained model's prediction behavior can be controled with malicious desires.
    \end{itemize}

    \subsubsection{Attackers' Knowledge}\label{concept:atk_know}
    To conduct adversarial attacks, the knowledge that attackers are allowed to access target recommender systems can heavily affect the attacking strategies and performance to achieve the adversarial goal. Typically,  auxiliary knowledge on a target recommender system includes the target model's architecture and parameters, and datasets, etc. 
    In general, adversaries can conduct three different types  of attacking strategies according to their accessibility to the target recommender systems' knowledge, including white-box, grey-box, and black-box attacks.
    
    \begin{itemize}
        \item \textbf{White-box Attacks}: In this setting, attackers can get all information about the target recommendation system, including training data, recommendation architecture and  parameters. One widely used strategy is to utilize the gradient to assist in generating adversarial perturbations.
        Since it is difficult for attackers to obtain such complete knowledge in the real world,  this type of attacks cannot  poses severe threats.
        However, researchers can utilize this type of attack to analyze the robustness of the target system in the worst case.
       
        \item \textbf{Grey-box Attacks}: In this setting, attackers can only get partial information about the target recommender system to conduct attack. 
         Compared with white-box attacks, grey-box attacks are more practical and dangerous, since such limited information is easy for attackers to obtain.
         For instance,  users' reviews and items' ratings information  on Amazon  are easily  collected, which motivates adversaries to take advantage of such available training dataset to train a surrogate model and perform white-box attack subsequently.

        \item \textbf{Black-box Attacks}: In this case, it is challenging for attackers  to access the target models of recommender systems and their training data. 
        This setting is more realistic and practical for existing adversarial attacks and have attracted increasing attentions recently. 
        Typically, black-box attacks tend to perform query the target recommender systems for updating the attacking strategies. 
    \end{itemize}

    \subsubsection{Adversarial Perturbation Type}\label{concept:perb}

    With  malicious goals, adversaries can add adversarial perturbations in different ways by considering various scenarios. In general,  
    such data perturbations are implemented via adding fake user profiles into user-item interactions,  modifying users attributes information (e.g., age,  gender, occupation, etc.), and modifying item side information such as the attributes and description of movies. 
    
    \subsubsection{Countermeasure Strategies Against Adversarial Attacks (Defense Methods)}\label{concept:defense}
    To prevent the harm from adversarial attacks, its countermeasure strategies in   recommendation systems can be divided into two categories: \emph{Perturbation Detection} and \emph{Adversarial Training}.
    \begin{itemize}
        \item \textbf{Perturbation Detection.} This kind of defense strategies is to  identify  perturbations data and remove them for resisting adversarial attacks in recommender systems.
        \item \textbf{Adversarial Training.} This is a widely used strategy to resist adversarial attacks by enhancing the robustness of recommender systems. 
    \end{itemize}

    \begin{table}
        \centering
        \caption{Taxonomy of related methods.}
        \begin{tabular}{c|l|l}
            \toprule
            & \textbf{Taxonomy} & \textbf{Related Research} \\
            \hline
            \multirow{3}{*}{\textbf{Attack}} & Heuristic Methods & ~\cite{lam2004shilling, burke2005limited, burke2005segment, mobasher2007toward, williams2006profile, burke2015robust}\\
            \cline{2-3}
            & Gradient-based Methods & ~\cite{christakopoulou2019adversarial, fang2020influence, fang2018poisoning, li2016data, tang2020revisiting, christakopoulou2018adversarial, lin2020attacking, wu2021triple}\\
            \cline{2-3}
            & Reinforcement Learning-based Methods & ~\cite{song2020poisonrec,fan2021attacking, chen2022knowledge}\\
            \hline
            \multirow{2}{*}{\textbf{Defense}} & Detection Methods& ~\cite{burke2006classification, zhang2014hht, karthikeyan2017prevention, mehta2007unsupervised, mehta2009unsupervised, lee2012shilling, gao2020shilling, zhang2020gcn, shahrasbi2020detecting}\\
            \cline{2-3}
            & Adversarial Robust Training Methods & ~\cite{he2018adversarial, tang2019adversarial, wang2019adversarial, yuan2019adversarial, chen2019adversarial}\\
            \bottomrule
        \end{tabular}
        \label{tab:robust_summary}
    \end{table}

    \subsection{Representative Attack Methods}
    \label{atk_method}
    In this subsection, we mainly introduce poisoning attack (i.e., shilling attack), which is the most widely studied mainstream attacks in recommendation systems. First, we give a unified formulation of poisoning attacks. Then, we present representative methods from various aspect, including heuristic methods, gradient-based methods, and reinforcement learning-based methods.
    
        \subsubsection{A Unified Formulation of Poisoning Attack.} 
\begin{figure}
        \centering
        \includegraphics[width=12cm]{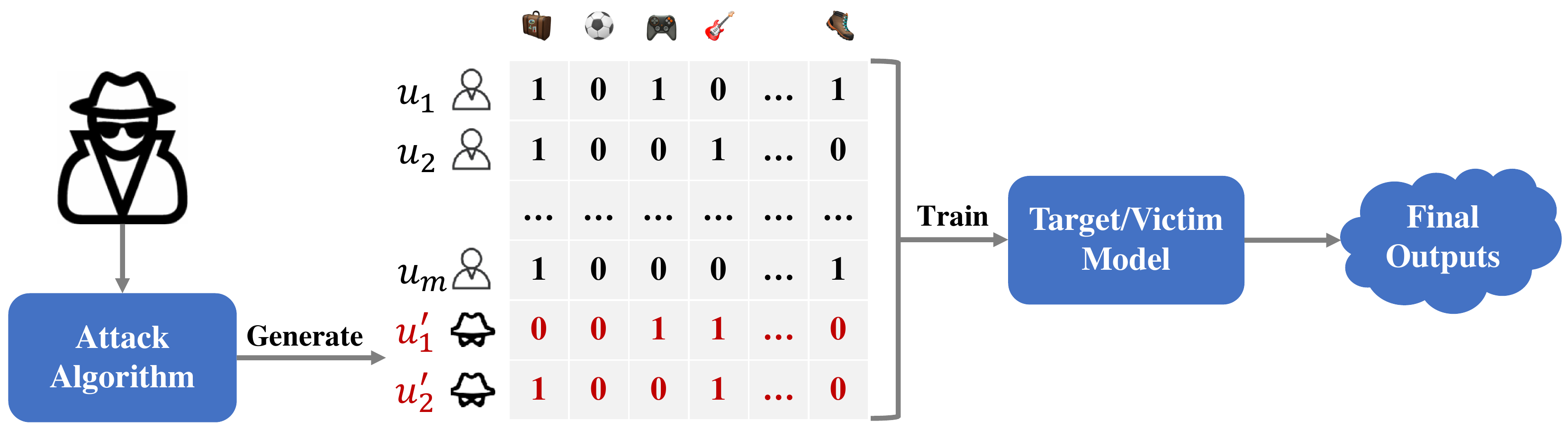}
        \caption{The poisoning attack. Attackers inject well-designed faker users into training data to manipulate target model's behaviors.}
        \label{fig:robust1}
    \end{figure}

The attackers' goal is to inject well-designed fake user profiles into recommender systems to manipulate the recommendation's output, as illustrated in  Figure ~\ref{fig:robust1}. In general, this attacking process can be formulated as a bi-level optimization problem.
        \label{atk_method:uniform}
        Mathematically, given a set of users $U=\{u_{1},u_{2},...,u_{|U|}\}$, a set of items $I=\{i_{1},i_{2},...,i_{|I|}\}$, and  user-item interactions matrix $\bm{R} \in \mathbb{R}^{|U|\times|I|}$, attackers aims to design a set of fake user $\widehat{U}=\{\widehat{u}_{1},\widehat{u}_{2},...,\widehat{u}_{|\widehat{U}|}\}$ with the fake user-item interaction data $\widehat{\bm{R}} \in \mathbb{R}^{|\widehat{U}|\times|I|}$ to achieve  their adversarial goals,  which can be formulated as $\mathcal{L}_{adv}$:
        \begin{equation}\label{eq:1}
            \min _{\widehat{U}} \mathcal{L}_{adv}(\theta^{*}),\quad \text { s.t. } \quad \theta^{*}=\underset{\theta}{\arg \min} (\mathcal{L}_{rec}(\bm{R},\bm{O}_{\theta})+\mathcal{L}_{rec}(\widehat{\bm{R}},\bm{O}_{\theta})),
        \end{equation}  
        where $\theta$ is the recommendation system's parameters, $\bm{O}_{\theta}$ is the system's predictions with parameters $\theta$, 
        and $\mathcal{L}_{rec}$ can be a general recommendation objective. 
        The objective function $\mathcal{L}_{adv}$ can be determined by the specific goal, such as promoting/demoting some items or destroy the target system's  utility. 
        Generally, the fake users $\widehat{U}$ cannot be arbitrarily designed, since defenders can easily detect such fake users who have a large discrepancy with normal users. 
        In addition, the perturbations will be constrained by pre-defined budget for attacking, like the number of fake users and their interactions.
        
        \subsubsection{Heuristic Methods.}\label{atk_method:heuristic}
        Some early studies are based on hand-engineered fake user profiles to poison a recommendation system. 
        A straightforward solution is that attackers can design fake users that assign high scores to target items and a low score to random others to promote target items ~\cite{lam2004shilling}. 
        Burke et al. ~\cite{burke2005limited}  consider to  interact with some popular items rather than random items,  so that the fake users can have more impact on normal users. 
        Some works~\cite{burke2005segment,mobasher2007toward} introduce low-knowledge promotion attack methods such as the random attack, average attack, bandwagon attack, and segment attack. 
        Similarly,  Williams and Mobasher ~\cite{williams2006profile} propose several demotion attack methods: love/hate attack and reverse bandwagon attack.

        However, these attacking methods have several limitations. First, fake user profiles generated by heuristics tend to have distinct characteristics, sometimes even forming self-established clusters~\cite{burke2015robust}, making them easy to be detected~\cite{mehta2008survey}.
        In addition, heuristic methods strongly rely on prior knowledge to generate fake profiles that influence normal users.
        Finally,  adversarial goals cannot be optimized by most existing heuristic methods, leading to low attack success rates.

        \subsubsection{Gradient-based Methods.}\label{atk_method:gradient}
        Unlike heuristic methods,  the process of poisoning attacks can be formulated as the optimization problem  as shown in Eq. ~\ref{eq:1}, where the generation process of fake user profiles can be formulated as a bi-level optimization problem,  consisting of a constraint called inner objective and the main goal called outer objective. 
        Here we take Matrix Factorization (MF) as the target recommendation model and then introduce it with  gradient-based methods.
        Given the user-item interaction data $\bm{R} \in \mathbb{R}^{|U|\times|I|}$, the MF model aims to learn the user embedding matrix $\bm{P} \in \mathbb{R}^{|U| \times d}$ and item embedding matrix $\bm{Q} \in \mathbb{R}^{|I| \times d}$, such that the inner product $\bm{P}\bm{Q}^{T}$ can approximate the interaction $\bm{R}$ under observed values (i.e., non-zero values in $\bm{R}$).
        And then, the well-learned $\bm{PQ}^{T}$ can be used to predict unobserved user-item pairs. 
        Mathematically, when considering fake users, the objective of MF learning can be formulated as:
        \begin{equation}\label{eq:2}
            \begin{array}{l}
                \underset{\theta}{\min} ~\mathcal{L}_{rec}(\bm{R},\bm{O}_{\theta})+\mathcal{L}_{rec}(\widehat{\bm{R}},\bm{O}_{\theta}), \\ 
                \rightarrow  \underset{\bm{P},\bm{Q},\widehat{\bm{P}}}{\min} ~\sum_{u,i}(\bm{R}_{u,i}-\bm{P}_{u}\bm{Q}_{i}^{T})+\sum_{\widehat{u},i}(\widehat{\bm{R}}_{\widehat{u},i}-\widehat{\bm{P}}_{\widehat{u}}\bm{Q}_{i}^{T})  + \lambda(\|\bm{P}\|^{2}+\|\bm{Q}\|^{2}+\|\widehat{\bm{P}}\|^{2}),
            \end{array}
        \end{equation}
        where $\widehat{\bm{P}}$ denotes the  embedding matrix of fake users, and $\lambda$ is a hyper-parameter to  regularize model for addressing overfitting.

        The adversarial objective function $\mathcal{L}_{adv}$ varies for different malicious purposes. Here, we focus on promoting target item $k$ to all normal users, which can be formulated as follows:
        \begin{equation}\label{eq:3}
            \begin{array}{c}
                \underset{\widehat{\bm{R}}}{\min} ~\mathcal{L}_{adv}(\theta^{*})=-\sum_{u \in \mathcal{U}} \log \left(\frac{\exp \left(r_{u k}\right)}{\sum_{i \in \mathcal{I}} \exp \left(r_{u i}\right)}\right), \\
                 \text {s.t.} \quad \theta^{*}=\underset{\theta}{\arg \min} ~\mathcal{L}_{rec}(\bm{R},\bm{O}_{\theta})+\mathcal{L}_{rec}(\widehat{\bm{R}},\bm{O}_{\theta}).
            \end{array}
        \end{equation}
        In this case, the model's parameters $\theta$ are embedding matrices $\bm{P}$ and $\bm{Q}$. 
        The goal is to minimum the loss, hoping that   all normal users' prediction on target item $k$ is greater than other items.
        Finding optimal fake users $\widehat{U}$ in Eq. ~\ref{eq:1} is equivalent to optimizing its rating matrix $\widehat{\bm{R}}$ in Eq. ~\ref{eq:3}, which can be achieved by  the projection gradient descent method as follows:
        \begin{equation}\label{eq:4}
            \widehat{\bm{R}}^{t+1} = \operatorname{Proj}_{\mathbb{R}}(\widehat{\bm{R}}^{t}-\alpha\cdot\nabla_{\widehat{\bm{R}}}\mathcal{L}_{adv}(\theta^{*})),
        \end{equation}
        \begin{equation}\label{eq:5}
            \nabla_{\widehat{\bm{R}}}\mathcal{L}_{adv}(\theta^{*}) = \nabla_{\widehat{\bm{R}}}\theta^{*} \nabla_{\theta^{*}}\mathcal{L}_{adv},
        \end{equation}
        where $\operatorname{Proj}_{\mathbb{R}}$ denotes the projection operator under the feasible region $\mathbb{R}$, and $\alpha$ is the step size. 
        Note that the second gradient $\nabla_{\theta^{*}}\mathcal{L}_{adv}$ can be easily obtained, while it is challenging to obtain the first gradient term (i.e., $\nabla_{\widehat{\bm{R}}}\theta^{*}$) because of involving the minimization term $\theta^*$.

        Li et al. ~\cite{li2016data} pioneer a gradient-based attack method for factorization-based recommendation systems. The key technique of their method is to approximately compute the Eq. ~\ref{eq:5} based on first-order KKT conditions. 
        Christakopoulou et al. ~\cite{christakopoulou2019adversarial} use the zero-order optimization method in evolutionary algorithms to find the gradient's direction. Specifically, they make several minor changes on fake user profiles $\widehat{\bm{R}}$ to evaluate the adversarial loss $\mathcal{L}_{adv}$, and then update   user profile by the change that makes the loss smaller.
        In ~\cite{fang2020influence}, Fang et al. propose the differentiable hit ratio loss to generate fake users for top-N recommendation systems and leverage first-order stationary condition to approximately compute the Eq. ~\ref{eq:5}.
        To improve imperceptibility, they select filler items relying on the value of the final solution of Eq. ~\ref{eq:4} and give ratings sampled from the distribution of normal users' interactions to these filler items. Moreover, they only select a subset of critical users with the influence function for efficient computation. Different from the previous approximation methods, Tang et al. ~\cite{tang2020revisiting} compute the exact solution based on the high-order gradient, while the method requires more computing resources.

        In addition, some studies take advantage of Generative Adversarial Networks (GAN) to approximate real users behaviors for attacking, so that the generative fake users are undetectable.
        For instance, Christakopoulou and Banerjee ~\cite{christakopoulou2018adversarial,christakopoulou2019adversarial} first train a GAN on real user profiles so that the generator of the GAN can generate faker users having the same distribution as normal ones;
        then the generator's outputs  are treated  as the initialization for gradient-based attacks.
        Different from methods~\cite{christakopoulou2018adversarial,christakopoulou2019adversarial}, Lin et al. ~\cite{lin2020attacking} propose an end-to-end GAN- based attacking method  AUSH by directly training a GAN with a loss that can include attacks as well. 
        Further, Wu et al. ~\cite{wu2021triple} propose a TripleAttack method, where an extra influence module provides the guideline of the generator outputting influential fake users.

        \subsubsection{Reinforcement Learning-based Methods.} 
        The gradient-based poisoning attacks have achieved good performance under the white-box setting, which cannot be directly applied into attacking black-box recommender systems due to extremely limited knowledge towards target systems accessed by adversaries.
        Recently, some studies have leveraged Deep Reinforcement Learning (DRL) to learn attacking policy strategies via query rewards under the black-box setting.
        More specifically, the DRL based attacking process can be formulated as a  Markov Decision Process (MDP) to learn a policy $\pi(s_{t}) \rightarrow a_{t}$ for outputting the action $a_{t}$ under  state $s_{t}$.
        
        To perform attacks under black-box setting, PoisonRec ~\cite{song2020poisonrec} proposes a model-free reinforcement learning based framework for generating fake user profiles. More specifically, a Biased Complete Binary Tree (BCBT) is constructed to model the item sampling process, which can help significantly reduce the time complexity in a hierarchical  action space. 
        Moreover, in order to improve the quality of fake user profiles, Chen et al. propose a knowledge-enhanced black-box attacks for recommendations (KGAttack) ~\cite{chen2022knowledge}, which takes advantage of items' attribute features (treated as Knowledge Graph) to enhance the process of sampling items. 
        More specifically, a graph neural networks and a recurrent neural network are introduced to  model knowledge graph for enhancing state representation learning.
        Meanwhile,  in order to effectively  select items from the large-scale discrete action space (i.e., the massive item sets),  hierarchical policy networks are proposed  to decompose the selection process into two actions, including  anchor item selection and next item picking.   
        
        Furthermore, instead of generating fake user profiles from scratch,  Fan et al. propose a novel copy mechanism  to obtain real user profiles for black-box recommender systems (CopyAttack)~\cite{fan2021attacking}. In detail,   cross-domain user profiles in source domain, which can share similar online behaviors with target recommender systems, are copied into target domain for promoting a set of items. 
        However, selecting real user profiles in source domain based on reinforcement learning is challenging due to the large-scale user profiles (i.e.,  discrete action), resulting in inefficiency and ineffectiveness. To address such challenges,  CopyAttack proposes hierarchical-structure policy gradient in balanced hierarchical clustering tree over cross-domain user profiles to search a path from the root to a certain leaf of the tree, where each non-leaf node represents as a policy gradient network and a leaf node represents a user profile. In addition, masking mechanism is introduced to exclude user profiles which does include target items,  so as to further reduce the action space.  At last, a crafting policy gradient network is introduced to refine the raw user profiles, so as to decrease the attacking budgets and reduce some noise. 
        It is worth noting that KGAttack and CopyAttack use some spy users as proxy to obtain reward for optimizing the proposed DRL based attacking framework, while PoisonRec uses the number of Page View on the target item as the reward.  
    
    \subsection{Representative Defense Methods}
    Various attacking methods expose the high vulnerability of modern recommendation systems, which motivates researchers to design countermeasure strategies  against adversarial attacks. In this subsection, we introduce some representative defense methods that improve the robustness of recommendation systems. 
    Generally, there are two pathways to defend against an adversarial attack in recommendation systems: (1) \emph{Detection} methods to localize anomalies (e.g., fake user profiles); (2) \emph{Adversarial  Robust Training}  to make recommender systems more robust against adversarial attacks.
        \subsubsection{Detection.}
        In early year research, some works utilize machine learning-based classifiers, e.g., SVM and KNN, to detect anomalies and outliers in recommender systems.
        These methods train a classifier using  specific attributes of user profiles, which works well against heuristic attacks. 
        For example, Burke et al. ~\cite{burke2006classification} study generic attributes of user profiles and exploit them to conduct defense. 
        In particular, they propose three variants strategies to measure discrepancies between user's ratings and item's average ratings.
        Zhang et al. ~\cite{zhang2014hht} propose a hybrid detection method that combines SVM and Hilbert–Huang transform, 
        where Hilbert–Huang transform is used to capture spectrum-based features of series rate values of each user and  then use the features train an SVM classifier to distinguish fake users. 
        Besides, some studies  explore unsupervised learning approaches to cluster outlier data, relying on statistical attributes of the whole dataset. 
        For instance, Mehta ~\cite{mehta2007unsupervised} finds that the soft-cluster method based on Probabilistic Latent Semantics Analysis is effective to determine fake users. Bhaumik et al. ~\cite{bhaumik2011clustering} use k-means method to cluster instances and identify fake users from small clusters.

        More recently, researchers have adopted deep learning models to develop more effective defense strategies.
        Gao et al. ~\cite{gao2020shilling} propose a LSTM-based model to encode a series of user behaviors to indicate whether  user profiles are suspicious. 
        Zhang et al. ~\cite{zhang2020gcn} propose a unified framework for both recommendation and attack detection based on GNNs, which can adaptively detect fake users in the process of learning users and items representations. 
        Specifically, the detection component is proposed to dynamically adjust the users' weights for representation learning according to their probability of being fake.
        Shahrasbi et al. ~\cite{shahrasbi2020detecting} propose a semi-supervised algorithm to detect fake user profiles using SeqGAN ~\cite{yu2017seqgan} that can deal with discrete sequential data compared with vanilla GAN, which can learn the distribution of normal users' behaviors over a partial dataset that is definitely normal, so as to  identify anomalies in recommender systems.

        \subsubsection{Adversarial  Robust Training.} Adversarial robust training endows a model with the ability to tolerate adversarial perturbations instead of detecting anomalies and  outliers. 
        In  general, adversarial training contains two alternating processes: (1) generating perturbations that can confuse a recommendation model; (2) training the recommendation model along with generated perturbations. Mathematically, this process can be formulated as a min-max game as follows:
        \begin{equation}
            \min_{\theta}\max_{\eta} \mathcal{L}(\mathcal{X} + \eta, \theta),
        \end{equation}
        where $\theta$ is the recommendation model's parameters, $\eta$ indicates perturbations, and $\mathcal{X}$ denotes the original normal dataset.

        In ~\cite{he2018adversarial}, He et al. propose an adversarial training method - Adversarial Personalized Ranking (APR) to enhance the robustness of BPR based Matrix Factorization method, which  aims to perturb the embeddings of users and items by leveraging adversarial training strategy, instead of perturb raw data input. 
        Mathematically, the optimization objective can be formulated as: 
        \begin{equation}
            \underset{\theta}{\min}~\mathcal{L}_{APR}(\theta)=\underset{\theta}{\min}~\underset{\eta}{\max} ~\mathcal{L}_{BPR}(\theta)+\lambda\mathcal{L}_{BPR}(\theta+\eta),
        \end{equation}
        where $\theta=\{\bm{P},\bm{Q}\}$  denotes the parameters (i.e., users and items embeddings) in MF based recommendation methods.
        $\eta$ is perturbations added to model parameters $\theta$. They demonstrated that APR is not only an effective defensive strategy but also boosts generalization performance. 
        Further, by extending APR to multimedia recommender systems,  Tang et al. ~\cite{tang2019adversarial} propose  Adversarial Multimedia Recommendation (AMR) framework by optimizing visual-aware BPR (VBPR) objective. 
        More specifically, adversarial perturbations are incorporated to visually-aware item space extracted by CNN encoder, so as to enhance the robustness of multimedia recommendation.  In addition, by considering the robustness of tensor-based recommendations,  Chen and Li ~\cite{chen2019adversarial} incorporate adversarial training to enhance the robustness of pairwise interaction tensor factorization ~\cite{rendle2010pairwise} for context-aware recommendations.
        
    \subsection{Application}
    In this section, we introduce robustness issues in two real-world applications to demonstrate the necessity of building adversarial robust recommendation systems.
    \begin{itemize}
        \item \textbf{E-health recommendation.} 
        Recent work shows that more than 42\% of clinical misdiagnoses are caused by inadequate doctors who are unfamiliar with certain drugs ~\cite{bao2016intelligent}. 
        In order to reduce the misdiagnosis rate and the burden on doctors  in such safety-critical scenario, intelligent systems are developed to assist doctors in making clinical diagnoses and perform  drug package recommendations~\cite{zheng2022interaction}.     
        People can trust  drug recommender systems only if the systems can obtain high accuracy and resist potential attacks without any vulnerability. Thus, it is important to investigate the vulnerability of recommender systems, so as to the enhance their robustness for trustworthy recommender systems.  

        \item \textbf{E-commercial recommendation.} Online e-commercial platforms, e.g., Amazon, Taobao, etc., dominate people's daily shopping needs. The prevalence of such platforms relies on a reliable recommender system to continuously recommend products of interest to users by exploiting users' previous purchases.
        Due to the property of openness in such online services, adversaries can easily  generate fake users and  reviews to maliciously mislead people behaviors  when they shop online ~\cite{zhang2018enhancing}, which  can damage to shoppers and businesses. 
        Thus, enhancing the robustness against such perturbations is becoming more and more important for building trustworthy recommender systems.
    \end{itemize}

    \subsection{Surveys and Tools}
    In this subsection, we sort out the existing surveys on safety \& robustness in  recommender systems and a useful toolkit evaluating the robustness of recommender systems to facilitate researchers in this field.

        \subsubsection{Surveys.} The robustness of  recommendation systems has been widely studied for a long time.
        Zhang et al. ~\cite{zhang2009survey}  gives a comprehensive taxonomy about shilling attack strategies, evaluation metrics, and defense methods in recommender systems. 
        In ~\cite{gunes2014shilling}, Gunes et al. provide a summary of attacks against various collaborative filtering recommendations and detection methods. They also introduce cost/benefit analysis, a new attribute for classification shilling attacks, and discussions on future directions. 
        Similarly, Si and Li ~\cite{si2020shilling} summarize shilling attacks and defenses from their style and discuss future directions to improve the robustness of recommendation systems. 
        In ~\cite{truong2018adversarial}, Truong et al. study the effect of adversarial training on recommendation systems and analyze its properties and designs. 
        Recently, Deldjoo et al. provide surveys ~\cite{deldjoo2021survey,deldjoo2020adversarial} about adversarial machine learning in recommender systems (AML-RecSys). They review AML methods in the traditional machine learning field and further survey AML in recommendation systems from two perspectives: adversarial strategy and the GAN-based model. 

        \subsubsection{Tools.} While various toolkits have been developed for researchers to build recommendation systems and evaluate the systems' performance conveniently, there are not too many toolkits for the robustness of recommendation systems.
        The only one we can find out is RGRecSys ~\cite{ovaisi2022rgrecsys}, which allows researchers to easily evaluate recommender system robustness with respect to attacks and other dimensions.

    \subsection{Future Directions}
    Robustness in recommendation has always been an important research topic; however, many open problems and challenges are still not well explored. In this section, we point out the  potentially valuable research directions.
        The main attack and defense research are aimed at the collaborative filtering-based model and consider manipulate user-item interactions (i.e., generating faker user profiles or perturbing real user profiles). 
        In practice, modern recommendation systems can incorporate many sources data in various scenarios,  such as social connections~\cite{fan2018deep,fan2019deep_daso} and knowledge graph~\cite{chen2022knowledge}, which motivates  to develop recommender systems based on various techniques, such as reinforcement learning~\cite{zhao2018deep} and graph neural networks~\cite{derr2020epidemic,fan2021jointly}.  Thus,  an important issue is to investigate the vulnerability of different target recommender systems, so as to improve their trustworthiness from robustness aspect.
        For adversarial robust training  in defense methods, instead of raw data space, most existing methods works on parameters space by adding adversarial perturbations, which may limit the robustness improvement. Another direction is to generate adversarial perturbations on user-item interactions to perform adversarial robust training. 
\section{Non-discrimination \& Fairness}
\label{sec:fairness}

To be widely deployed in high-stakes scenarios such as finance and healthcare~\cite{veale2018fairness}, a trustworthy recommender system should avoid exhibiting discriminatory behaviors in human-machine interactions and guarantee to make fair decisions for users from certain groups. Unlike the general machine learning tasks such as classification~\cite{castelnovo2022clarification, mehrabi2021survey}, fairness in recommendation algorithms has several unique characteristics: first, multi-sided fairness needs to be considered~\cite{burke2017multisided} since  recommender systems serve users and item providers as a two-sided platform;
second, discriminatory bias might exist everywhere in the dynamic feedback loop between human and recommender systems~\cite{mansoury2020feedback} and even get amplified without appropriate interventions. 
The manifestation of bias and prejudice severely hampers the promotion of trustworthiness, and affects the long-term benefits of  recommender systems. 
For example, in a job recommendation platform like LinkedIn, if the platform exhibits gender discriminatory bias~\cite{geyik2019fairness}, e.g., women are being recommended with fewer job opportunities or lower-payment jobs compared to men, it will cause detrimental effects from both the ethical and legal aspects.
In  movie recommender systems with popularity bias,  popular movies would always be over-recommended, which will not only intensify the homogenization of users but also reduce the exposure opportunities of other equally qualified but less popular movies~\cite{abdollahpouri2021user}. To alleviate these issues, it is necessary to analyze the potential bias and mitigate the unfairness in recommender systems~\cite{ekstrand2018all, islam2021debiasing}. 

In this section, we will first introduce fundamental definitions and concepts regarding fairness in recommender systems, where we provide a detailed taxonomy of the causes of unfairness, the definitions of fairness criterion and the evaluation metrics of fairness. Then, we review and categorize existing bias mitigation methods, which can enhance the fairness performance from different perspectives in recommender systems. Lastly, we discuss the applications and future directions in this field. We hope that researchers can benefit from the broad overview of bias and fairness issues in recommender systems and reach a consensus on pushing for further advances in this field.

    \subsection{Concepts and Taxonomy}
    In this subsection, we first introduce the origins of unfairness in recommender systems, then present a taxonomy of fairness definitions and several standard fairness evaluation metrics. It is worth noting that there exists a large number of fairness definitions and evaluation metrics due to the great magnitude of related works. Therefore, we summarize the categories from several typical perspectives~\cite{li2021tutorial, wang2022survey}. 

    \subsubsection{Bias.}
    The discriminatory bias in recommender systems often leads to unfairness issues ~\cite{liu2021trustworthy, ekstrand2021fairness, ashokan2021fairness}, which means that the system unfairly treats certain individuals or protected groups by providing poorer recommendation quality. Though the sources of bias in recommender systems can be various~\cite{olteanu2019social}, we can divide the recommender systems' feedback loop into three parts from a bird's-eye view ~\cite{chen2020bias}: $\text{user} \rightarrow \text{data}$ (\textsl{data collection}), $\text{data} \rightarrow \text{recommendation model}$ (\textsl{model training}), and $\text{recommendation model} \rightarrow \text{user}$  (\textsl{model serving}), and categorize the potential bias as follows:
    \begin{itemize}
        \item \textbf{Data Bias} is the distribution difference between the collected training data and the ideal test data. It pre-exists in the data generation process~\cite{castelnovo2022clarification} and may come from many aspects. Following~\cite{chen2020bias}, data bias can be further categorized into the following four groups:
        \begin{itemize}
        
        \item \emph{Selection Bias} refers to that users' selective rating behavior~\cite{marlin2012collaborative}and the observed ratings do not fully reveal the true ratings. As a consequence, the collected data is missing not at random (MNAR).
          \item \emph{Exposure Bias} means that unobserved interactions in implicit feedback do not necessarily disclose users' disliked items since users are merely exposed to a small portion of items. 
          \item \emph{Conformity Bias} indicates that users behave similarly to other group members, even if what they do goes against their judgment~\cite{chen2020bias}.
          \item \emph{Position Bias} refers to the observation that items in the higher positions of a recommendation list are more likely to receive interaction no matter how highly relevant they are to users~\cite{chen2020bias}.
        \end{itemize}
        In addition, collected feedback data can be biased by other factors, such as \emph{marketing bias}~\cite{wan2020addressing}, indicating that consumers' interactions may be affected by the human model's profile in a product image (a reflection of a product's marketing strategy) and result in the under-representation of particular niche markets.
        
        \item \textbf{Model and Result Bias} refers to the bias in the algorithm design and model results~\cite{baeza2018bias}, in which recommendation algorithms tend to exhibit bias and generate unfair recommendation results (e.g., popularity bias), when optimizing without any fairness constraints.
        \begin{itemize}
          \item \emph{Popularity Bias} happens when popular items are over-recommended compared to what their popularity warrant~\cite{chen2020bias}.
        \end{itemize}
        
        \item \textbf{Feedback Loop Bias} refers to the reinforced bias introduced by the RS feedback loop mechanism~\cite{chouldechova2018frontiers}. Unfair recommendations would influence users' behaviors in the online serving process, which makes the observed feedback encode biases. Moreover, biased users' behavior data would enlarge the model's discrimination when collected for model training. To be specific, popular items attract a large traffic volume in recommender systems~\cite{abdollahpouri2020multi} and a biased model will provide such popular items with better recommendation quality (i.e., precision) than those unpopular items. Consequently, online serving will lead to a greater traffic volume gap between popular and unpopular items. 
     
    \end{itemize}
    There are other causes of unfairness, such as conflicts between different fairness requirements~\cite{chouldechova2017fair,kleinberg2016inherent}. In this case, fulfillment of one fairness criterion would violate some other fairness requirements. In the subsequent part, we will introduce details of different fairness definitions.
    
\subsubsection{Fairness}
    Previous works~\cite{li2021tutorial,narayanan2018translation} have presented various fairness metrics to quantify the effects of discriminatory bias in recommender systems. Though there is still no consensus on a general definition of fairness, it can usually be classified into procedural fairness and outcome fairness~\cite{wang2022survey}. 
    
   \noindent \textbf{Procedural Fairness} represents procedural justice in decision-making processes, which is critical since it affects people’s trust and cooperation of recommender systems. Most works~\cite{grgic2018beyond,lee2019procedural} mainly focus on whether the usage of input features in decision-making processes is fair. 
   For example, in~\cite{grgic2018beyond}, new scalar fairness measures are introduced to explicitly account for individuals’ moral sense of whether it is fair to use the input features in decision processes. 
    
   \noindent  \textbf{Outcome Fairness} holds that fairness-aware models ought to exhibit fair outcome performance~\cite{fang2020achieving}, which is also called as distributive fairness~\cite{grgic2018beyond}. 
   Since there are large amounts of definitions falling under this category, we group related concepts as follows:
    \begin{itemize}
        \item \textbf{Group by subject.} 
        Since recommender systems are multi-stakeholders connecting users with items,  fairness requirements from different sides need to be considered. 
        \textsl{User fairness} refers to whether different users obtain fair recommendation, such as equal recommendation accuracy~\cite{ekstrand2018all} or equal recommendation explainability~\cite{fu2020fairness}. 
        \textsl{Item fairness}, also called provider fairness, refers to whether different item groups are fairly treated, such as equal prediction errors for ratings~\cite{rastegarpanah2019fighting}, equal recommendation probabilities of different item groups given the items are truly liked by users~\cite{zhu2020measuring}, and equal ranking position that is proportional to relevance score~\cite{biega2018equity}. 
        \textsl{Joint-sided fairness} considers both users and items, such as the fairness of recommendation results' quality on the user side and the fairness of providers' exposure on the item side~\cite{wu2021tfrom,patro2020fairrec}. 
        
        \item \textbf{Group by granularity.}
        When looking into the granularity of resource allocation processes, outcome fairness can be classified as: \textsl{Group Fairness} and \textsl{Individual fairness}. Group Fairness refers to the performance parity among different sensitive groups, which generally specified by user/item sensitive attributes (i.e., gender or race))~\cite{xiao2017fairness};
        \textsl{Individual fairness}~\cite{fleisher2021s} requires that similar individuals should be treated equally~\cite{joseph2016rawlsian}. In general, individual fairness can be achieved when two similar individuals always have similar predictions in the output space. 
        
        \item \textbf{Others.}
        There are quantities of works defining fairness from other perspectives, and here we choose the following five representative definitions.
        \textsl{Causal fairness} aims to eliminate the causal relation between sensitive attributes and model predictions. 
        By incorporating additional structural knowledge regarding how variables propagate on a causal graph, causal fairness is achieved when recommendation results remain the same in the factual and counterfactual world for each possible user~\cite{li2021towards}. 
        \textsl{Personalized fairness} takes personalized demands from users into consideration, where users are free to select sensitive attributes they care about~\cite{wu2022selective}, and recommendation models provide flexible support in filtering out any chosen sensitive attributes from original feature representations. 
        \textsl{Explainable fairness} aims to explain why the recommendation model is unfair and provide insights for further improvement. For instance, in~\cite{ge2022explainable}, they develop an explainable counterfactual framework to explain which input features significantly influence the fairness-utility trade-off in recommendations. Then by alleviating the negative influence of the detected features when doing fair learning, the recommendations can achieve a better fairness-utility trade-off.
        \textsl{Rawsal max-min fairness} focuses on maximizing outcome performance of the worst individual or group~\cite{zhu2021fairness}.
        \textsl{Dynamic fairness}  requires guaranteeing fairness under dynamic factors' influence, such as users' evolving preferences or item's popularity degree change as a result of user interactions throughout the recommendation process~\cite{zhang2021recommendation}.
    \end{itemize}

\subsubsection{Fairness Evaluation Metrics.}
     Next, we present corresponding evaluation metrics for the fairness definitions mentioned above.
    
\begin{itemize}
\item \textbf{Absolute Difference} (AD)  measures utility differences between the disadvantaged group $G_0$ and the advantaged group $G_1$, which can be formulated as:  
      \begin{equation}\label{ad}
       AD=\left | u(G_0)-u(G_1) \right |, 
      \end{equation}
      where  $u(G)$ denotes as the group utility function, which is used to calculate the average rating prediction scores or the average ranking performance (i.e., NDCG or F1-score) of the group $G$.  
      A low group utility difference value indicates fair recommendation performance.

\item  \textbf{Variance} measures the performance dispersion at the group-level or individual-level~\cite{rastegarpanah2019fighting}.
It can be calculated by adding up and averaging the performance difference between any two different groups/individuals, e.g., $v_i,v_j \in \mathcal{V}$. Here, $\mathcal{V}$ represents the whole set of individuals or groups.
      \begin{equation}\label{va}
      \text{Variance}=\frac{1}{|\mathcal{V}|^{2}}\sum_{v_i\ne v_j}\left ( u(v_i)-u(v_j) \right )^ 2.
      \end{equation}
      
\item   \textbf{Min-Max Difference} (MMD) measures the difference between the maximum and the minimum score value of all allocated utilities, which can be adopted to reflect the disparity of multiple item groups' exposure opportunities~\cite{gupta2021online}.
      \begin{equation}\label{mmax}
      MMD=\max\left \{ u\left ( v \right ),\forall v\in \mathcal{V} \right\} - \min\left \{ u\left ( v \right ),\forall v\in \mathcal{V} \right\}.
      \end{equation}

\item \textbf{Entropy} usually reflects the uncertainty and disorder of a system, which can also be adopted to evaluate the inequality of item exposure opportunities in recommendations~\cite{patro2020fairrec}.
      \begin{equation}\label{entro}
       \text{Entropy} = - \sum_{v\in \mathcal{V}} p\left ( v \right ) \cdot \text{log} ~p\left ( v \right ). 
      \end{equation}

\item       \textbf{KL-Divergence} measures the difference between two probability distributions. 
This metric can be used to calculate the difference between the item groups' exposure distribution $p$ and their historical exposure $q$ in recommendations~\cite{ge2022toward}. 
A lower KL-divergence value indicates fairer recommendations. Note that the JS-divergence can be viewed as the symmetrical version of KL-divergence. 
      \begin{equation}\label{kl}
        D_{KL}\left(p,q\right)= -\sum_{v\in V}\frac{p\left(v \right )}{q\left(v \right )}. 
      \end{equation}

\end{itemize}

\subsection{Methods}
    In this subsection, we introduce some representative fair recommendation methods. 
    Based on the specific stage that these methods can be applied in the whole recommendation pipeline, we categorize existing  methods into the following three types: \textbf{\textsl{Pre-processing}, \textsl{In-processing}}, and \textbf{\textsl{Post-processing}}.

    \begin{table}[htbp]
    \centering
    \caption{Taxonomy of related methods}
    \scalebox{0.90}{
    \begin{tabular}{c|l|lllll}
    \hline
    \multicolumn{1}{l|}{\textbf{Taxonomy}} & \textbf{Method type}   & \multicolumn{5}{l}{Related research} \\ \hline
\multirow{2}{*}{Pre-processing}        & Data Re-sampling            & \multicolumn{5}{l}{\cite{ekstrand2018all}}                 \\ \cline{2-7} 
                                       & Adding Antidote Data      & \multicolumn{5}{l}{\cite{rastegarpanah2019fighting}}                 \\ \hline
\multirow{5}{*}{In-processing}         & Regularization \& Constrained Optimization         & \multicolumn{5}{l}{\cite{beutel2019fairness, yao2017beyond, xiao2017fairness, wan2020addressing, zhu2018fairness}}                 \\ \cline{2-7} 
                                       & Adversarial Learning   & \multicolumn{5}{l}{\cite{bose2019compositional,li2021leave, li2021towards, li2022fairgan, wu2021fairness, qi2022profairrec, wu2021learning}}                 \\ \cline{2-7} 
                                       & Reinforcement Learning & \multicolumn{5}{l}{\cite{ge2021towards,ge2022toward,liu2020balancing}}                 \\ \cline{2-7} 
                                       & Causal Graph           & \multicolumn{5}{l}{\cite{ge2022explainable, huang2022achieving, wu2018discrimination, zheng2021disentangling}}                 \\ \cline{2-7} 
                                       & Others                 & \multicolumn{5}{l}{\cite{farnadi2018fairness, borges2019enhancing, islam2021debiasing, li2022contextualized}}                 \\ \hline
\multirow{3}{*}{Post-processing}       & Slot-wise Re-ranking    & \multicolumn{5}{l}{\begin{tabular}[c]{@{}l@{}}\cite{karako2018using, geyik2019fairness, kaya2020ensuring, liu2019personalized, morik2020controlling,  sacharidis2019top, sato2022enumerating}\\ \cite{serbos2017fairness, sonboli2020opportunistic, steck2018calibrated,yang2021maximizing, zehlike2017fa}\end{tabular}}  \\ \cline{2-7} 
                                       & User-wise Re-ranking    & \multicolumn{5}{l}{\cite{biega2018equity, mehrotra2018towards, sarvi2022understanding, singh2018fairness}}                 \\ \cline{2-7} 
                                       & Global-wise Re-ranking  & \multicolumn{5}{l}{\cite{do2021two, fu2020fairness, li2021user, mansoury2021graph, patro2020fairrec, surer2018multistakeholder, zhu2021fairness, wu2021tfrom}}                 \\ \hline
     \end{tabular}
     }
     \end{table}

\subsubsection{Pre-processing Methods.}
     Pre-processing methods usually directly modify the training data,  aiming to remove data bias before training recommendation models.

     The advantage of these methods is their flexibility, since they are decoupled with recommender systems. 
     However, there are multiple steps between the data and the final output, which also indicates that performance gains in the pre-processing step may not be maintained by the following steps (i.e., re-ranking).
     There are two typical pre-processing methods as follows:
    
     \begin{itemize}
     \item      \textbf{Data Re-sampling.}
     Data re-sampling aims to balance data sets so that the data size of each sensitive group or individual is close.
     Ekstrand et al.~\cite{ekstrand2018all} conduct an empirical study on the effectiveness of several collaborative filtering algorithms across multiple datasets, which are stratified by the users' sensitive attributes.
     Experiment results show that different demographic groups obtain different utilities due to the imbalanced data distribution.
     Based on this observation, they propose to balance the ratio of various user groups via a re-sampling strategy and then re-train recommendation algorithms, while this approach achieves minor fairness improvement.  
    
     \item      \textbf{Adding Antidote Data.} 
     Augmenting input with additional data is another alternative for pre-processing, where the augmented data is designed to promote recommender systems' fairness and thereby can be viewed as antidote data. Rastegarpanah et al.~\cite{rastegarpanah2019fighting} design data augmentation strategies to address the unfairness issues, where additional antidote data is optimized via gradient descent methods for satisfying the fairness objective function. This method mitigates unfairness more effectively than the re-sampling method but consumes more time.
     \end{itemize}
     
\subsubsection{In-processing Methods.}  In-processing methods aim to mitigate bias in the model training process. Based on their optimization perspectives, we categorize relevant methods as follows:

    \begin{itemize}
     \item \textbf{Regularization and Constrained Optimization.} The in-processing fairness methods~\cite{yao2017beyond, beutel2019fairness, xiao2017fairness, wan2020addressing, zhu2018fairness} are primarily based on regularization and constrained optimization, where various fairness criterion are formulated as constraints or regularizers for guiding model optimization. 
     Formally, the loss function consists of a traditional recommendation loss $\mathcal{L}_{rec}$ and a fairness-related regularization loss $\mathcal{L}_{fair-reg}$ as follows: 
     \begin{equation}\label{reg_summary}
     \underset{\theta}{\min } ~ \mathcal{L}_{rec} \left ( \theta \right ) + \mathcal{L}_{fair-reg} \left ( \theta \right ).
     \end{equation}
    
    In general, this direct regularization approach is to integrate fairness metrics into the overall loss function. In~\cite{xiao2017fairness}, several fairness metrics are specifically designed for the group recommendation scenario, which can then be transformed into model regularizers and constitute a multi-objective optimization problem from the perspective of Pareto Efficiency. 
    Wan et al.~\cite{wan2020addressing} design two fairness metrics for evaluating marketing bias, where \emph{rating prediction fairness} measures the global parity of prediction errors across different user-product market segments, and \emph{product ranking fairness} measures the KL-divergence of the frequency distribution of market segments between real and predicted interactions.
    By regularizing the correlation between prediction errors and market segment distribution, the recommendation model can explicitly calibrate prediction errors' equity across different market segments. 
    In~\cite{yao2017beyond}, four new metrics are proposed to address different forms of unfairness in collaborative filtering methods. For example, the absolute unfairness metric is denoted as follows: 
       \begin{align}
     \label{reg_method1}
     U_ {abs} = \frac {1}{n}\sum_{i=1}^{n} 
     \bigg|  \left|E_{adv}[y]_{i} - E_{adv}[r]_{i}\right|- \left| E_{\neg adv}[y]_{i} - E_{\neg adv}[r]_{i}\right |  \bigg|, 
     \end{align}
    where $E_{adv}[r]_{i}$ and $E_{\neg adv}[r]_{i}$ represent the average ratings for the i-th item from the advantaged and disadvantaged user groups, $E_{adv}[y]_{i}$ and $E_{\neg adv}[y]_{i}$ represent the average predicted score for the i-th item from advantaged users and disadvantaged user groups. 
    By integrating these fairness terms into the overall learning objective function, fairness metrics can be jointly optimized with recommendation quality metrics. 
    However, some fairness metrics such as equal exposure opportunity or statistical parity are non-differentiable. 
    Thus, several methods are developed to impose indirect regularizers in recommendation models. 
    For example, a fairness-aware tensor-based recommendation method (FATR) is proposed to encourage isolating sensitive features from the original latent factor matrix by adding an orthogonal regularization term~\cite{zhu2018fairness}. 
    Beutel et al.~\cite{beutel2019fairness} propose a novel pairwise regularizer for pairwise ranking fairness, which decouples the residual between clicked and unclicked items with clicked item's group membership. The overall loss function can be formulated as:
     \begin{equation}\label{reg_indirect}
     \min _{\theta} ~\left(\sum_{(\mathbf{q}, j, y, z) \in \mathcal{D}} \mathcal{L}_{rec} \left(f_{\theta}\left(\mathbf{q}, \mathbf{v}_{j}\right),(y, z)\right)\right)+\left|\operatorname{Corr}_{\mathcal{P}}(A, B)\right|,
     \end{equation}
     where $\mathbf{q}$ is the query that contains user feature and context feature, $\mathbf{v}_{j}$ is the feature vector of item $j$, $f_{\theta}\left(\mathbf{q}, \mathbf{v}_{j}\right)$ is the prediction for query $\mathbf{q}$ on item $j$, $(y, z)$ represents the user click  feedback and post-click engagement, $\mathcal{L}_{rec}$ is the recommendation model training loss (i.e., squared error loss), $\mathcal{P}$ is a gathered dataset that includes random pairs of relevant items shown to the user and is recorded when the user clicks on one of the items, and $A,B$ represent random variables over pairs from $\mathcal{P}$.
     
     \item \textbf{Adversary Learning.} 
     The adversary learning methods are also potential choices for mitigating unfairness in recommender systems. A fair recommender system ought to make equal prediction outcomes toward different sensitive groups. In other words, model predictions should be independent from sensitive attributes. Adversary learning is hereby proposed to learn truly fair representations or predictions, from which the sensitive attribute cannot be inferred. 
     Formally, adversary learning can be formulated as a min-max optimization between the main recommendation model $\mathcal{L}_{rec} \left ( \theta \right )$ and the adversarial discriminator $\mathcal{L}_{adv}\left (\Psi \right )$ for predicting sensitive features as follows:
     \begin{equation}
        \label{reg_summary}
        \underset{\theta}{\min}\,\underset{\Psi }{\max} ~\mathcal{L}_{rec} \left ( \theta \right ) + \alpha ~\mathcal{L}_{adv}\left (\Psi \right ),     
     \end{equation}
    where $\alpha$ is the adversarial coefficient parameter that controls the trade-off between recommendation quality and fairness performance.
    A general architecture to mitigate bias in such kinds methods can be illustrated in~\autoref{fig4.1}, where sensitive attribute filters are integrated into the original recommendation model for removing sensitive information, and discriminators are additionally incorporated to predict corresponding sensitive attributes from filtered representation. 
    The learned feature representation can be viewed as fair until the discriminator fails to predict sensitive attributes from filtered representation.
    
    \begin{figure}[htbp]  
    \centering  
    \includegraphics[height=3cm, width=10.5cm]{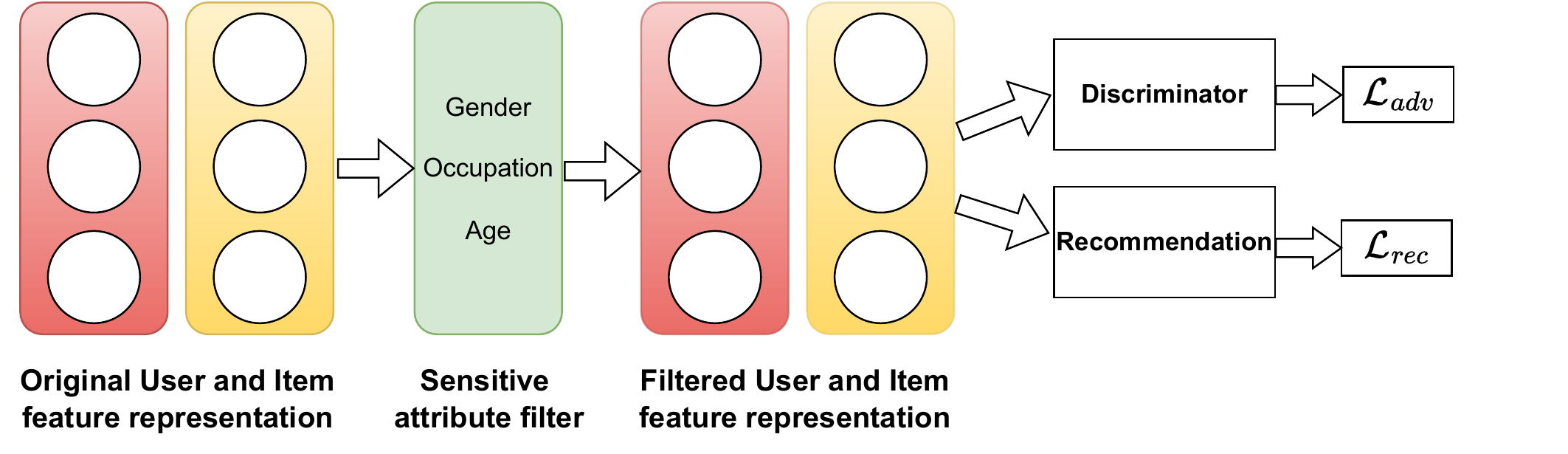}  
    \caption{General framework of fairness-aware adversarial learning in the recommendation field.}
    \label{fig4.1}
    \end{figure}
    
    In~\cite{bose2019compositional}, a set of filters are trained to generate graph node embeddings invariant to corresponding sensitive attributes.
    After training, these filters can be composed flexibly to fulfill different compositional fairness requirements. 
    Wu et al.~\cite{wu2021learning} consider that each user's embedding is composed of ego-centric graph embedding and original embedding learned from the user-item bipartite graph, and propose to utilize discriminators to regularize both embeddings. In this way, node-level fairness and ego-centric fairness can be satisfied simultaneously. 
    Li et al.~\cite{li2021towards} propose a counterfactual fair recommendation framework to generate sensitive feature-independent user embeddings via adversary learning, which naturally implements individual-level personalized fairness.
    In~\cite{wu2021fairness,qi2022profairrec}, they study user and provider fairness in news recommendation with adversarial learning methods.
    Instead of using sensitive attribute filters, they propose to learn biased user/provider embedding and bias-free user/provider embedding simultaneously and encourage them to be orthogonal to each other.
    
    Adversarial learning can also be used for fairness objectives other than learning fair representations. 
    In~\cite{zhu2018fairness}, a discriminator is designed to predict item-sensitive attributes based on the recommendation model's predicted score, which alternatively helps to learn fair prediction scores. 
    In~\cite{li2021leave}, a discriminator is incorporated to reconstruct user and item textual information from learned feature representations, which forces the representations to preserve their unique properties before being used for rating prediction, thereby reducing the unfairness between mainstream users and non-mainstream users. 
    Li et al.~\cite{li2022fairgan} propose a GAN-based model to achieve fair learning without utilizing negative implicit feedback, consisting of a ranker and a controller. Both ranker and controller include a generator and a discriminator.
    More specifically, the ranker aims to learn user preferences, and its discriminator distinguishes users' real interactions from model-generated interactions. The controller provides fairness signals to the ranker, and its discriminator identifies the generated exposure distribution from the exposure distribution calculated based on the ranker's predictions.
    
    \item \textbf{Reinforcement Learning.}
    Since the recommendation feedback loop is both dynamic and sequential, fairness issues in this dynamic process can be modeled as a Markov Decision Process (MDP), which can be addressed by reinforcement learning techniques.
    In~\cite{liu2020balancing}, a two-fold reward that measures the system's accuracy and fairness gain is newly designed, facilitating reinforcement learning with the actor-critic architecture. Specifically, time-varying recommendations are performed by an actor network, considering both the system's fairness status and user preferences; 
    the critic network estimates the output of the actor network, which can provide information about whether item groups are over-presented or under-presented. 
    Ge et al.~\cite{ge2021towards} propose a multi-objective reinforcement learning framework, where the conditioned network can seek the Pareto frontier of fairness and utility, and thereby facilitate decision-makers to control the fairness-utility trade-off. 
    In~\cite{ge2022toward}, the dynamic long-term fair recommendation is modeled as a constrained Markov decision process, where the model can dynamically adjust recommendation policy to satisfy fairness requirements when the environment changes, such as the popularity of different item groups due to users' interaction.

    \item \textbf{Causal Graph.} 
    Causal methods have recently been applied to eliminate causal effects between sensitive variables and decisions. 
    In~\cite{wu2018discrimination}, a causal graph is built to identify and remove discrimination in ranked data.
    Both direct and indirect discrimination would be removed once detected, thereby guaranteeing to reconstruct a fair ranking. In~\cite{zheng2021disentangling}, causal inference is applied to solve popularity bias in recommender systems.  In order to mitigate popularity bias's effect, a general framework DICE is proposed to disentangle user interests and popularity bias through learning interest embedding and popularity embedding separately. 
    Huang et al.~\cite{huang2022achieving} propose to incorporate causal inference into bandits for achieving counterfactual fairness for users in online recommendations. Specifically, they incorporate soft intervention to model the arm selection strategy and adopt the d-separation set identified from the underlying causal graph for developing a fair causal bandit algorithm. Such design can promote fairness by choosing arms that satisfy the fairness constraint. 
    Causality-based methods can also be used to enhance model transparency. In~\cite{ge2022explainable}, they propose to use counterfactual reasoning to explain which features can cause item exposure unfairness in recommendations. 
    
     \item \textbf{Others.} There are other methods for promoting fairness in recommendations. 
     Borges et al.~\cite{borges2019enhancing} propose randomness variational autoencoders, which incorporate randomness into the regular operation to alleviate the position bias in multiple-round recommendation. The experimental results indicate that adding noise to VAE in the latent representation sampling process can promote long-term fairness in recommendations with a tolerable trade-off between recommendation quality and fairness.
     Wu et al.~\cite{wu2022big} empirically demonstrate that big recommendation models elicit unfairness issues to cold users. Moreover, they propose a self-distillation framework called BigFair, where model predictions on original user data serve as a teacher to regularize predictions on augmented user data generated by randomly dropping historical behaviors. Experimental results indicate that BigFair can encourage big recommendation models to improve cold-start users' performance and achieve better fairness.
     \end{itemize}
     
\subsubsection{Post-processing Methods.} The post-processing methods aims to promote fairness based on recommendation models' output, which are primarily in the form of re-ranking. Existing re-ranking methods can be categorized into the subsequent three categories: 

    \begin{itemize}
      \item  \textbf{Slot-wise Re-ranking.} Slot-wise re-ranking methods are typically implemented by adding items sequentially to empty slots of a recommendation list according to specific rules or re-ranking scores. In~\cite{serbos2017fairness}, to promote fairness in the package-to-group recommendation, a greedy algorithm is proposed to add items to the package gradually by considering item category and distance constraints. In order to balance the relevance of items to group members for each prefix of the top-N, Kaya et al.~\cite{kaya2020ensuring} present a new and rank-sensitive definition of fairness (GFAR) for top-N group recommendations,  in which a greedy algorithm is designed to find top-N group recommendations according to the GFAR definition. The work of ~\cite{sato2022enumerating} propose an efficient method that enumerates all fair packages, which allows users to select their favorite packages in their own way. Liu et al.~\cite{liu2019personalized}  propose a personalized re-ranking algorithm to achieve a fair microlending recommendation system, where the objective is formulated as a combination of personalization score $P(v\mid u)$ and a fairness term, together with a trade-off controlling hyper-parameter $\lambda$ as follows: 
      \begin{equation}
        \max _{v \in R(u)} \underbrace{(1-\lambda) P(v \mid u)}_{\text {personalization }}+\underbrace{\lambda \sum_{c} P\left(\mathcal{V}_{c}\right) \mathbb{1}_{\left\{v \in \mathcal{V}_{c}\right\}} \prod_{i \in S(u)} \mathbb{1}_{\left\{i \notin \mathcal{V}_{c}\right\}},}_{\text {fairness }}
      \end{equation}
      where $R(u)$ denotes the initial ranking list, $\mathcal{V}_{c}$ represents a group of loans with attribute $c$, $P\left(\mathcal{V}_{c}\right)$ represents the importance of $\mathcal{V}_{c}$, $\prod_{i \in S(u)} \mathbb{1}_{\left\{i \notin \mathcal{V}_{c}\right\}}$ indicates the coverage of $\mathcal{V}_{c}$ for the current generated re-ranked list $S(u)$.
    
     \item  \textbf{User-wise Re-ranking.} User-wise re-ranking aims to find the most appropriate recommendation list for each user guided by the overall optimization objective. In~\cite{biega2018equity}, to implement amortized fairness, which refers to attention being fairly distributed across a series of rankings, the optimization is formalized as an integer linear programming problem and solved by an efficient heuristic algorithm Gurobi. Besides, Mehrotra et al.~\cite{mehrotra2018towards} combine fairness with recommendation utility by adopting an interpolation strategy and a probabilistic strategy to generate the candidate recommendation lists. 

     \item  \textbf{Global-wise Re-ranking.} Global-wise re-ranking aims to re-rank several recommendation lists at the same time. In~\cite{li2021user}, an integer programming-based approach is proposed for solving user unfairness problems in commercial recommendations, and the overall optimization objective is defined as follows:
     \begin{equation}
        \begin{array}{ll}
        \underset{\mathbf{W}_{i j}}{\max} & \sum_{i=1}^{n} \sum_{j=1}^{N} \mathbf{W}_{i j} \mathrm{~Y}_{i, j} \\
        \text { s.t. } & \operatorname{UGF}\left(Z_{1}, Z_{2}, \mathbf{W}\right)<\varepsilon \\
        & \sum_{j=1}^{N} \mathbf{W}_{i j}=K, \mathbf{W}_{i j} \in\{0,1\},
        \end{array}
     \end{equation}
     where $\mathbf{W}_{i j}$ represents the binary variable that indicates whether item $j$ is recommended to user $i$, $\mathrm{~Y}_{i, j}$ represents the preference score of user $i$ to item $j$, $Z_{1}$ and $Z_{2}$ refer to the advantaged and disadvantaged user groups, $\operatorname{UGF}$ represents user unfairness evaluation metric, $K$ denotes the total length of the recommendation list and $\varepsilon$ is a hyper-parameter that controls the strictness of fairness requirements.
     \end{itemize}
    
\subsection{Applications}
When recommender systems are deployed in the large-scale and resource-allocated platform, unfairness becomes a severe threat to the platform's trustworthiness. In the following subsection, we illustrate the potential unfairness in two real-world applications from different domains, which featured the necessity of building fair recommender systems.
\begin{itemize}
\item \textbf{Job Recommendation.}
 Job recommendation systems are widely deployed in the job recruitment market, such as LinkedIn~\cite{geyik2019fairness} and Indeed~\cite{sainju2021job}. These systems are expected to guarantee fair opportunities for all qualified candidate users,  since the recommendations can be viewed as social resource allocations. If recommendation models fail to meet fairness requirements, over/under-representation of specific user groups or racial/gender stereotypes in the recommendation results will inadvertently occur in practice, raising severe legal and societal issues. Therefore, building fair job recommendation systems is vital to both systems' own benefits and societal benefits.
      
\item \textbf{E-commercial Recommendation.}
 E-commercial recommendation systems, i.e., Amazon, Taobao, etc., are prevalent for their effectiveness in connecting consumers and the relevant products. Users' satisfaction and platforms' interests highly depend on the quality of the generated recommendation results. However, previous studies~\cite{li2021user} indicate that most users are disregarded by commercial recommendation engines when we categorize users into groups according to their different activity levels. Such issues also exist on the provider side~\cite{naghiaei2022cpfair}. Mitigating unfairness issues on both sides are essential to commercial recommender systems' long-term benefits. 
      
\end{itemize}

\subsection{Surveys and Tools}
In this subsection, we sort out the existing surveys and tools on fairness in recommender systems  to facilitate researchers in this field.

\subsubsection{Surveys}
    There have been growing concerns regarding fairness in recommender systems in recent years. Chen et al.~\cite{chen2020bias} gives a detailed summary of bias existing in recommender systems, provides a comprehensive taxonomy to organize current recommendation debiasing works, and discuss the strengths and weaknesses of different debiasing methods. 
    In~\cite{zehlike2021fairness}, they propose a survey that connects related approaches across various fields, aiming to motivate fairness-enhancing interventions in ranking.
    Pitoura et al.~\cite{pitoura2021fairness} provide a more technical view of definitions and methods used to guarantee fairness in rankings and recommendations, which has a much broader content coverage and structured content comparison.
    Li et al.~\cite{li2022fairness} present a comprehensive survey of the foundations for fairness. The content ranges from fairness in general machine learning tasks to more sophisticated ranking and recommender systems. In~\cite{wang2022survey}, they mainly focus on fairness in recommender systems.

\subsubsection{Tools}
    Various toolkits have been developed to evaluate or mitigate bias in machine learning models. IBM Fairness 360~\cite{bellamy2019ai} is a comprehensive tool that provides more than 70 metrics for quantifying individual or group fairness and nine bias mitigation algorithms. Furthermore, it enables metric explanations to help users understand the fairness evaluation results.
    Google What-if~\cite{wexler2019if} tool allows users to test model performance in a what-if way, where users can edit the values of data points and see their effects on model performance. In addition, the What-If tool supports comparing multiple models in the same workflow and testing several algorithmic fairness constraints.
    Fairkit-Learn~\cite{johnson2022fairkit} is an interactive python toolkit that supports developers to reason about and understand model fairness. By comparing different machine learning models concerning quality and fairness metrics, Fairkit-learn can find a model that is both high-quality and fair models. However, evaluation and analysis toolkits for recommender systems are still blank, which motivates further development.

\subsection{Future directions}

    Fairness  has attracted intensive attention for achieving trustworthy recommender systems. 
    However, many essential open problems and challenges are still not well addressed. In this subsection, we discuss some critical future directions.
    
    \begin{itemize}
    \item \textbf{Consensus on fairness definitions.}
      The fairness definition usually varies in different application scenarios, and the biases that lead to unfairness are also typically multi-sourced. Then the fairness demands can arise from multiple sides (i.e., user/provider side) and different perspectives. Some fairness definitions may even conflict with each other~\cite{kleinberg2016inherent}, such as the calibrated fairness and Rawlsian max-min fairness. Given these situations, it is vital to achieving consensus on fairness definitions. Specifically, there are three key challenges. 
      First, to enhance a recommender system's fairness, how to determine the priority of multiple fairness objectives and make an appropriate balance if there is a conflict? 
      Second, how to determine the most suitable fairness metric for a specific scenario? 
      Third, how to simultaneously incorporate multiple fairness notions into one general framework for achieving the fairness for trustworthy recommender systems? To address these questions, it is highly desired to evaluate different fairness metrics on benchmark datasets and achieve a consensus on their relationships from a unified view.

    \item \textbf{Fairness-aware algorithm design.}
      There have been extensive studies conducted on improving fairness of recommender systems. Nevertheless, it is still unclear whether the existing model design implicitly inherits or induces bias. 
      For example, graph neural networks have shown great potential in recommender systems~\cite{wu2020graph,fan2019graph}, a recent study indicates that given a biased graph topology as input, the information propagation mechanism of graph neural networks may induce bias to the node embeddings. 
      Besides, recent studies have shown that fairness and causality are closely related, and causality-based methods for mitigating bias have become a new trend~\cite{nilforoshan2022causal}. Since traditional  recommendation models tend to capture spurious associations during collaborative filtering, sensitive features may be encoded into feature representation even if not explicitly used. To solve this problem, causality-based methods for unbiased recommendation are worth exploring to address the model-induced bias.
      
    \item \textbf{Trade-off between fairness and utility}.
      Previous studies on fairness in different fields have revealed a trade-off between fairness and utility. While in the industrial recommender system~\cite{corbett2017algorithmic}, overall performance degradation is unacceptable due to the revenue loss. Therefore, extensive research needs to be conducted to figure out the trade-off mechanism so that the decision-makers can make a better balance. 
      
\end{itemize}

\section{Explainability}
\label{sec:interpret}

Recommendation with explainability, or to say explainable recommendations, refers to the recommendation algorithms focusing on providing interpretation for recommendation results. 
In fact, it represents the intersection of explainable AI and recommendation algorithms for enhancing the trustworthiness of recommender systems. 
In recent years, with the introduction of many black-box modules such as Multilayer Perceptron (MLP) \cite{li2022mlp4rec,zhou2022filter}, Transformer \cite{sun2019bert4rec,wu2020sse} and Reinforcement Learning (RL) \cite{ie2019slateq,zheng2018drn} in recommender systems, the working mechanisms of advanced recommender systems are obscure without explainability.
This problem makes the models hard to be fully trusted and to put into safety-critical applications \cite{benbasat2005trust,nilashi2016recommendation}. Therefore, building a trustworthy recommender system requires explainable recommendation modules.

This section will first introduce the concepts and taxonomy of explainable recommendations. Then, we provide detailed descriptions of some representative methods. Finally, we provide discussions on some applications and open topics in this direction.

\subsection{Concepts and Taxonomy}
In this subsection, we first introduce the basic concepts about explainable recommendations and then gives taxonomies of research on explainable recommendations.

\subsubsection{Concepts}
The \textit{explainability} of a system can be described as \textit{the ability to explain or to present in understandable terms to a human} \cite{doshi2017towards}. When such explainability exists in a recommendation model, this framework is called an explainable recommendations model. 
Explainable recommendations not only provide users with their recommendations but also give reasons why to recommend them. 
Although the research of explainable recommendations can be dated back to the 2000s \cite{bilgic2005explaining,herlocker2000explaining}, the formal analysis \cite{zhang2014explicit} and the wide attention from the academia \cite{zhang2020explainable} on it have just started recently.

\subsubsection{Taxonomy}

In this subsection, we introduce the taxonomy of research on explainable recommendations. There are two main parts: the taxonomy for models and the taxonomy for evaluations.

\noindent \textbf{Taxonomy for models.} Explainable recommendation models can be classified according to the following two criteria.
    
\begin{itemize}
    \item \emph{How to produce explanations: model-intrinsic based or post-hoc.} If a technique seeks to derive explanations from the intrinsic structure of the model, it is called model-intrinsic based explanation. Most of the current studies about explainable recommendations fall into this category, and these techniques are usually important components of relevant recommendation models. \cite{kamishima2016model}. Depending on the model structure, these techniques often pay more attention to the reasoning process of the model. In contrast, another class of techniques called post-hoc methods \cite{wang2020personalized} provide explanations based only on the inputs, outputs and extrinsic conditions of the model \cite{verma2020counterfactual}. In this case, the recommender system is treated as a completely black-box model \cite{jesus2021can}. 
    Note that post-hoc methods in recommender systems are more flexible than model-intrinsic methods, since post-hoc methods are model-agnostic and can be applied to any recommendation methods. 
    In addition, such  post-hoc explanation methods are widely utilized to explain deep recommender systems (with millions of parameters) which are too complicated to be understood.

    \item \emph{How the explanations are presented: structured or unstructured}. Structured methods present explanations in the form of logical reasoning based on some particular structures, such as a graph \cite{park2017uniwalk}, or a knowledge graph \cite{zhu2021faithfully}. The explanations they provide are highly organized and generally characterized by strong logic, good visualization, and comprehensive reasons. Unstructured methods, on the other hand, do not rely on, or explicitly rely on logical reasoning to give explanations. They tend to generate easy-to-understand sentences, ratings, or features directly from a black-box model \cite{sharma2013social}. The explanations they provide are often fragmented and structurally uncertain. These two types are not completely opposite. Generally speaking, structured methods focus more on explicit associations between users and items, while unstructured methods focus more on implicit expressions such as emotions and overall evaluations.
\end{itemize}
    
Table \ref{tab5.1} lists the different categories of explainable recommendations models with some recent representative studies. It also summarizes some common characteristics and focuses of different categories. A detailed description of explainable recommendations model techniques is given in subsection \ref{sec: methodsofmodels}.

    \begin{table}[t]
      \centering
      \caption{Taxonomy of existing explainable recommendations methods and some representative studies of different aspects.}
        \scalebox{0.9}{
        \begin{tabular}{llll}
        \toprule
              & \textbf{Model-intrinsic based} & \textbf{Post-Hoc} & \textbf{\textit{Characteristics}} \\
        \midrule
        \textbf{Structured} & \cite{chen2021temporal,xie2021explainable,fu2020fairness,xian2020neural,wang2019explainable,xian2019reinforcement}  & \cite{peake2018explanation,singh2018posthoc}  & Logical, Visible \\
        \midrule
        \textbf{Unstructured} & \cite{chen2021towards,ren2017social,chen2019co}  & \cite{li2021personalized,tan2021counterfactual,shmaryahu2020post} & Diversified, Fragmented \\
        \midrule
        \textbf{\textit{Focus}} & Model's reasoning process & Instances' relationship & -   \\
        \bottomrule
        \end{tabular}}
      \label{tab5.1}
    \end{table}

\noindent \textbf{Taxonomy for evaluations.}
Evaluations of explainable recommendations can be classified according to two main criteria. It is worth noting that most of the current model-intrinsic based explainable models consider the characteristic of explainability and the improvement of the recommendation effect. These two aspects are usually evaluated together. In this case, the improvement of the model itself sometimes can also serve as an evaluation of the explainability.
    
\begin{itemize}
    \item \emph{Evaluation perspectives: Effectiveness, Transparency and Scrutability.} 
    Explainable recommendations can be used for different purposes to produce different effects. Therefore, when evaluating explainable recommendations, different perspectives should be taken into full consideration. Tintarev et al. \cite{tintarev2011designing,tintarev2015explaining} and Balog et al. \cite{balog2020measuring} propose seven useful perspectives in the previous research \footnote{i.e. Transparency, Scrutability, Trust, Effectiveness, Persuasiveness, Efficiency and Satisfaction}. According to our summary of previous papers, we summarize that three of them: Effectiveness, Transparency and Scrutability, are the most representative perspectives that are frequently considered. The characteristics and representative papers of these perspectives have been summarized in Table \ref{tab5.2}.

    \item \emph{Evaluation form: Quantitative metrics, Case study, Real-world performance, and Ablation Study.}
    There are several evaluation forms available for explainable recommendations. According to previous studies, most of them can be divided into the following four forms: Quantitative metrics, Case study, Real-world performance, and Ablation Study. Quantitative metrics focus on the quantification of explainability in mathematics. Case study focuses on examining whether the explanation conforms to human logic. Real-world performance aims at examining the practical effects of the explanation. And ablation study intends to illustrate how algorithmic modules provide explanations and how these modules enhance the recommendation model. A detailed discussion of evaluation techniques is given in subsection \ref{sec: methodsofevas}. Research summaries of various forms of evaluations and their most corresponding perspectives are shown in Table \ref{tab5.3}.
\end{itemize}

\begin{table}[t]
  \centering
  \caption{Summary of evaluation perspectives}
    \scalebox{0.95}{
    \begin{tabular}{cp{6cm}l}
    \toprule
    \textbf{Evaluation perspective} & \textbf{Evaluation criteria} \centering  & \textbf{Related research} \\
    \midrule
    Effectiveness & Whether the explanations are useful to users? (e.g. Decision making, Recommendation results) & \cite{chen2019dynamic,takami2022educational,ai2021model} \\
    \midrule
    Transparency & Whether the explanations can reveal the working principles of the model? & \cite{li2021you,balog2019transparent,he2015trirank} \\
    \midrule
    Scrutability & Whether the explanations contribute to the prediction of the model? & \cite{wang2018tem,song2019explainable,tsai2019evaluating} \\
    \bottomrule
    \end{tabular}}
  \label{tab5.2}
\end{table}

\begin{table}[t]
  \centering
  \caption{Taxonomy of evaluation forms}
  \scalebox{0.95}
{

    \begin{tabular}{ccl}
    \toprule
    \textbf{Evaluation form} & \textbf{Corresponding perspectives} & \textbf{Related research} \\
    \midrule
    Quantitative metrics & Effectiveness; Scrutability & \cite{takami2022educational,tan2021counterfactual} \\
    \midrule
    Case study & Effectiveness; Transparency & \cite{wang2018tem,li2021you,xie2021explainable}\\
    \midrule
    Real-world performance & Effectiveness; Scrutability; Transparency & \cite{chen2019dynamic,tsai2019evaluating,xian2021ex3} \\
    \midrule
    Ablation Study & Effectiveness; Transparency & \cite{li2021personalized,song2019explainable,chen2019co} \\
    \bottomrule
    \end{tabular}}
  \label{tab5.3}
\end{table}

\subsection{Methods}\label{sec: methodsofmodels}

In this subsection, we introduce some representative methods of explainable recommendations models according to two taxonomies in Table \ref{tab5.1} and summarize some representative characteristics and focuses of these categories. These characteristics and focuses are also shown in Table \ref{tab5.1}.

\subsubsection{Different methods of producing explanations: model-intrinsic based and post-hoc}
This category is mainly defined by the underlying principle of the explainable recommendations. Based on this perspective, the explainable recommendations methods can be grouped into model-intrinsic based methods and post-hoc methods.

\begin{itemize}
    \item \emph{Model-intrinsic based Methods.} Methods in this category are also called model-intrinsic methods. These methods embed explicable ability into the recommendation model and provide explanations for the model that they are attached to by generating explanatory graphs, paths, parameters and other contents in the process of recommendation. In general, such methods are only effective for embedded models and cannot simply be reused in the recommendation process of other models. 
    For example, in Co-Attentive Multi-task Learning (CAML) \cite{chen2019co}, Chen et al. design an encoder-selector-decoder architecture for multi-task learning to realize a model that considers recommendation and explanation at the same time using transferred cross knowledge. As shown in Figure \ref{fig:model-intrinsic based}, they first encode the information and reviews of users and items into implicit factors ($h$ for information, $d$ and $D$ for reviews, $u$ for users, $v$ for items), then they feed the review implicit factors $d$ to a selector and generate new factors represented by $e$. By using different $e$ blocks and $h$ blocks in the following layers, the model can use partially shared knowledge $C$ and partially proprietary knowledge $X$ for multi-task prediction of explanation generation and rating prediction.

    \begin{figure}
        \centering
        \includegraphics[width=0.8\textwidth]{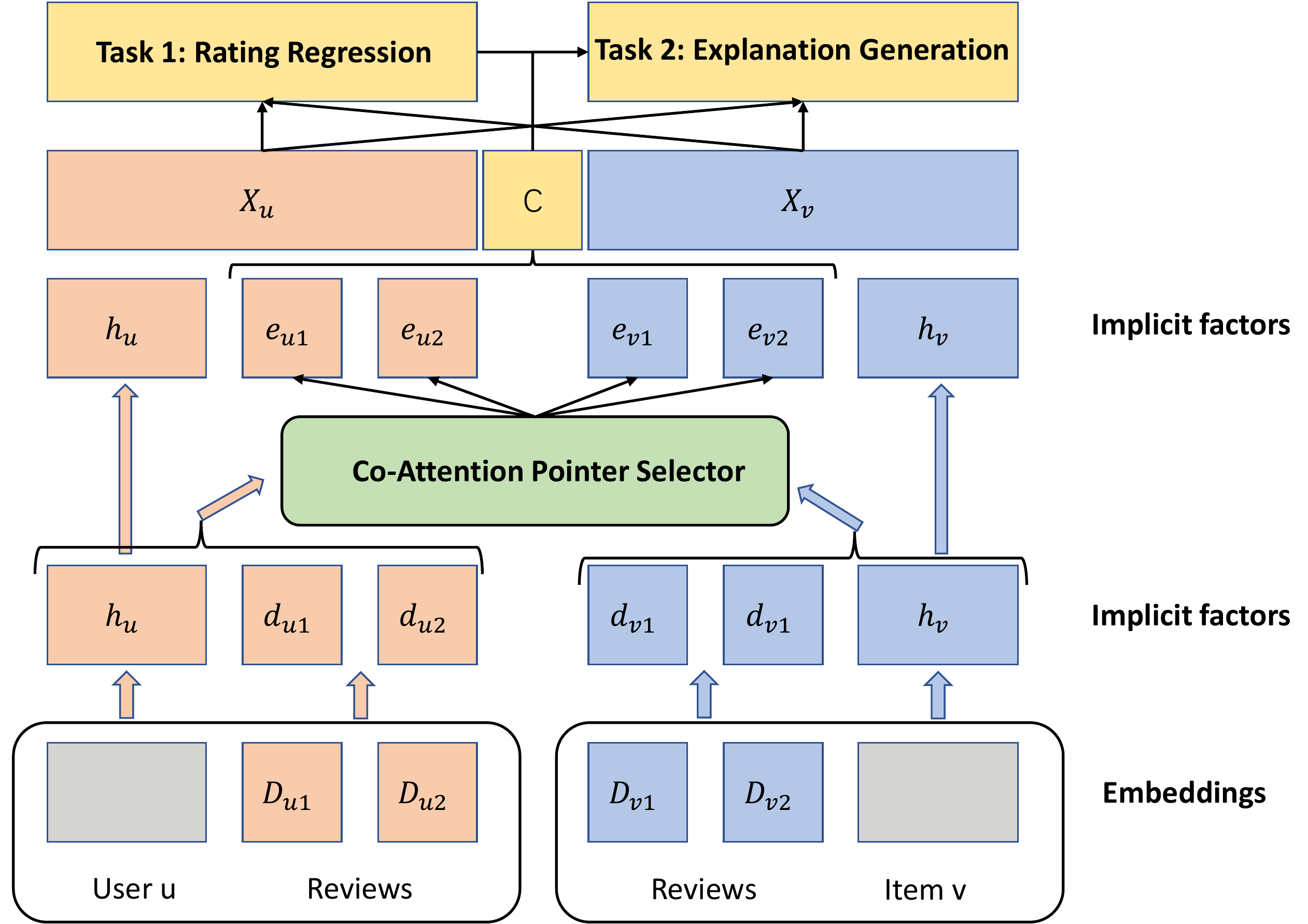}
        \caption{An example (CAML) of the model-intrinsic based methods. In such explainable recommendations, explanations are often generated along with recommendations by different tasks at the end of the model.}\label{fig:model-intrinsic based}
    \end{figure}

    Different from multi-task learning, Aspect-aware Latent Factor Model (ALFM) \cite{cheng2018aspect} and Multi-Modal based Aspect-aware Latent Factor Model (MMALFM) \cite{cheng2019mmalfm} perform explainable rating prediction by modeling aspect importance to generate aspect ratings in the latent factor model. Specifically, the user-item rating modeled according to the latent factor algorithm can be defined as: 
    \begin{equation}
        \label{equ: 5.3.1.1}
        \hat{r}_{u, i}=\sum_{a \in \mathcal{A}} \overbrace{\rho_{u, i, a}}^{\text {aspect importance}} \underbrace{r_{u, i, a}}_{\text {aspect rating}},
    \end{equation}
    where $\rho_{u, i, a}$ represents the aspect importance and $r_{u, i, a}$ represents the aspect rating for a user $u$, an item $i$ and a related aspect $a$ in aspect set $\mathcal{A}$. ALFM proposes Aspect-aware Topic Model (ATM) to model the relationship between the user, item, aspect and latent topic through probability distribution, and calculates the parameters needed by $\rho_{u, i, a}$ and $r_{u, i, a}$ according to the corpus composed of text words by ATM method. MMALFM further proposes Multi-modal Aspect-aware Topic Model (MATM) to add the input and modeling of visual words for parameter estimation. Finally, the acquired parameters related to aspects such as $\rho_{u, i, a}$ can explain the user's inclination towards objects in different aspects, and the top words about a certain aspect of a user or an item can also explain the user's inclinations or object characteristics.

    Another example comes from Neural-Symbolic explainable recommendations (NSER) \cite{xian2020neural}. In this paper, the author introduces Knowledge Graph (KG) to their model and uses neural symbolic reasoning methods to narrow down the explainable path search area in KG and then generate coarse-to-fine explanations for their recommendation results.
    Moreover, Temporal Meta-path Guided explainable recommendations (TMER) \cite{chen2021temporal} uses the dynamic KG rather than static KG technique to model the Temporal Meta-path Guided explainable recommendations. Specifically, it uses the multi-head attention module to learn the combinational features from multiple path instances:  
    \begin{equation}
        \label{equ: 5.3.1.2}
        \begin{aligned}
        &\text {Attention}\left(Q_{\phi}, K_{\phi}, V_{\phi}\right)=\operatorname{Softmax}\left(\frac{Q_{\phi} K_{\phi}^{T}}{\sqrt{d_{k}}}\right) V_{\phi}, \\
        &\operatorname{MultiHead}\left(Q_{\phi}, K_{\phi}, V_{\phi}\right)=\text {Concat}\left(\text{head}_{1}, \ldots, \text {head}_{m}\right) W,
        \end{aligned}
    \end{equation}
    where $W$ is the weight, ${ head }_{i}$ is the Attention module with dimensionality $d_{k}$ of Query $Q$, key $K$ and value $V$ related to path $\phi$. After that, the model captures sequential dependencies through time-ordered links for recommendation. Finally, it will be able to provide explanations for recommendations by aggregating multiple item-item instance paths generated by user history.

    Although the techniques used by the above methods and other methods in this class may differ greatly, the explanations they provide commonly rely on the recommendation process and serve the recommendation models. Therefore, they mainly focus on the reasoning process of the recommendation models. This characteristic often helps these methods produce more detailed explanations, but also makes them difficult to migrate to different recommendation models.
    
    \item \emph{Post-hoc Methods.} Methods in this classification are also called model-agnostic methods. Such methods regard the recommendation model as a black-box or do not even consider the recommendation model. Instead, they only use the known input and prediction information of the recommender system to train a new model similar to the system for explanation generation, or to directly generate explanations for predictions or features viewed by users. An example process of these methods is shown in Figure \ref{fig:post-hoc}. Such approaches are more general than model-intrinsic based approaches, since in most cases they require only the specific form of input and prediction data. Note that because of their versatility, these methods can also be applied to model-intrinsic based explainable recommendations models.

    \begin{figure}
        \centering
        \includegraphics[width=0.9\textwidth]{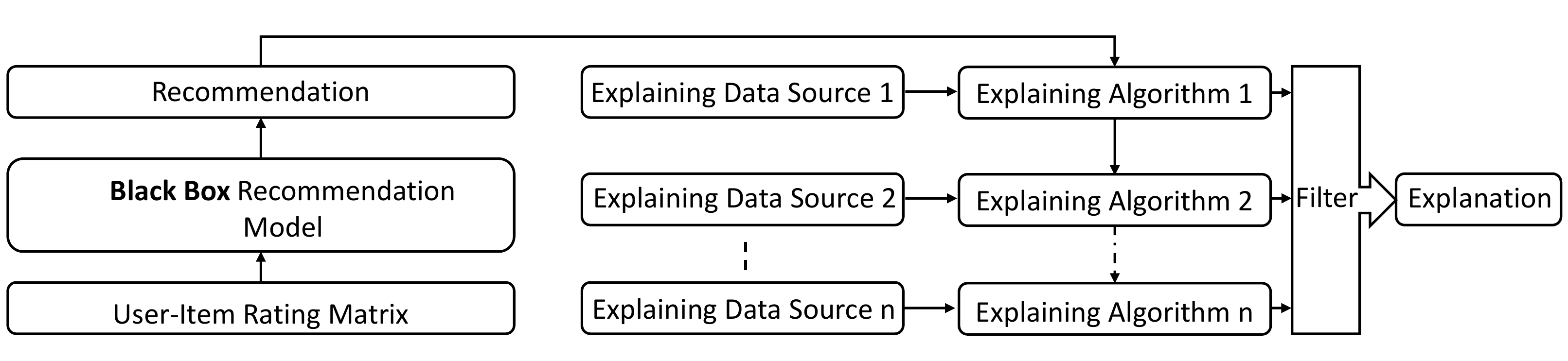}
        \caption{An example (\cite{shmaryahu2020post}) process of the post-hoc methods. In this work, the original recommendation model is treated as a black-box model. And their explainable model can generation explanations for recommendations from any recommendation model.}\label{fig:post-hoc}
    \end{figure}

    As an example, Singh and Anand \cite{singh2018posthoc} approximate the black-box ranker model used in the web search area by modeling an explainable tree-based model to obtain explanations of instances. Similarly, Shmaryahu et al. \cite{shmaryahu2020post} provide explanations by approximating complex algorithms using a set of simple explainable recommendations algorithms.
    To get better explanations, Peake and Wang \cite{peake2018explanation} treat a matrix factorization recommendation model in their paper as a black-box. Their model takes the rating prediction matrix of the recommendation model as input and tries to extract association rules for explanation generation while approximating the black-box model.

    Another example comes from Ai et al. \cite{ai2018learning}, where they propose to provide personalized recommendation using KG embedding-enhanced Collaborative Filtering (CF) methods, and try to generate explanations using similarity matching between the user and item embeddings. 
    Mathematically, they generate explanation paths with probability $P$ for an arbitrary user $e_{u}$ and item $e_{i}$ with relation sets $R_{\alpha}$, $R_{\beta}$ and the aggregation operation represented by $trans(\cdot)$ through any intermediate $e_{x}$, as follows: 
     \begin{equation}
        \label{equ: 5.3.1.3}
        P\left(e_{x} \mid e_{u}, R_{\alpha}, e_{i}, R_{\beta}\right)=P\left(e_{x} \mid \operatorname{trans}\left(e_{u}, R_{\alpha}\right)\right) P\left(e_{x} \mid \operatorname{trans}\left(e_{i}, R_{\beta}\right)\right). 
    \end{equation}
    After that, they choose the best path with top probability for final explanation. Since the explanation is independent of the model reasoning process, this approach is also considered as a post-hoc approach.

    In general, most of these methods obtain explanations by approximating simple explainable models to complex models, or by summarizing inputs, outputs and visible features using different techniques. Since they are mostly independent of specific recommendation models, they can generally generate explanations for different recommendation algorithms. However, since they cannot fully obtain the internal reasoning logic of the recommender system, most of the explanations generated by post-hoc methods rely on the association between instances, and the explanation effect is commonly weaker than model-intrinsic based methods.
\end{itemize}

\subsubsection{Different methods of presenting explanations: structured and unstructured}

This category is mainly defined by the presentation of explanation. Based on this perspective, explainable recommendations methods can be divided into structured methods and unstructured methods.

\begin{itemize}
    \item \emph{Structured Methods.} These approaches mainly provide structured explanations with strong logic, and the explanation logic is often visible. To generate structured explanations, these methods usually use structured data as the main part of their input. The most representative technique of this category is explainable recommendations based on knowledge graph \cite{xian2019reinforcement,wang2019explainable,zhao2020leveraging,wang2022multi}, and the explanations provided by this kind of explainable recommendations method mainly include explanation path graph \cite{xian2019reinforcement,wang2019explainable,zhao2020leveraging}, association rules \cite{peake2018explanation}, decision tree \cite{singh2018posthoc}, etc.

    \begin{figure}
        \centering
        \includegraphics[width=0.85\textwidth]{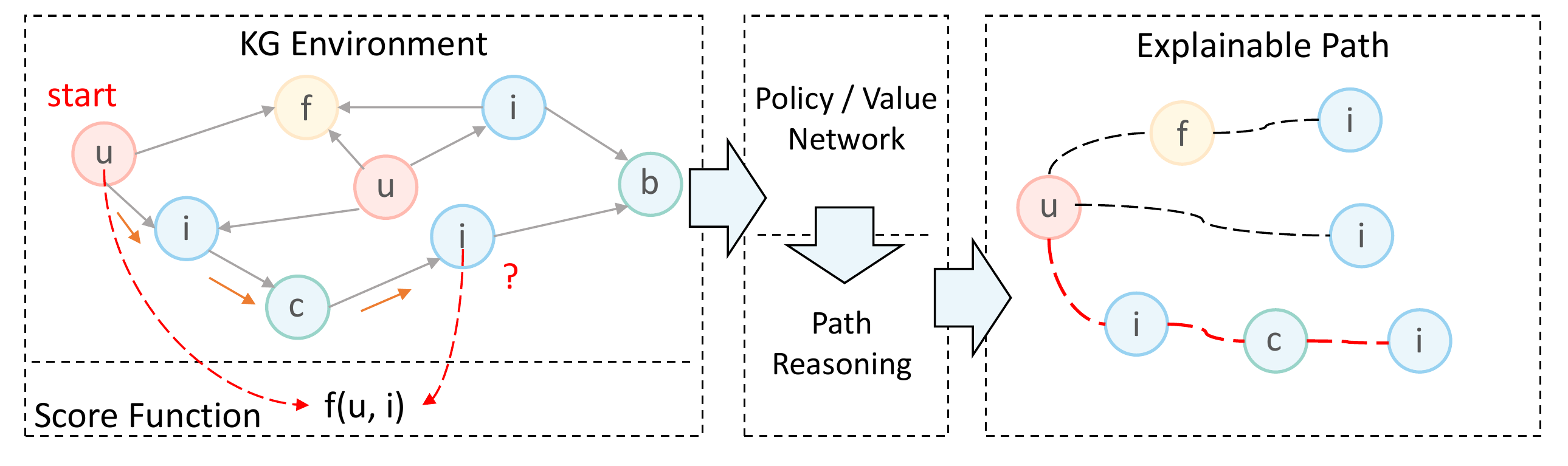}
        \caption{An example (PGPR) of the reasoning path for structured methods. $\text{u}$ for users, $\text{i}$ for items, $\text{b}$ for brands, $\text{c}$ for categories, and $\text{f}$ for features in the KG environment.}\label{fig:structured}
    \end{figure}

    One of the most representative works in this category is Policy-Guided Path Reasoning (PGPR) \cite{xian2019reinforcement} whose structure is shown in Figure \ref{fig:structured}. In this model, a path $p_{k}$ with $k$ entities in KG is defined as $p_{k}\left(e_{0}, e_{k}\right)=\left\{e_{0} \stackrel{r_{1}}{\leftrightarrow} e_{1} \stackrel{r_{2}}{\leftrightarrow} \cdots \stackrel{r_{k}}{\leftrightarrow} e_{k}\right\}$, where $e_{i}$ represents entity $i$ and $r_{i}$ represents relation $i$. This model simulates path reasoning in KG with Markov Decision Process (MDP) with reward function as follows:  
    \begin{equation}
        \label{equ: 5.3.2.1}
        R_{T}= \begin{cases}\max \left(0, \frac{f\left(u, e_{T}\right)}{\max _{i \in I} f(u, i)}\right), & \text { if } e_{T} \in \mathcal{I} \\ 0, & \text { otherwise }\end{cases}
    \end{equation}
    where $f(\cdot)$ is a scoring function. $u$ is a user,  $i$ is an item of item set $I$,  and  $e_{T}$ is the terminal entity.
    Then, REINFORCE algorithm \cite{williams1992simple} is used to learn a path finding policy $\pi\left(\cdot \mid \mathbf{s}, \tilde{\mathbf{A}}_{u}\right)$, where $\tilde{\mathbf{A}}_{u}$ is a binarized vector of pruned action space for user ${u}$ and ${s}$ is the current state. Finally, it can couples recommendation and explainability by providing paths in KG using beam search.

    Another representative work is Knowledge-aware Path Recurrent Network (KPRN) \cite{wang2019explainable}. In this study, the path can also be represented as $p_{k}\left(e_{0}, e_{k}\right)=\left\{e_{0} \stackrel{r_{1}}{\leftrightarrow} e_{1} \stackrel{r_{2}}{\leftrightarrow} \cdots \stackrel{r_{k}}{\leftrightarrow} e_{k}\right\}$. Differently, KPRN aggregates entity, entity type, and relation type pairs for each possible path and enters them sequentially into a Long Short-Term Memory (LSTM) model to predict a score for each path. Ultimately, these paths provide explanations for the recommendation based on their contribution scores.

    Moreover, in ADversarial Actor-Critic (ADAC) \cite{zhao2020leveraging}, to model imperfect paths and relations in KG more efficiently and improve the persuasion of path explainability, an actor-critic model is used to search for persuasive paths for final explanation. In the Reinforcement learning framework for Multi-level recommendation Reasoning (ReMR) \cite{wang2022multi}, to solve the lack of explainability of high-level abstract categories in previous knowledge graph reasoning processes, Wang et al. use their Cascading Actor-Critic module-based multi-level reasoning over KGs to better infer and represent user interests. Different from using KG to generate explanations during the recommendation process, Singh and Anand \cite{singh2018posthoc} try to train a structured tree-based model for post-hoc explanations of learning-to-rank-models, which are widely used in most information retrieval and recommender systems.

    In general, the explanations provided by structured methods usually tend to have  inference paths or rules, which makes them highly logical. This feature also partly limits the use scenarios of structured methods, because in most cases they need a large number of structured data with many features or connections between users and items to provide the information that they need for reasoning.
    	
    \item \emph{Unstructured Methods.} Unlike structured approaches,  unstructured methods do not require explanations to be organized, and thus have few restrictions on how explanations can be presented, as long as they are intuitive to human beings. Because of the loose restrictions on interpretive logic, these approaches can focus on different aspects of a problem and their inputs are diversified. The most representative technique of unstructured methods is the sentence generation model based on RNNs or Transformer \cite{chen2019co,li2021personalized}, and the explanations provided by this kind of method mainly include sentences, scores, important features, etc.
    An example of sentence generation for unstructured methods (PETER~\cite{li2021personalized}) is illustrated in Figure \ref{fig:unstructured}.

    \begin{figure}
        \centering
        \includegraphics[width=1\textwidth]{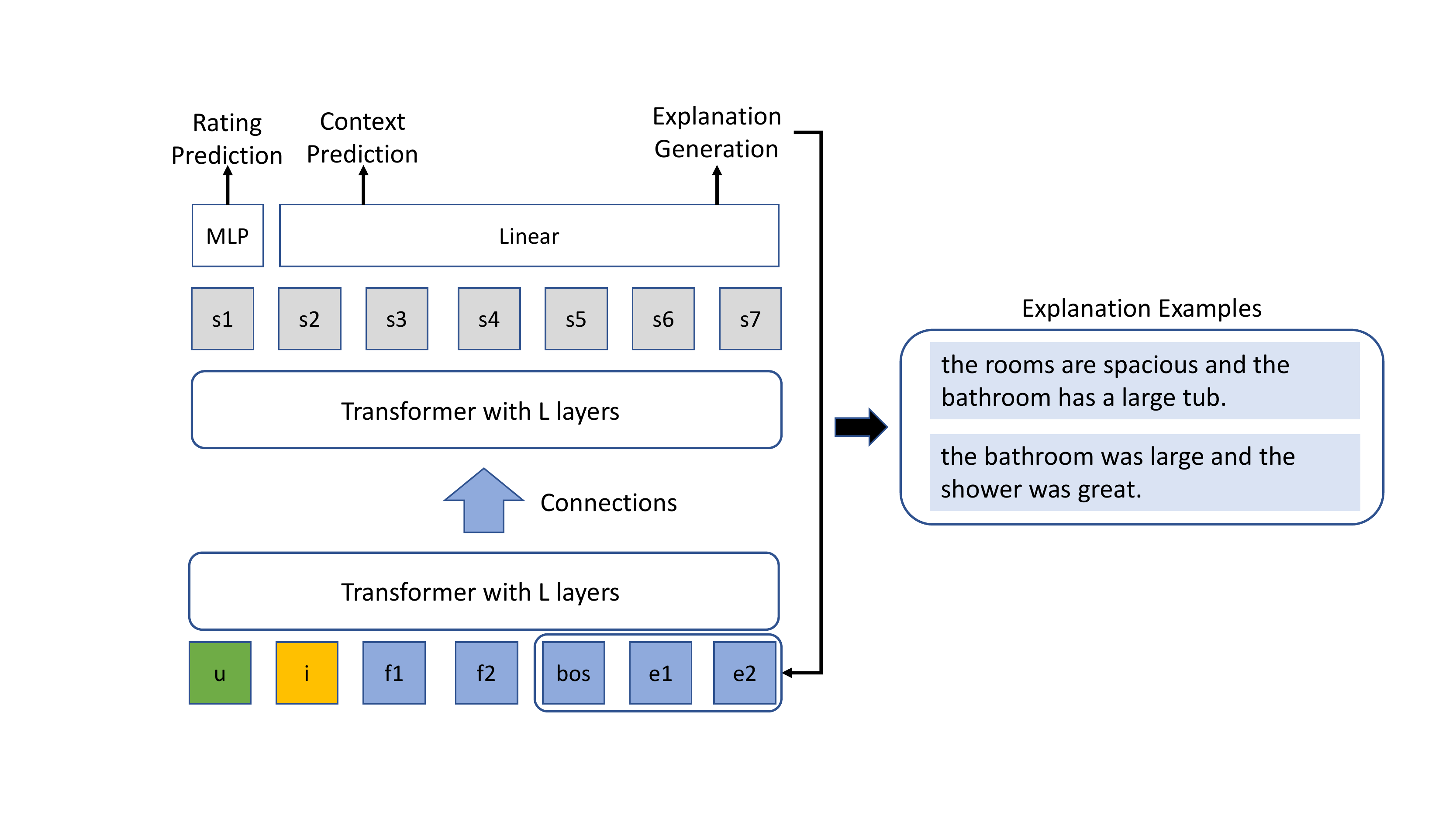}
        \caption{An example (PETER) of sentence generation for unstructured methods. $\text{u}$ for users, $\text{i}$ for items, $\text{f1}$, $\text{f2}$ for features, $\text{bos}$, $\text{e1}$ and $\text{e2}$ for explanation words where $\text{bos}$ is the ``start" representation.}\label{fig:unstructured}
    \end{figure}

    One work using unstructured method comes from Explainable Conversational Recommendation (ECR) \cite{chen2021towards} for explainability in conversational recommendation. This model uses incremental multitask learning and conversation feedback to improve the effectiveness of both recommendation and explanation results from conversational recommendations. The pre-trained loss function constructed in this model is designed as:
    \begin{equation}
        \label{equ: 5.3.2.3}
        \mathcal{L}^{\Omega}=\sum_{u \in U}\left(\mathcal{L}_{r}^{\Omega_{u}}+\lambda_{n} \mathcal{L}_{n}^{\Omega_{u}}+\lambda_{c} \mathcal{L}_{c}^{\Omega_{u}}\right)+\lambda_{\theta}\|\Theta\|_{2}^{2}
    \end{equation}
    where $\mathcal{L}_{r}^{\Omega_{u}}$ is Factorization Machine (FM) \cite{rendle2010factorization} based BPR loss \cite{rendle2012bpr}, $\mathcal{L}_{n}^{\Omega_{u}}$ is the negative log-likelihood loss of two Gated Recurrent Unit models (GRUs), $\mathcal{L}_{c}^{\Omega_{u}}$ is the concept relevance loss \cite{chen2019co}, and $\Theta$ represents the model parameters. $\lambda$ represents the weight. 
    In this loss function, $\mathcal{L}_{n}^{\Omega_{u}}$ and $\mathcal{L}_{c}^{\Omega_{u}}$ are highly related to the generation task of explanation. Finally, the model can make reliable explainable recommendations in the form of conversations. 
    Another example comes from PErsonalized Transformer for explainable recommendations (PETER) \cite{li2021personalized}, which also uses multitask learning for explainable recommendations. Differently, this model mainly uses Transformer as its core module and follows a linear layer to generate personalized explanations for different IDs.

    Different from  explanations from sentences, Counterfactual explainable recommendations (CountER) \cite{tan2021counterfactual} attempts to construct counterfactual items for recommended items to provide explanations. Specifically, this model tries to use small changes in item aspects to reverse the decision. These small changes in item aspects constitute the explanation of the recommended item. The mathematical representation of the method can be written as follows: $Minimize~Explanation~Complexity, ~s.t.,~Explanation~Strength~is~higher~enough$. Here Explanation Complexity and Strength should be predefined  according to the properties of these concepts.

    Generally, the explanations provided by this kind of method are mostly applied in the directions of comments, conversations, sentence generation and other fragmented explanation generations. Their forms are diversified, and the content is often fragmented. Therefore, such models are suitable for a wide variety of data. However, since they usually focus on a few aspects of the data and lack clear reasoning logic, the validity of the explanations they generate depends more on human experts' intuitive evaluation.
    
\end{itemize}

\subsection{Methods of Evaluations}\label{sec: methodsofevas}

In this subsection, we introduce some representative methods of explainability evaluations   in recommendations, which can be  summarized  in Table \ref{tab5.2} and Table \ref{tab5.3}. 

\subsubsection{Quantitative Metrics} 
To better evaluate the explanation, many quantitative metrics \cite{lin2004rouge,papineni2002bleu,li2020generate,li2020generate,liu2020explainable} have been designed to numerically approximate the general evaluation of some aspects of the explanations ~\cite{li2017neural,lin2019explainable,kang2019recommendation}. Most of these evaluations are designed for natural language generation (NLG) tasks \cite{chen2022measuring}.
Since sentences can be split into words, it is convenient for the metrics to perform mathematical association matching and numerical accumulation operations. 
The most common quantitative metrics are ROUGE score \cite{lin2004rouge} and BLEU metric \cite{papineni2002bleu} as follows:
\begin{itemize}
\item  ROUGE score:
\begin{equation}
    \label{equ: 5.4.1.1}
    \text{Rouge-N}=\frac{\sum_{S \in \text{Reference Summaries}} \sum_{gram_n \in S} Count_{match }\left(gram_n\right)}{\sum_{S \in \text{Reference Summaries}} \sum_{gram_n \in S} Count\left(gram_n\right)},
\end{equation}
where $n$ represents the length of the n-gram, $gram_n$ is the maximum number of n-grams co-occurring in a candidate summary and $Count_{match}(gram_n)$ is a set of reference summaries.

\item BLEU metric:
\begin{equation}
    \label{equ: 5.4.1}
    \mathrm{BLEU}=\mathrm{BP} \cdot \exp \left(\sum_{n=1}^{N} \frac{1}{N} \log p_{n}\right),
\end{equation}
where $BP$ represents the brevity penalty. $p_{n}$ represents the average value of the modified n-gram precision.  $N$ represents the length of a sentence. 

\end{itemize}

In addition, there are also some other quantitative matrices such as Unique Sentence Ratio (USR) \cite{li2020generate}, Feature Matching Ratio (FMR) \cite{li2020generate}, shift scores \cite{liu2020explainable}, etc.

\subsubsection{Case Study} 
This approach provides examples to illustrate explanations and even part of the explanation generation process~\cite{xian2020cafe, chen2020try,chen2019personalized},  so that people can intuitively judge their validity. Specifically,  researchers usually lists one or several examples of explanations, and summarizes the advantages of the proposed explanations by comparing them with each other or with explanations generated by other models. 
For most structured explanations, researchers can further verify the logic of the explanation by showing the detailed features of the explanation or the path of reasoning. Since explanation is oriented towards people's actual judgment, which includes logic, experience, intuition and other factors, providing concrete examples for people can help improve the credibility of an explanation.  
 
\subsubsection{Real-world Performance} This approach focuses on collecting people's actual feedback or carrying out online experiments for evaluation. This approach is usually evaluated by running the model online and gathering feedback, or by recruiting annotators to carefully evaluate the explanations~\cite{chen2019personalized, xian2021ex3}. 
This approach focuses on the practical effects of the explanation, i.e., whether the explanation has a positive impact on the recommendation in the application. Moreover, the different evaluation details will further reflect different perspectives of the explanation.

\subsubsection{Ablation Study} This approach is often used to analyze the role of different modules in recommender systems ~\cite{li2021personalized,chen2019co}.
Specifically, the researchers verify the impact of important modules on the overall model by removing some modules or drastically adjusting some parameters. In explainable recommendations, this approach can analyze the specific contributions that explainable modules bring to the model, or specific modules that contribute most to generating the explanation.

\subsection{Applications}

The explainable recommendations have been widely applied in various scenarios in our daily lives.

\begin{itemize}
\item \textbf{E-commercial Recommendation.}
The main purpose of explainable recommendations in various e-commerce platforms is to recommend different products to each customer for justifiable reasons, so as to help improve the attractiveness of the recommended products to customers and thus to stimulate consumer confidence.
For example, Zhang et al. \cite{zhang2014explicit} use a big commercial e-commerce platform JD.com to evaluate the effect of their explainable recommendations in real situations. In \cite{chen2019personalized}, Chen et al. combine visual and textual features to provide visual explanations for fashion recommendations on platforms including Amazon.

\item \textbf{Social Media.} 
In  social media such as Facebook and TripAdvisor,  recommendations with certain explanations can be used to strengthen people's understanding of the recommended content, serving people's life and friendship.  
For example, explainable Point-of-Interest (POI) recommendations are designed to discover places of interest and provide explanations for people~\cite{zhao2015sar,wang2018tem}.
Zhao et al. \cite{zhao2015sar} propose a joint sentiment-aspect-region model based on the Yelp dataset. This model could investigate whether people like a certain aspect of a place. Wang et al. \cite{wang2018tem} propose a tree-enhanced embedding method for transparent and explainable recommendations in tourist attractions and restaurant recommendations.

\end{itemize}

\subsection{Surveys and Tools}
In this subsection, we describe the existing surveys on explainability in recommender systems and tools on explainability in AI to facilitate researchers in this field.

\subsubsection{Surveys}
There have been growing concerns regarding explainability in the recommender system in recent years. Zhang et al. \cite{zhang2020explainable} provide a detailed survey of explainable recommendations, and distinguish explainable recommendation models with different algorithmic techniques and explanatory forms.
Tintare and Masthoff et al. \cite{tintarev2011designing,tintarev2015explaining} provide seven useful perspectives for evaluating explanations in the recommender system.
Balog and Radlinski \cite{balog2020measuring}  discuss the relevance of the seven perspectives and develop two novel explainable evaluation metrics.
Moreover, Chen et al. \cite{chen2022measuring} focus on the evaluation part of explainability in the recommender system, and provide main evaluation perspectives for different evaluation methods.

\subsubsection{Tools}
In this section, we provide commonly-used tools for explainable models. AIX360 \cite{arya2020ai} is a useful open-source toolkit for explainable models and evaluation metrics in Python environment. In addition, Quantus \cite{hedstrom2022quantus} provides guidance for the evaluation of explainable methods and a comprehensive set of evaluation metrics. For deployment, XAITK \cite{hu2021xaitk} provides an open-source collection of explainable AI tools and resources to meet the critical needs of deploying and testing explainable AI systems. It is worth noting that the above tools are mostly oriented to general explainable AI models and methods. Currently, the toolkit for explainable recommendations is still limited, which is also one of the future research directions.

\subsection{Future Directions}

In this subsection, we discuss future directions for explainable recommendations.

\begin{itemize}

\item \textbf{Natural Language Generation for Explanation.}
Most existing explainable recommendations  aim to generate sentence explanations based on predefined templates. 
However, a more user-friendly explanation could be a natural sentence that is automatically generated. Recently, there have been several studies that have attempted to provide explanations using natural language generation methods \cite{chen2019co, li2021personalized}. Although they have achieved considerable results, there is still a gap between them and the ideal detailed and natural language explanation, which is one of the important directions of explainable recommendations in the future.

\item \textbf{Explainable recommendations in more fields.}
In addition to methodology, another potential future direction for explainable recommendations is development in more fields, such as medical care and education. At present, explainable recommendations are concentrated mainly in e-commerce and social media, while relevant research in academic support, medical care, education and other fields is still limited. 
Fortunately, some researchers have noticed the potential value of explainable recommendations in these fields\cite{jin2022explainable, takami2022educational}. In the future, the promotion and development of explainable recommendations in these fields will better benefit people's lives.

\end{itemize}

\section{Privacy}
\label{sec:privacy}

Human beings have entered the era of big data, in which data can effectively characterize users' profiles via online behaviors (e.g.,  browsing history and online shopping) and inevitably contain users' private information (e.g., email addresses and gender)~\cite{jainBigDataPrivacy2016, huangPrivacyProtectionRecommendation2019}.
Modern recommender systems, especially deep learning-based strategies, heavily rely on big data and even private data to train algorithms for obtaining high-quality recommendation performance~\cite{liu2021trustworthy,aghasianUserPrivacyRecommendation2018}.
This raises huge concerns about the safety of private and sensitive data when recommendation algorithms are applied to safety-critical tasks such as finance and healthcare.
To build trustworthy recommender systems, protecting data privacy has become increasingly important. Therefore, it is necessary to investigate how to perform privacy attack methods to steal knowledge from the target recommender systems, and then develop privacy-preserving countermeasures to protect data privacy.

In this section, we will focus on privacy attacks and their corresponding strategies with regard to privacy protection for the trustworthiness of recommender systems. 
We first give some brief concepts and the taxonomy of privacy attacks and privacy protection. Then,  representative methods of attacks and privacy-preserving methods on recommender systems are detained, followed by some surveys and tools related to the privacy of recommender systems. 
At last, we introduce some real-world applications and describe future directions to explore in achieving trustworthy recommender systems.

    \subsection{Concepts and Taxonomy}\label{con_and_tax}
    This subsection briefly introduces the widely-received concepts in trustworthy recommender systems, specifically focusing on privacy attacks and privacy-preserving.

        \subsubsection{Privacy Attacks}\label{attacks-definition}
         In recommender systems, privacy attacks aim to steal knowledge that is not intended to be shared, such as the sensitive information of users and model parameters.
         Depending on how much knowledge attackers know about target victim recommender systems, privacy attack methods can be classified into white-box attacks, black-box attacks, and grey-box attacks.
         More specifically, in the white-box setting, attackers are allowed to access all the information about recommender systems, such as the model's architecture, parameters, gradients, training data, etc. 
         In contrast to the white-box attack, attackers in the black-box setting could have access to minimal knowledge about the victim model.
         Grey-box attacks, a combination of white-box and black-box attacks, are able to access partial knowledge about the target recommender system, such as users' public reviews and attributes on Amazon.
         
         Moreover, the attack methods can also be divided into four categories based on the information stolen by the attacker: membership inference attacks, property inference attacks, reconstruction attacks, and model extraction attacks.
            \begin{itemize}
            \item \textbf{Membership Inference Attacks} (MIA) aim to identity whether the target user is used to \emph{train} the target recommender system. 
            When certain training data is known to the attacker, it can result in a privacy breach.
            For example, an intelligent system in the healthcare domain recommends treatments for patients with schizophrenia. If an attacker knows a certain person in the training set for building the intelligent system, it is likely to infer that this person is suffering from  schizophrenia~\cite{shokriMembershipInferenceAttacks2017}, where the patient's information is sensitive and may not intend to be published.

            \item \textbf{Property Inference Attacks} (PIA) aim at stealing global properties of the training data in the target recommender system. These properties are usually not directly reflected in any specific user but in the global features of the training set, such as the gender ratio and the total amount of Tiktok users~\cite{heSecurityThreatsDeep2019a}. 
            
            \item \textbf{Reconstruction Attacks} (RA), or Model Inversion Attacks, aim to infer private information or labels on training data. Unlike property inference attacks, RA focuses on the sensitive properties of users, such as a user's gender or occupation in Facebook~\footnote{https://www.businessinsider.com/stolen-data-of-533-million-facebook-users-leaked-online-2021-4}.
            
            \item \textbf{Model Extraction Attacks} (MEA), also known as Model Stealing Attacks ~\cite{zhangThiefBewareWhat2021}, aims to steal the parameters and structure of a target model by creating a new replacement model that behaves similarly to the target model ~\cite{rigakiSurveyPrivacyAttacks2021}. After that, a successful model extraction can transform the setting into a more manageable white-box attack ~\cite{liu2021trustworthy}.
            \end{itemize}
            
        \subsubsection{Privacy Preserving}\label{defnition_privacy_preserving}
            In order to defend against privacy attacks,  privacy-preserving methods have been proposed based on different strategies, which can be broadly divided into five categories: differential privacy, federated learning, adversarial learning, and anonymization \& encryption.

            \begin{itemize}
            \item \textbf{Differential Privacy} (DP) is a common way to preserve users' privacy, which can provide strict statistical guarantees for data privacy ~\cite{dworkCalibratingNoiseSensitivity2006}. 
            The main idea is to add random noise into data to protect actual data while preserving recommendation accuracy ~\cite{dworkAlgorithmicFoundationsDifferential2014, mullner2022reuseknn}.

            \item \textbf{Federated Learning} (FL) keeps the training data and recommendation models decentralized, which isolates users' data and the cloud server by only transferring the parameters between them. 
            Thus, it avoids privacy leakage during the data transfer ~\cite{zhangTrustworthyGraphNeural2022,yangFederatedMachineLearning2019}.
            Specifically, every client uses their data to train a decentralized recommender system locally. Then, the server collects these models' parameters and aggregates them into a new set of recommendation parameters for update~\cite{liFederatedLearningChallenges2020, kairouzAdvancesOpenProblems2021}. The mathematical model can be defined as follows:
            \begin{equation}\label{FL}
            \mathop {\min }\limits_\theta  \mathcal {L} (\theta ) = \mathop {\min }\limits_\theta  \sum\limits_{k = 1}^n {{p_k}{\mathcal {L} _k}(\theta )},
            \end{equation}
            where $\theta$ is the global recommender system's parameter, $n$ is the number of the decentralized devices, $p_k$ and $\mathcal {L}_k$ represent the weight and the loss function on the $k$-th device, respectively.

            \item \textbf{Adversarial Learning} (AL) is a relatively general method for privacy-preserving recommendations, which can be formulated as the minimax simultaneous optimization of recommendation and privacy attacker models~\cite{huangAdversarialMachineLearning2011}.
             The mathematical model can be formulated as follows:
            \begin{equation}\label{Adversarial}
            \mathop {\min }\limits_\theta  \mathop {\max }\limits_\Psi  {{\mathcal {L}}_{rec}}(\theta ) - \alpha {{\mathcal {L}}_{adv}}(\Psi ),
            \end{equation}
            where $\theta$ and $\Psi$ represent the recommender system's parameters and adversary model parameters, respectively, maximizing the loss of adversarial could enhance the attack ability. The minimization of learning loss could provide better performance. $\alpha$ is a hyperparameter to trade-off the contribution of adversarial to the training.
            
            \item \textbf{Anonymization \& Encryption.} Both of them protect the user's privacy by adding noise. Anonymization ~\cite{sweeneyKANONYMITYMODELPROTECTING2002, machanavajjhalaLdiversityPrivacyKanonymity2006} obscures the privacy attributes of users. Then make it impossible to correlate the privacy attributes with individual identities of people.
            Encryption techniques prevent people who do not have the authorization from any useful information  ~\cite{ghadirliOverviewEncryptionAlgorithms2019, yuPrivacyPreservingMultiTaskFramework2020}.
            \end{itemize}
            
    \subsection{Methods}\label{methods}
    
    \subsubsection{Privacy Attacks}
\begin{figure}[htbp]
        \centering
        \includegraphics[width=12cm,height=3.2cm]{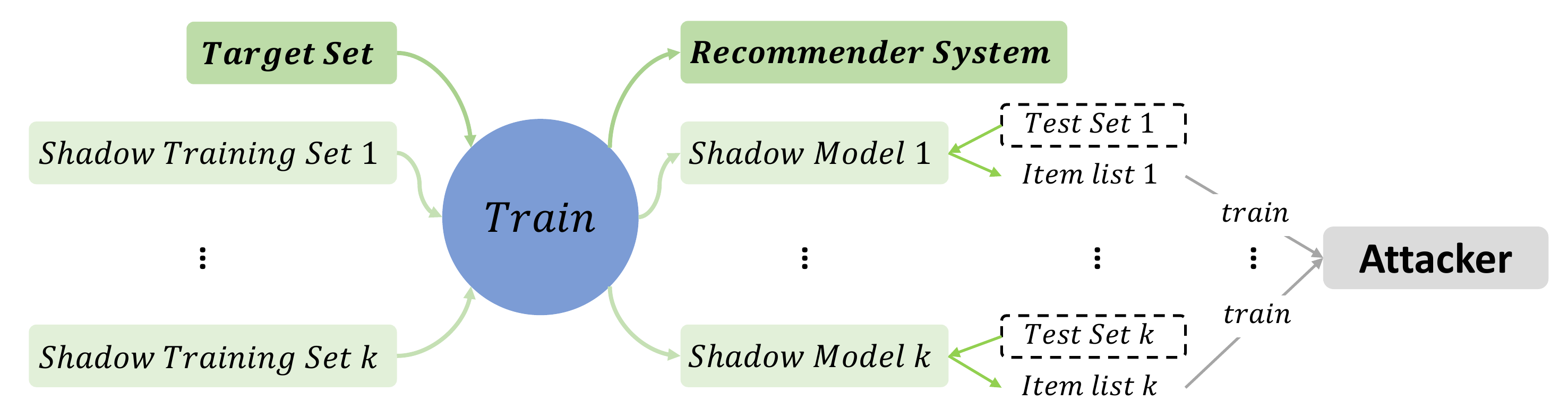}
        \caption{Shadow training strategy in privacy attacks for recommender systems. 
        The training process consists of two steps: training shadow recommendation models with public accessible auxiliary data and using the recommendation predictions to train the attacker.
        }\label{fig:ShadowTraining}
    \end{figure}

    In this subsection, we will introduce some representative methods of privacy attacks in recommender systems, which are summarized in  Table~\ref{tab-attacks}. 
    It is worth mentioning that many attack methods utilize shadow training (i.e., building a surrogate system) ~\cite{shokriMembershipInferenceAttacks2017} to generate the training data for the attacker. Shadow training can be divided into two steps: training shadow models and using the predictions of the models to train the attacker~\cite{huMembershipInferenceAttacks2022}. First, use the shadow data (e.g., public auxiliary data) to train a series of shadow models that could mimic the behavior of the target recommender system. Then, the attack model can be well trained using the predictions of the shadow models. The framework of shadow training in recommender systems is shown in Figure~\ref{fig:ShadowTraining}.

        \begin{table}
            \centering
            \caption{The categories of privacy attack methods on recommender systems}
            \begin{tabular}{l|l|l}
            \hline
             ~                                      &  \textbf{Taxonomy}                     & \textbf{Related methods} \\ \hline
            \multirow{4}*{Privacy Attacks}          & Membership Inference Attacks           &  ~\cite{zhangMembershipInferenceAttacks2021, danhierFidelityLeakagesApplying2020}  \\ \cline{2-3}
                                                    & Property Inference Attacks             &  ~\cite{InferenceAttacksGraph2022, ganjuPropertyInferenceAttacks2018a, atenieseHackingSmartMachines2013, parisotPropertyInferenceAttacks2021}\\ \cline{2-3}
                                                    & Reconstruction Attacks                 &  ~\cite{mengPrivacyPreservingSocial2019, mengPrivacyPreservingSocial2019, calandrinoYouMightAlso2011, salemUpdatesLeakDataSet2020, dudduQuantifyingPrivacyLeakage2020, hidanoExposingPrivateUser2020}\\ \cline{2-3}
                                                    & Model Extraction Attacks               &  ~\cite{yueBlackBoxAttacksSequential2021}  \\ \hline
            \end{tabular}
            \label{tab-attacks}
        \end{table}
        
        \begin{itemize}

        \item{\textbf{Membership Inference Attacks (MIA).}} The goal of MIA is to identify whether the target user is utilized to \emph{train} the target recommendation model.
        The general procedure of membership inference attacks is illustrated in Figure ~\ref{fig:MIA}. By querying the recommender system, the attacker will obtain a recommendation list for the target user.
        Then the attacker can utilize the distribution difference between the recommended item list and the user's historically interacted items to infer whether the target user is used in the training process.
        Furthermore, some attacking models utilize the temporal differences of the items ranking list to conduct MIA without the label~\cite{calandrinoYouMightAlso2011}.
        For instance, to quantify the privacy leakage in recommender systems, Zhang et al.~\cite{zhangMembershipInferenceAttacks2021} attempt to conduct MIA by measuring the similarity between recommended item list and the user historically interacted item list to infer whether the data of the target user is used by the target recommender system. 
        
        \begin{figure}[t]
            \centering
            \includegraphics[width=12cm,height=2.2cm]{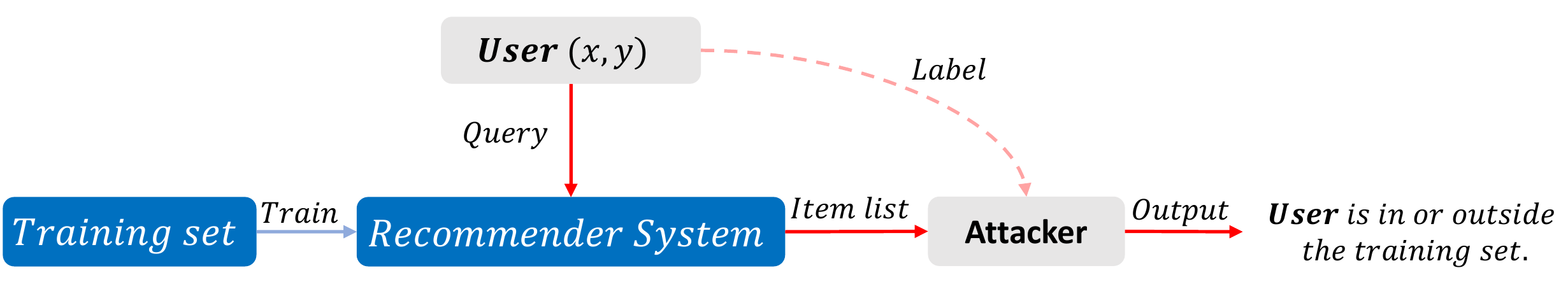}
            \caption{Membership inference attack. The attacker queries the recommender with a user and obtains the corresponding item list. Afterward, the attacker can infer if the user is in the training set based on the discrepancy between recommended item list and label (i.e., the historically interacted items).}
            \label{fig:MIA}
        \end{figure}

        \item{\textbf{Property Inference Attacks (PIA).}} 
        Different from the MIA aiming to infer whether the target user is in the training set or not, property inference attacks mainly focus on the global sensitive information in the training set. To get useful information from machine learning classifier, Ateniese et al.~\cite{atenieseHackingSmartMachines2013} train a series of classifiers to form a meta-classifier, which could recognize the unexpected but useful properties of the target training set.
        To attack graph-structured data, the work of ~\cite{InferenceAttacksGraph2022} conducts PIA on graph neural networks to infer the global information of the input graph, such as the number of nodes and links.
        In addition, the attacker training also adopts the same strategies as shadow training to build shadow models by utilizing auxiliary graphs. Recent works~\cite{ganjuPropertyInferenceAttacks2018a, parisotPropertyInferenceAttacks2021}  adopt shadow training to conduct property inference attacks against the full connected neural network and convolutional neural networks classification tasks, respectively.

        \item{\textbf{Reconstruction Attacks.}}
        Recommender systems also face a privacy risk similar to that exists in statistical database queries, i.e., using the user's publicly available attributes and multiple queries to reconstruct their sensitive information ~\cite{ramakrishnanPrivacyRisksRecommender2001}.
        For instance, reconstruction attacks can use the non-sensitive rating information to reconstruct the sensitive features of a certain user by matrix factorization~\cite{mengPrivacyPreservingSocial2019}. 
        The work of ~\cite{calandrinoYouMightAlso2011}  utilizes the temporal changes in the recommendation item list to conduct a reconstruction attack. 
        With the help of shadow training, recent works ~\cite{salemUpdatesLeakDataSet2020,dudduQuantifyingPrivacyLeakage2020} leverage the differences in predictions when a new user is added to train the target recommender system to reconstruct the information of the new user.
        Moreover, the work of ~\cite{hidanoExposingPrivateUser2020}  reconstructs recommended sensitive items by using data poisoning attacks, in which malicious users with certain items are used to link with the user's private information.

        \item{\textbf{Model Extraction Attacks.}} Another research field close to this attack is Knowledge Distillation  (KD) ~\cite{hintonDistillingKnowledgeNeural2015}. KD uses a complex "teacher" model to compress into a "student" model that can mimic the behavior of "teacher" but with a simpler structure. KD can be considered as a special case of model extraction attack with strong assumptions, such as accessing the parameters of the "teacher" model and the training data ~\cite{orekondyKnockoffNetsStealing2018}. In contrast, the model extraction attack is usually conducted under the black-box setting.
        Differing from the other three privacy attack methods that seek to infer the users' information, the model extraction attack targets the parameters and structure of recommender systems. 
        The work of ~\cite{yueBlackBoxAttacksSequential2021} attempts to steal parameters of sequential recommender systems by utilizing the specific autoregressive regimes of sequential recommender systems.
        Specifically, the attacker uses the recommended items to generate a series of item sequences, which act as the training data for the attacker. These sequential data can be used to train a victim recommender system which can imitate the behavior of the target recommender system.

        \end{itemize}

        \subsubsection{Privacy-preserving Methods}
        In this subsection, we investigate the privacy-preserving approaches in recommender systems. The structure will follow the taxonomy in section~\ref{defnition_privacy_preserving}, namely differential privacy, federal learning, adversarial learning, and anonymization \& encryption. 
        The representative methods for each part are summarized in Table~\ref{tab-preserving}.
            \begin{table}[!htb]
                \centering
                \caption{The categories of the privacy-preserving in recommendation systems.}
                \resizebox{\linewidth}{\height}{
                \begin{tabular}{l|l|l}
                \hline
                 ~                                      &  \textbf{Taxonomy}                   & \textbf{Representative Methods} \\ \hline
                \multirow{4}*{Privacy-preserving Methods}       & Differential Privacy                   &  ~\cite{zhuDifferentialPrivacyCollaborative2016, chenDifferentialPrivateKnowledge2022, chenPracticalPrivacyPreserving2020, zhangGraphEmbeddingRecommendation2021, xiaoDeepReinforcementLearningBasedUserProfile2021, zhangEnablingProbabilisticDifferential2021}\\ \cline{2-3}
                                                        & Federated Learning                     &  ~\cite{wuFedGNNFederatedGraph2021, wuFedCLFederatedContrastive2022, qiPrivacyPreservingNewsRecommendation2020, liFederatedRecommendationSystem2020, guoPREFERPointofinterestREcommendation2021, huangFederatedMultiViewDeep2020, flanaganFederatedMultiviewMatrix2021}\\ \cline{2-3}
                                                        & Adversarial Learning                   &  ~\cite{beigiPrivacyAwareRecommendationPrivateAttribute2020, resheffPrivacyFairnessRecommender2018, liAdversarialPrivacyPreservingGraph2021a, liaoInformationObfuscationGraph2021, wangPrivacyPreservingRepresentationLearning2021}\\ \cline{2-3}
                                                        & Anonymization \& Encryption                                 &  ~\cite{chenThwartingPassivePrivacy2014, wangProtectingMultipartyPrivacy2019, pengEPRTEfficientPrivacyPreserving2021, zhangPrivacyPreservingOptimizationNeighborhoodBased2021,  yuPrivacyPreservingMultiTaskFramework2020, huoPrivacypreservingPointofinterestRecommendation2021, salamatianManagingYourPrivate2015, yangPrivacyPreservingSocialMedia2019}\\ \hline
                \end{tabular}}
                \label{tab-preserving}
            \end{table}

        \begin{itemize}
        \item{\textbf{Differential Privacy.}} 
        These privacy-preserving methods can be effective in resisting membership inference attacks by adding random noise~\cite{ chenDifferentialPrivacyProtection2020}.  
        Xue and Sun ~\cite{zhuDifferentialPrivacyCollaborative2016} propose two differential privacy-based methods to preserve users' privacy information, namely differentially private item-based recommendation and differentially private user-based recommendation.
        To perform privacy-preserving cross-domain recommendation, a two-stage based method (PriCDR) is proposed to firstly preserve the data privacy in the source domain via differential privacy techniques and then transfer cross-domain knowledge to enhance recommendation performance in the target domain~\cite{chenDifferentialPrivateKnowledge2022}. 
        To defend against  privacy attacks in various scenario, researchers have successfully applied differential privacy to point-of-interest recommendations ~\cite{chenPracticalPrivacyPreserving2020}  and GNNs based recommendations~\cite{zhangGraphEmbeddingRecommendation2021}.
        In view of the trade-off between recommendation performance and privacy protection, Xiao et al. ~\cite{xiaoDeepReinforcementLearningBasedUserProfile2021} propose a deep reinforcement learning (RL)-based privacy protection method. Differential privacy is utilized to protect users' privacy of the recommender system, and RL is responsible for the choice of privacy budget, which can optimize the privacy budget over time based on the estimated privacy loss.
        The work of ~\cite{zhangEnablingProbabilisticDifferential2021}   proposes a differential privacy-based privacy-preserving framework (PLORE) to protect the users' privacy in location recommendation systems. PLORE  can balance the recommendation performance and the privacy-preserving using the probabilistic differential privacy mechanism.

        \item{\textbf{Federated Learning.}} 
        For federated learning-based privacy-preserving methods, the user's data can be stored in the local devices for training a local recommendation model, while a server in the cloud is responsible for aggregating the distributed model parameters. 
        With such special architecture, federated learning-based recommender systems can avoid the leakage of users' privacy data naturally. 
        For example, FedGNN is proposed to combine federal learning and GNNs based recommendation systems to protect the users' privacy~\cite{wuFedGNNFederatedGraph2021}, where each client device stores a user-item graph and a local GNN recommendation model. In addition,   differential privacy techniques are further used on the local gradients to protect user privacy. 
        In addition, recent works~\cite{wuFedCLFederatedContrastive2022, qiPrivacyPreservingNewsRecommendation2020} also incorporate local differential privacy for the gradients to ensure the privacy protection on news recommendation. 
        Since these methods may balance the trade-off between privacy protection and the recommendation performance, Li et al. ~\cite{liFederatedRecommendationSystem2020}  theoretically analyze such trade-off and provide a privacy error bound. 
        Guo et al.~\cite{guoPREFERPointofinterestREcommendation2021} propose an edge-accelerated framework PREFER, which aggregates the decentralized parameters on the edge server (e.g., base station) rather than the cloud server, so as to meet the real-time recommendation needs.
        The works of ~\cite{huangFederatedMultiViewDeep2020, flanaganFederatedMultiviewMatrix2021} introduce federated learning-based multi-view recommendation frameworks to address the cold-start issue in recommender systems.

        \item{\textbf{Adversarial Learning.}} 
        Adversarial learning is also an effective technique to protect privacy with an attacker discriminator~\cite{huangAdversarialMachineLearning2011,kurakinAdversarialMachineLearning2017,millerAdversarialLearningTargeting2020}.
        The work of ~\cite{beigiPrivacyAwareRecommendationPrivateAttribute2020} proposes an adversarial learning-based recommendation framework to defend against reconstruction attacks, which consists of two main components, bayesian personalized ranking recommendation systems, and a reconstruction attacker. In general, the recommendation task is proposed to model the user's preferences, while  attackers' gain is minimized to protect users' privacy, which can be formulated a min-max game. 
        On top of that, the user representations themselves may be used together with external data to recover users' sensitive information. To address this, Resheff et al. ~\cite{resheffPrivacyFairnessRecommender2018} propose an adversarial recommendation method, which adds the loss of demographic prediction tasks together with the recommendation tasks to the parameters optimization process.
        Then, the elimination of demographic information about the user can be controlled by hyperparameters.
        Additionally, increasing attention has been paid to develop adversarial privacy-preserving methods on GNNs ~\cite{liAdversarialPrivacyPreservingGraph2021a, liaoInformationObfuscationGraph2021, wangPrivacyPreservingRepresentationLearning2021}, which can contribute to enhance the design of  privacy-preserving recommender systems.

        \item{\textbf{Anonymization \& Encryption.}} Similar to differential privacy, the goal of anonymization and encryption is to protect sensitive information by adding noise or mapping data to another feature space.
        In general, anonymization techniques aim to prevent the public data from being linked to individual identities of people, 
        while encryption techniques make data unreadable to those who do not have the key to decrypt it\footnote{https://media13.connectedsocialmedia.com/intel/01/9768/Using\_Data\_Anonymization\_Enhance\_Cloud\_Security.pdf}.
        In ~\cite{chenThwartingPassivePrivacy2014},  Chen et al. propose a suppression and permutation-based anonymous method for collaborative filtering, which can reserve users' privacy by limiting the probability of a successful passive privacy attack. 
        More specifically,  the suppression operation is used to suppress an item from a related item list, while the permutation operation aims to permute an item that has climbed in a related item list to a lower position.
        
        Besides,  for the encryption techniques in recommender systems, PLAS ~\cite{wangProtectingMultipartyPrivacy2019} and EPRT ~\cite{pengEPRTEfficientPrivacyPreserving2021} propose homomorphic encryption technologies to protect users' sensitive information. 
        Zhang et al. ~\cite{zhangPrivacyPreservingOptimizationNeighborhoodBased2021} utilize the BGN Cryptosystem to conduct privacy protection. 
        By adding noise to the original data against privacy attacks, Yu et al. ~\cite{yuPrivacyPreservingMultiTaskFramework2020} combine knowledge graph enhancement techniques with multi-task learning to improve recommendation performance while adding Gaussian noise to sensitive data to protect privacy. 
        Huo et al. ~\cite{huoPrivacypreservingPointofinterestRecommendation2021} add Laplacian distributed noise to fuse the users' social relationships.
        
        In addition, there are also distortion-based methods ~\cite{salamatianManagingYourPrivate2015, yangPrivacyPreservingSocialMedia2019}, where they map the original data into a new feature space through a probabilistic mapping function and achieve data privacy protection under a certain distortion budget constraint. 
        \end{itemize}

        \subsection{Applications}\label{applications}
    This subsection presents some representative examples of using privacy-preserving techniques to protect sensitive information in real systems.  
    \begin{itemize}
        \item{\textbf{Private medical recommender systems.}} 
        In view of the fast development of eHealthcare, the privacy protection of the users' sensitive information attracts great attention. Xu et al.~\cite{xuPPMRPrivacyPreservingOnline2019} propose a privacy-preserving medical service recommendation method based on the modified Paillier cryptosystem, truth discovery technology, and the Dirichlet distribution. The work of ~\cite{katzenbeisserPrivacyPreservingRecommendationSystems2008}  employs cryptographic privacy-enhancing protocols to deal with the privacy challenges in health service recommender systems.
        To recommend the trusted physician for patients without privacy leakage, Hoens et al. ~\cite{hoensReliableMedicalRecommendation2013} propose two privacy-friendly recommender frameworks SPA and ACA, where they use secure multiparty computation techniques and anonymization to protect the users' private data, respectively.
    
        \item{\textbf{Location-private recommender systems.}}  
        Recent years have witnessed the increased need for location-based services due to the prevalence of smart mobile devices. For example,  Facebook and Google Map will collect the location information of the users to conduct more accurate recommendations.
        It is important to protect sensitive location information. In~\cite{zhaoPrivacyawareLocationPrivacy2014}, Zhao et al. obfuscate the rating vectors of the recommender to keep the location information secure. 
        In ~\cite{zhangEnablingProbabilisticDifferential2021}, a recommendation framework PLORE is proposed to address the location privacy challenges, where PLORE uses differential privacy to give sensitive location data privacy guarantees.
        To hide the real location data, Gao et al.~\cite{gaoPrivacypreservingCrossdomainLocation2019} adopt the differential privacy to obfuscate the historically visited locations data. 
    \end{itemize}

    \subsection{Survey and Tools}\label{survey}
    This subsection collects some existing surveys and tools regarding privacy in recommender systems to facilitate researchers in this field.

        \subsubsection{Surveys} 
        Recent comprehensive surveys on privacy-preserving recommendations are summarized in  ~\cite{aghasianUserPrivacyRecommendation2018,huangPrivacyProtectionRecommendation2019}. Specifically, \cite{aghasianUserPrivacyRecommendation2018} investigates privacy attacks and privacy-preserving methods in large-scale social recommendation systems and discusses the main issues in the privacy protection of the online social network.  
        The work of ~\cite{huangPrivacyProtectionRecommendation2019} gives the taxonomy of recommender systems and the definition of privacy leakage, in which the characteristics and specific measures of different privacy-preserving methods are compared. 
        On top of that, privacy in machine learning and deep learning is comprehensively reviewed in   ~\cite{rigakiSurveyPrivacyAttacks2021,mireshghallahPrivacyDeepLearning2020}.

        \subsubsection{Tools} 
        For differential privacy, there are some popular tools, such as Facebook Opacus~\footnote{https://opacus.ai/}, TensorFlow-Privacy ~\footnote{https://github.com/tensorflow/privacy}, OpenDP ~\footnote{https://opendp.org/home}, Diffpriv ~\cite{rubinsteinDiffprivPackageEasy} and Diffprivlib ~\cite{holohanDiffprivlibIBMDifferential2019}. For federated learning, popular tools  include TFF ~\footnote{https://github.com/tensorflow/federated}, FATE ~\footnote{https://github.com/FederatedAI/FATE}, FedML ~\cite{heFedMLResearchLibrary2020}, and LEAF ~\cite{caldasLEAFBenchmarkFederated2019}. Popular tools in Homomorphic Encryption are Awesome HE~\footnote{https://github.com/jonaschn/awesome-he} and TF Encrypted~\footnote{https://github.com/tf-encrypted/tf-encrypted}.

    \subsection{Future Directions}\label{future}
    Regarding privacy attacks and preservation in recommender systems, we have introduced many methods above. However, the need for privacy-preserving always comes and goes. There are many unresolved privacy issues in the field of recommender systems, including the following three points.
        \begin{itemize}
            \item \textbf{Privacy and performance trade-off}. Whether it is differential privacy, anonymization, or encryption, the most prominent means of countering privacy attacks is adding noise to the original data or distorting it. These methods protect privacy while reducing the utility. Because the sensitive information obscured is often the critical information that affects the performance of the recommendation. Therefore, depending on different task requirements, how to protect privacy with minimal performance cost may be a continuous research direction.
            \item \textbf{Comprehensive privacy protection}. The privacy protection methods usually protect against only one kind of privacy attack. However, the actual usage scenario is that a recommender system is subject to multiple privacy attacks. It is still challenging research to combine these privacy protection approaches without degrading the recommendation performance or to propose a comprehensive privacy protection framework.
            \item \textbf{Defence against shadow training}. Among the four privacy attack methods mentioned in this paper, membership inference attack, attribute inference attack, and reconstruction attack all use shadow training methods to train attackers. The training method provides vital support to the privacy attacks but is indeed trained under reasonable assumptions. Therefore, investigating how to defend against such training methods is crucial for privacy protection.
        \end{itemize}

\section{Environmental Well-being}
\label{sec:environment}
The application of deep learning brings great success to recommender systems \cite{wu2021survey}, which elevates recommending precision to a high level. However, more sophisticated models also result in longer training time and larger energy consumption. For example, Alibaba, an e-commerce site, has to consume several hours to train its model, which contains tens of billions of parameters on hundreds of servers \cite{jiang2019xdl}. Adnan et al. \cite{adnan2021accelerating} study that training a model on the Taobao dataset needs 621 minutes with 4 GPUs, whose average GPU power consumption is 56.39W per hour. As recommender systems have been adopted in many aspects of society, demands for large recommendation models will constantly increase. 
If we do not control the huge source consumption by recommender systems, the environmental damage will force humans to distrust and even give up the recommendation technique. Therefore, how to build environmental-friendly recommendation models is one essential component of trustworthy recommendation. 

In this section, we will conclude existing works on energy saving in the recommendation research field. Firstly, we introduce the concepts and taxonomy of environmental well-being techniques. Next, we summarize some methods to reduce the storage and energy consumption of recommender systems, including model compression and acceleration methods. Then, some applications in real systems are listed. Finally, we summarize some surveys and tools of this topic and give some promising directions.

\subsection{Concepts and Taxonomy}
In this subsection, we will refer to the concepts and taxonomy of techniques that benefit environmental well-being.

    \subsubsection{Concepts} As the development of recommender systems, the requests for storage and computation resources increase rapidly. To tackle the problem of intensive natural resources, model compression and acceleration techniques are proposed. The model compression \cite{cheng2017survey} aims to shrink the size of recommendation models to save storage resources. The acceleration techniques \cite{mittal2019survey,chen2020survey} focus on reducing training or inference time to save computation resources.
     
    \subsubsection{Taxonomy} According to the characteristics of recommendation models, model compression methods are devised for embedding layers or middle layers specifically, while acceleration techniques focus on the training or inference stage.
     
    \begin{itemize}
        \item \textbf{Model Compression}. Different from general deep models, the embedding layers always account for most of the parameters in recommendation models \cite{yi2018factorized,lui2021understanding,wan2021flashembedding}, so we categorize model compression techniques for recommendation into two types: (1) \textit{Embedding Layer}, which is always used to map discrete feature, such as ID feature and categorized feature, into a more expressive dense vector. (2) \textit{Middle Layer}, which extracts the user's preference or relations between features, such as the self-attention layer. 
        
        \item \textbf{Acceleration Techniques}. As for acceleration, those techniques can be divided for \textit{training} and \textit{inference} purposes because models are always placed on different platforms and devices when training or inference.
    \end{itemize}

\subsection{Methods}

In this subsection, we summarize storage-saving and energy-saving methods for recommendation models, i.e., model compression and acceleration techniques.

\newcommand{\tabincell}[2]{\begin{tabular}{@{}#1@{}}#2\end{tabular}}

\begin{table}[htbp]
\centering
\caption{Classification of Model Compression Methods for Recommendation}
\scalebox{0.90}{
\begin{tabular}{cl|l|l}
\hline
\multicolumn{2}{c|}{}           & \multicolumn{1}{c|}{\textbf{Embedding Layer}} & \multicolumn{1}{c}{\textbf{Middle Layer}} \\ \hline
\multicolumn{2}{c|}{Hash}       & \tabincell{l}{\cite{das2007google,li2011scene,zhou2012learning,zhang2014preference,serra2017getting}, \\ \cite{wang2019adversarial,lian2019discrete,shi2020beyond,kang2020learning,zhang2020model}}       
& \cite{serra2017getting,wang2019adversarial} \\ \hline
\multicolumn{2}{c|}{Quantization}               &  \tabincell{l}{ \cite{lian2020lightrec,lian2020product,jiang2021xlightfm,xiao2021matching,liu2020online,wu2021linear}, \\ \cite{wang2021compressing,liu2021online,zhang2022anisotropic,shi2020compositional,chen2021learning,li2021lightweight,hang2022lightweight}}               
&  \cite{wu2021linear,wang2021compressing,li2021lightweight}            \\ \hline
\multicolumn{2}{c|}{Knowledge Distillation}     &  \tabincell{l}{ \cite{tang2018ranking,chen2018adversarial,lee2019collaborative,kang2020rrd,wang2020next},\\ \cite{zhu2020ensembled,kweon2021bidirectional,kang2021topology,chen2021scene,xia2022device}}          &   \tabincell{l}{ \cite{tang2018ranking,chen2018adversarial,lee2019collaborative,kang2020rrd,wang2020next},\\ \cite{zhu2020ensembled,kweon2021bidirectional,kang2021topology,chen2021scene,xia2022device}}           \\ \hline
\multicolumn{2}{c|}{Neural Architecture Search} &   \tabincell{l}{ \cite{yan2021learning,liu2021learnable,zhaok2021autoemb,liu2020automated,zhao2021autodim,cheng2020differentiable}, \\ \cite{joglekar2020neural,chen2021learning,liu2021automated,wang2022autofield,lin2022adafs}}              &   \cite{song2020towards,chen2021scene}           \\ \hline
\multicolumn{2}{c|}{Others}     &  \cite{shen2020umec,sun2020generic,ginart2021mixed}               &    \cite{shen2020umec,sun2020generic,chen2021quaternion}             \\ \hline
\end{tabular}}
\label{tab:modelcompression}
\end{table}

        \subsubsection{Model Compression} Model compression \cite{cheng2017survey} aims to cut down model size for more efficient training and inference, which is environmental-friendly. While general model compression techniques are mainly designed for various neural network layers in a model, we categorize the recommendation model only into two parts: the embedding layer and the middle layer. Meanwhile, there are five types of methods to achieve model compression in recommendation: hash, quantization, knowledge distillation, neural architecture search, and others. Representative works are listed in Table~\ref{tab:modelcompression} for a clear description.
        
    \begin{itemize}
        
        \item \textbf{Hash}. The embedding table is always extremely large in the recommendation model because of plenty of items and users. Therefore, how to compress the embedding layer is a focus in the academic and industrial fields. Hash has been proved an efficient method. The hash function $h(\cdot)$ maps original concrete features $\textbf{x} \in \{0,1\}^n$, such as ID, into relatively short binary codes $\textbf{y} \in \{0,1\}^m$, where $n$ and $m$ are the vocabulary size of original and hashed feature respectively \cite{weinberger2009feature}. The embedding table is compressed much due to $m \ll n$, and thus the memory cost of the embedding table decreases. We cluster hash methods into two groups as follows: 
        
        \begin{itemize}
            \item \emph{Data-independent methods}. In this thread of methods, the hash function $h(\cdot)$ is pre-defined without considering the dataset. Locality Sensitive Hashing (LSH)  \cite{gionis1999similarity} is one representative method that can generate pairs of similar items. Two news recommendation works \cite{das2007google,li2011scene} make use of LSH to cluster similar news items and then find similar users to recommend relevant news.
            
            \item \emph{Data-dependent methods}.  Data-dependent methods always learn hash functions $h(\cdot)$ for specific dataset. For example, some works highlight the importance of users' preference over items and propose to preserve such information when learning hash function \cite{zhou2012learning,zhang2014preference,shi2020beyond}. Lian et al. \cite{lian2019discrete} combine with auxiliary information to learn better binary codes. 
             Works of  \cite{serra2017getting,zhang2020model} propose to adopt multiple hash functions to tackle  collision  problem in the hash. 
            Besides, DHE \cite{kang2020learning} gives a new idea to get embedding without using an embedding table, which adopts multiple hash functions and DNN to get embedding directly. Though most hash methods are designed for a lightweight embedding layer, Bloom Embeddings \cite{serra2017getting}, and ABinCF \cite{wang2019adversarial} also optimize the middle layer.
            
        \end{itemize}

        \item \textbf{Quantization}. Quantization is also an efficient technique to compress the embedding layer. In this type of method, the embedding of one feature will be clustered into several classes, and each embedding can be represented by the center of its cluster, named codeword, which greatly decreases the number of embeddings. To enhance the ability of representation, representation space is always decomposed, and an embedding is quantized to subvectors by several codebooks. An item's quantized representation vector $\textbf{q}_i \in \mathbb{R}^D$ can be computed as follow: 
        \begin{equation}
            \textbf{q}_i = f(c^1_{w_i^1}, c^2_{w_i^2}, ..., c^B_{w_i^B}),
        \end{equation}
        where $c^b_{w_i^b} \in \mathbb{R}^D$ is the $w$-th codeword in the $b$-th codebook. $f(\cdot)$ is the composing function of each subvector. According to the types of composing functions, quantization methods can be categorized into product quantization   \cite{jegou2010product}, additive quantization  \cite{Babenko_2014_CVPR} and residual quantization \cite{chen2010approximate}.  It is worth mentioning that product quantization   and additive quantization are often adopted in recommendation field. 
        Besides, some recent works using compositional embedding also belong to quantization. 
        
        \begin{itemize}
            \item \emph{Product Quantization (PQ)}. PQ is a type of quantization method that composes quantized vectors by product. LightRec \cite{lian2020lightrec} and pQCF \cite{lian2020product} adopt PQ to compress user and item embedding size for matrix factorization methods. Furthermore, Jiang et al. \cite{jiang2021xlightfm} propose a memory-efficient factorization machine based on PQ. Compared with traditional PQ methods, recent works \cite{lian2020lightrec,jiang2021xlightfm,xiao2021matching} integrate quantization into training process for optimal models. Liu et al. \cite{liu2020online} design an online optimized product quantization for the online recommendation model specifically. Besides, LISA \cite{wu2021linear} and MDQE \cite{wang2021compressing} even utilize PQ to lightweight and accelerate self-attention layer for sequential recommendation models.
            
            \item \emph{Additive Quantization (AQ)}. AQ uses add operation to compose vectors. For example, Liu et al. \cite{liu2021online} propose an online AQ, which is more efficient than online PQ \cite{liu2020online}. Zhang et al. \cite{zhang2022anisotropic} design a new loss for AQ and achieve a lower approximation error in contrast to PQ.
            
            \item \emph{Compositional Embedding}. Recently, another special thread of quantization methods prevailed, named compositional embedding. The main idea of these methods is to generate meta embedding for each feature based on their characteristics \cite{shi2020compositional}. Chen et al. \cite{chen2021learning} and Hang et al. \cite{hang2022lightweight} propose a lightweight compositional embedding for on-device and online recommendation models respectively. Besides, compositional embedding also can be used to create a lightweight self-attention layer in sequential recommendation \cite{li2021lightweight}. 
        \end{itemize}
        
        \item \textbf{Knowledge Distillation}. Knowledge Distillation (KD) is one of the most crucial techniques in model compression \cite{gou2021knowledge}, and is also adopted in the recommendation field for lightweight models. KD aims to use a smaller model (student model) to approximate the capacity of the original big model (teacher model). The key is distillation loss. Similar to \cite{gou2021knowledge}, existing KD methods for recommendation models can be categorized into two groups according to the distillation loss: response-based and feature-based.

        \begin{itemize}
            \item \emph{Response-based Methods}. Response-based methods transfer knowledge via the output layer of the teacher models, and the distillation loss can be formulated as:
            \begin{equation}
                \mathcal{L}_{res}=\mathcal{L}_R(z_t,z_s),
            \end{equation}
            where $z_t$ and $z_s$ are the logits of teacher and student models, respectively, and $\mathcal{L}(\cdot)$ refers to the divergence loss function.
            Tang et al. \cite{tang2018ranking} firstly adopt KD to ranking problems and propose the ranking distillation (RD) method. Then, Lee et al. \cite{lee2019collaborative} design a novel sampling technique and a new distillation loss function to improve RD. Based on these two methods, researchers further propose KD methods for some specific recommendation tasks, such as POI recommendation \cite{wang2020next}, sequential recommendation \cite{xia2022device} and CTR prediction \cite{zhu2020ensembled}. Besides, Kweon et al. \cite{kweon2021bidirectional} propose a bidirectional distillation method to elevate the accuracy of teacher and student recommendation models simultaneously.
            
            \item \emph{Feature-based Methods}. Compared to response-based methods, feature-based methods transfer the knowledge in intermediate layers of teacher models. The distillation loss of this thread is as follows:
            \begin{equation}
                \mathcal{L}_{feat}=\mathcal{L}_F(f_t(x), f_s(x)),
            \end{equation}
            where $\mathcal{L}_F(\cdot)$ is the similarity function. $f_t(x)$ and $f_s(x)$ are the output from the middle layers of teacher and student models.
            Recent studies \cite{kang2021topology,chen2021scene} propose to transfer the structure information of teacher models to student models. To fuse external knowledge into models, Chen et al. \cite{chen2018adversarial} design a novel training scheme with feature-based KD. Furthermore, Kang et al. \cite{kang2020rrd} combine response-based and feature-based KD, and propose a DE-RRD method.
        \end{itemize}
        
        \item \textbf{Neural Architecture Search}. Recently, applying automated machine learning (AutoML) techniques to design neural architecture for deep recommendation models has become a hotspot \cite{chen2022automated,zheng2022automl}. Neural Architecture Search (NAS) aims to search for the optimal architecture for deep models, which can prune the redundant parameters. The general idea of NAS is utilizing the validation loss to adjust the model architectures. Therefore, the objective of NAS can always be formulated into a bi-level optimization problem:
        \begin{equation}
            \min_\mathcal{A} \ \mathcal{L}_{valid}(\mathcal{W}^* (\mathcal{A}), \mathcal{A}),
        \end{equation}
        \begin{equation}
            s.t. \ \mathcal{W}^* (\mathcal{A})=arg \min_{\mathcal{W}} \mathcal{L}_{train} (\mathcal{W},\mathcal{A}),
        \end{equation}
        where $\mathcal{L}_{train}$ and $\mathcal{L}_{valid}$ are training and validation loss respectively. $\mathcal{A}$ and $\mathcal{W}$ are parameters and architectural weights of recommendation models.
        Most of the NAS works are beneficial to the embedding layer compression, but Song et al. \cite{song2020towards}, and Chen et al. \cite{chen2021scene} also utilize neural architecture search to compress the middle layers of recommendation models. The works that aim at the embedding layer can be categorized into two groups: embedding dimension search and automated feature selection.
        \begin{itemize}
            \item \emph{Embedding Dimension Search}. Some related studies focus on searching for optimal and minimal embedding size for each feature, which can compress the embedding layer efficiently. For example, ATML \cite{yan2021learning} and PEP \cite{liu2021learnable} propose a gradient-based and a pruning-based solution to search optimal embedding size for each feature. However, these two methods face the problem of a vast search space of embedding size. To tackle this problem, AutoEmb \cite{zhaok2021autoemb} and ESAPN \cite{liu2020automated} cut the embedding into several segments to reduce the search space, which can be called row-wise methods. In detail, AutoEmb designs a soft selection strategy to combine different segments with learnable weights. By contrast, ESAPN proposes a frequency-based hard selection strategy. Compared to AutoEmb and ESAPN, some other works group embedding with different values of a feature field to shrink search space named column-wise methods. AutoDim \cite{zhao2021autodim} is a special case of the column-wise method because it searches for a unified dimension for each feature field. In detail, AutoDim fuses the embedding of different dimensions by learnable weights, which is similar to AutoEmb. Cheng et al. \cite{cheng2020differentiable} group values of a feature field based on frequency. Furthermore, NIS \cite{joglekar2020neural} and RULE \cite{chen2021learning} propose to combine both row-wise and column-wise methods.
            
            \item \emph{Automated Feature Selection}. In addition to searching embedding dimensions, some works of automated feature selection \cite{liu2021automated,wang2022autofield,lin2022adafs} are also useful for lightweight embedding layers because they decrease the number of input features. Liu et al. \cite{liu2021automated} propose a method based on reinforcement learning, which regards feature selection as a multi-agent problem. AutoField \cite{wang2022autofield} equips with a controlling architecture to calculate the drop and select probability of each feature field and retrain the recommendation models after selection. Compared with the previous works selecting a fixed set of features for a dataset, Lin et al. \cite{lin2022adafs} propose to select various features for each user-item interaction to capture the dynamics of the practical recommender system.
        \end{itemize}
        
        \item \textbf{Others}. There are also some other novel techniques used to compress recommendation models. Shen et al. \cite{shen2020umec} utilize the alternating direction method of multipliers (ADMM) to jointly optimize feature selection and model compression. Sun et al. \cite{sun2020generic} propose an adaptive decomposition method for lightweight input and output layers and a parameter sharing scheme to compress middle layers. QFM and QNFM \cite{chen2021quaternion} adopt quaternion representations to decrease parameters of factorization machine models. Ginart et al. \cite{ginart2021mixed} design a mixed dimension embedding scheme to shrink the memory of the embedding layer and simplify neural architecture search into tuning  hyper-parameters.
        
    \end{itemize}
   
 \begin{table}[htbp]
\centering
\caption{Classification of Acceleration Techniques for Recommendation}
\scalebox{0.9}{
\begin{tabular}{c|l|l|l}
\hline

                        \multicolumn{2}{c|}{}                        & \multicolumn{1}{c|}{\textbf{Training}} & \multicolumn{1}{c}{\textbf{Inference}} \\ \hline
\multirow{3}{*}{Hardware-related} & Near/In Memory Computing &  \cite{kwon2021tensor}        &   \cite{kwon2019tensordimm,ke2020recnmp,dai2022dimmining,hwang2020centaur,wang2021rerec,wilkening2021recssd}        \\ \cline{2-4} 
                                  & Cache Optimization       &  \cite{zhao2019aibox,guo2021scalefreectr,yang2020mixed,ibrahim2021efficient}        &  \cite{eisenman2019bandana,xie2022fleche}         \\ \cline{2-4} 
                                  & CPU-GPU Co-design        &  \cite{zhao2020distributed,adnan2021accelerating,sethi2022recshard,zheng2021bips,adnan2022heterogeneous,kwon2022training}        &  -         \\ \hline
\multirow{2}{*}{Software-related} & Optimization             &  \cite{zheng2022cowclip,he2022metabalance,yin2021tt,ginart2021mixed,guo2021hyperrec}        &   \cite{gupta2020deeprecsys,gupta2021recpipe}        \\ \cline{2-4} 
                                  & Efficient Retrieval                & -         & \tabincell{l}{\cite{friedman1977algorithm,dasgupta2013randomized,keivani2018improved,ram2019revisiting}, \\ \cite{morozov2018non,liu2020understanding,tan2021norm,xu2022proximity}}          \\ \hline
\end{tabular}}
\label{tab:accelerationtechniques}
\end{table}
    
        \subsubsection{Acceleration Techniques}
        In addition to model compression, the acceleration is also a useful way to reduce training and inference time,  and thus can save energy and is environmental-friendly. Existing acceleration techniques \cite{mittal2019survey,chen2020survey} mainly focus on:  (1) memory-based challenges, which is the difficulty of data access by computation units, and (2) computation-based challenges, which means huge and complex computation. In the recommendation field, we conclude that hardware-related methods always aim at the memory-based challenge, and software-related methods are mainly for computation-based challenges. We will introduce hardware- and software-related methods, respectively, and the representative works are listed in Table~\ref{tab:accelerationtechniques}.
        
        \begin{itemize}
        
            \item \textbf{Hardware-related}. The advancements in computing units and hardware accelerators (e.g., GPUs) are huge. However, the memory techniques improve much slower. Such a growth gap leads to the problem of memory wall and hinders the improvement of acceleration techniques. Google \cite{boroumand2018google} has proved that data movement between memory and computing units accounts for 62.7\% of energy consumption across their many applications. As for the recommendation field, though large embedding tables bring high accuracy to recommendation models, they also cause severe storage and communication burdens \cite{yi2018factorized,lui2021understanding,zhao2019aibox}. Many hardware-related methods aim to optimize data moving between the storage device and computing units in model training or inference. We categorize hardware-related methods into three lines:
            
            \begin{itemize}
                \item \emph{Near/In Memory Computing (NMC/IMC)}. The main idea of NMC is to put computing units closer to the memory, which can lower the distance of data moving and thus reduce latency. Kwon et al. \cite{kwon2019tensordimm} firstly adopt NMC to accelerate the embedding layer for recommendation models. In detail, they design an NMC architecture, named TensorDIMM, based on general DRAMs to elevate bandwidth for embedding operations. RecNMP \cite{ke2020recnmp} is proposed for co-location operators for embedding vectors. DIMMining \cite{dai2022dimmining} is for Graph-based models specifically. TensorDIMM, RecNMP, and DIMMining are all tailored for DIMM, which is always for GPU-based systems. By contrast, Centaur \cite{hwang2020centaur} is designed for FPGAs in CPU-based systems. The methods mentioned above can accelerate model inference efficiently but are not fit for training. Tensor Casting \cite{kwon2021tensor} accelerates embedding layer training by design for tensor gather-scatter. Different from NMC, IMC puts memory and computing units together. ReRec \cite{wang2021rerec} and RecSSD \cite{wilkening2021recssd} design processing units on DRAMs and SSD according to embedding access frequency.
                
                \item \emph{Cache Optimization}. The cache mechanism is to specifically store some data that are accessed frequently on the memory device. Though cache provides low communication latency, its size is much limited. Therefore, efficient usage of cache is vital for acceleration. For instance, AIBox \cite{zhao2019aibox} caches embedding table on SSDs for training CTR models on a centralized system. Compared to AIBox, ScaleFreeCTR \cite{guo2021scalefreectr} is a distributed training system that contains a MixCache mechanism to accelerate embedding synchronization. Xie et al. \cite{xie2022fleche} propose a novel cache query mechanism to speed up embedding lookup on GPUs. Yang et al. \cite{yang2020mixed} design different precision for training on general memory and cache. Ibrahim et al. \cite{ibrahim2021efficient} and Eisenman et al. \cite{eisenman2019bandana} both propose an embedding placement strategy to elevate the efficiency of cache usage. It is worth noting that all of the methods mentioned are mainly for training recommendation models, except for Fleche \cite{xie2022fleche} and Bandana \cite{eisenman2019bandana}.
                
                \item \emph{CPU-GPU Co-design}. Hybrid CPU-GPU mode prevails in industrial recommendation systems due to huge embedding tables. General recommendation models can be partitioned into the embedding part and DNN part. The embedding part is always stored and processed on the CPU, and the DNN part is on GPU \cite{zhao2020distributed}. However, the communication of embedding vector between CPU and GPU is a big challenge, so the CPU-GPU co-design is in need. Adnan et al. \cite{adnan2021accelerating} propose to put highly accessed embeddings on GPU memory to reduce communication time. Similarly, Recshard \cite{sethi2022recshard} partition embedding into several parts according to the distribution of training data and store each part on hierarchical memory. BiPS \cite{zheng2021bips} focuses on the parameter update process during the model training. It contains a bi-tier parameter server to accelerate parameter synchronization. Recently, Adnan et al. \cite{adnan2022heterogeneous} and Kwon et al. \cite{kwon2022training} both propose a data pipeline to speed up parallel training.
                
            \end{itemize}
            
            \item \textbf{Software-related}. There are many studies \cite{reagen2016minerva,hegde2019extensor,pentecost2019maxnvm} on designing accelerators for DNN to tackle the computation challenges, but not all of them can fit recommendation models. Besides, embedding plays a vital role in the recommendation field, so how to accelerate embedding computation needs exploring. Based on these two observations, we group software-related methods into two groups: Optimization and Efficient Retrieval.
            
            \begin{itemize}
                \item \emph{Optimization}. In order to accelerate training recommendation models, some works focus on the training process, such as Cowclip \cite{zheng2022cowclip} and MetaBalance \cite{he2022metabalance}. Cowclip uses a large batch to train recommendation models and contains an adaptive clipping strategy to maintain training accuracy. Metabalance is proposed to adjust gradients of each loss for the multi-task recommendation dynamically and can reduce training epoch efficiently. Besides, some other works accelerate training by optimizing the embedding layer. Yin et al. \cite{yin2021tt} adopt tensor train decomposition to recommendation models and optimize the batched matrix multiplication. Ginart et al. \cite{ginart2021mixed} design mixed dimension embeddings for recommendation models. In HyperRec \cite{guo2021hyperrec}, the general embedding layer is replaced by high-dimensional binary vectors, whose computation is more efficient on various computing devices. In addition, DeepRecSched \cite{gupta2020deeprecsys} and RecPipe \cite{gupta2021recpipe} are two optimization methods for inference. DeepRecSched is an effective scheduler to optimize parallel data movement and can reduce latency for several recommendation models. RecPip partition recommendation models into multistage and then compute each stage in a parallel mode under the control of the designed scheduler.
                
                \item \emph{Efficient Retrieval}. In industrial, it is common that train user and item embeddings offline to represent their preference and attributes, then get recommending list by Embedding-Based Retrieval (EBR) online. Such a schema can reduce inference latency efficiently under the condition of a huge volume of users and items. Many studies \cite{covington2016deep,grbovic2018real} aim to refine the embedding for accuracy, but EBR is also important for efficiency. The main idea of efficient retrieval is to build an index for each embedding to achieve sub-linear searching times rather than conducting the inner product directly. According to the type of index, we categorize efficient retrieval methods into four clusters: tree-based, graph-based, hash-based, and quantization-based. Tree-based methods partition high-dimensional space into several sub-space and use leaf nodes on a tree to represent each sub-space. Due to the lower time complexity of the index, it has been used for efficient retrieval for a long time, such as KD-tree \cite{friedman1977algorithm}. To further improve efficiency, randomized-partition trees (RP-tree) \cite{dasgupta2013randomized,keivani2018improved} are proposed with minimal loss of accuracy. Ram et al. \cite{ram2019revisiting} combine KD-tree with RP-tree to reach a trade-off. Recently, graph-based methods have attracted much attention. This line of methods builds a similarity graph for retrieval, in which each vertex in the graph represents an embedding, and the edge shows the distance between embeddings. Morozov et al. \cite{morozov2018non} firstly utilize a Delaunay graph to construct an index graph and propose the theory of similarity graph. ip-NSW+ \cite{liu2020understanding} and NAPG \cite{tan2021norm} further improve the ip-NSW \cite{morozov2018non} by introducing an angular similarity graph and a hierarchical navigable small world graph respectively. Xu et al. \cite{xu2022proximity} propose the IPGM algorithm to tackle the problem of node deletion in the similarity graph. Besides, some of hash methods \cite{li2011scene,zhou2012learning,zhang2014preference,lian2019discrete} and quantization methods \cite{lian2020lightrec,lian2020product,zhang2022anisotropic} can also be utilized to accelerate EBR, which have been introduced in Model Compression part.
                
            \end{itemize}
            
        \end{itemize}

\subsection{Applications in Real Systems}
Many companies and researchers have devoted efforts to the environmental problems when developing modern recommender systems in many aspects of society. 
In this subsection, we will introduce these actions from the following three aspects: Big Model, Edge Computation, and Embedding-Based Retrieval Systems.

    \subsubsection{Big Model} In the field of NLP, Pretrained Language Models (PLMs) such as BERT \cite{devlin2018bert}, ERNIE \cite{sun2019ernie} and GPT-3 \cite{brown2020language} prevail, because PLMs with only low-cost fine-tuning can adapt to various tasks and possess high precision. Such a training paradigm can emit training a whole model for each specific task and thus save energy largely. Based on the success of PLMs in NLP, many studies aim to shift this scheme to the recommendation field. As the natural similarity between recommendation and NLP, several early works \cite{yuan2020parameter,yuan2021one,shin2021one4all} consider a user behavior history as a sentence, and then they can train a big model on datasets from different tasks. The pretrained model often generates a universal user representation, which can be used in many downstream tasks and get rid of the huge energy cost by retraining for each task. However, these works require all tasks use a unique item set because they are trained based on the item id. To tackle this problem, recent works \cite{hou2022towards,li2022personalized,geng2022recommendation,cui2022m6} utilize textual information of items to covert recommendation task to a text-to-text task, which is more suitable for PLMs. Besides, TransRec \cite{wang2022transrec} fuses multi-modality of items to avoid using item id. These works urge recommendation to step into the big model period and contribute to environmental well-being.
    
    \subsubsection{Edge Computation}  In the traditional service framework, most computations are conducted on the cloud, which may cause high latency of service and high cost of communication. To tackle these challenges, the edge-cloud scheme becomes a hotpot \cite{yao2022edge}. Due to the limited capacity of edge devices, the model on edge must be computing- and storage-efficient. Alibaba \cite{gong2020edgerec} firstly designs a recommender system for edge specifically named EdgeRec, and applies it to their e-commerce application-taobao. EdgeRec deploys the rerank module and user behavior module on edge to respond to users' dynamic preferences and capture real-time behaviors. To avoid storing the whole embedding table on edge, it fetches items and corresponding embedding in each request. Furthermore, Alibaba \cite{yao2021device} achieves their "thousands of people with thousands models" with edge-cloud collaboration. Unlike EdgeRec, it can also update the user's model on the edge device, which benefits long-tailed users. To further improve the edge-cloud collaboration, Yao et al. \cite{yao2022device} design a meta controller to optimize the recommending list from edge and cloud recommenders. Another recent work \cite{chen2021mc} regards cloud and edge as slow and fast components and designs a bidirectional collaboration mechanism to benefit both.
    
    \subsubsection{Embedding-Based Retrieval Systems} EBR has been widely adopted in  many aspects, such as question answering \cite{karpukhin2020dense,su2021whitening}, web search \cite{fan2019mobius,li2021adsgnn} and recommender system \cite{huang2020embedding,li2021embedding}.  EBR plays the main role in the recall stage in a recommender system, and a good EBR system should meet the trade-off of three key points: memory, latency, and accuracy. To achieve these goals, many companies build efficient EBR systems for their services. Airbnb \cite{grbovic2018real} introduces their EBR system mainly from the aspect of how to construct informative embedding to get higher accuracy. Facebook \cite{huang2020embedding} proposes a hybrid EBR system with boolean and KNN matching to elevate recall efficiency. Besides, Amazon \cite{nigam2019semantic} publishes their semantic product search model for embedding retrieval. Next, Alibaba \cite{li2021embedding} and JD \cite{zhang2020towards} also propose MGDSPR and DSPR respectively for their e-commerce applications.

\subsection{Surveys and Tools} In this subsection, we give out some surveys and tools about environmental well-being for readers who want to investigate this topic further.
    
    \subsubsection{Surveys} To our best knowledge, our survey is the first attempt to conclude works about the environmental well-being of the recommender system. Furthermore, a survey on model compression and acceleration technique for recommender systems is also few, so we list some surveys on these techniques in other fields, such as computer vision (CV). As for model compression, Cheng et al. \cite{cheng2017survey} introduce four types of methods to compress DNN and compare compression rate between several methods on the CIFAR dataset. Among techniques of model compression mentioned above, there is a survey \cite{wang2017survey} about learning to hash, and another survey \cite{luo2020survey} about deep hash methods. Besides, Gholami et al. \cite{gholami2021survey}, and Gou et al. \cite{gou2021knowledge} survey the techniques about quantization and KD for DNN, respectively. It is worth noting that two surveys \cite{chen2022automated,zheng2022automl} related to NAS in the recommender system came out recently. Chen et al. \cite{chen2022automated} categorize methods of NAS into four groups, in which feature selection and feature embedding are two types beneficial to shrink models. Another line of work is acceleration. Deng et al. \cite{deng2020model} survey many acceleration methods for DNN, especially hardware-related methods. Le et al. \cite{le2021efficient} summarize efficient retrieval methods of recommendation. This survey considers the retrieval pipeline and illustrate the methods from the aspects of two consecutive process: candidate generation and candidate ranking.

    \subsubsection{Tools} Faiss \cite{johnson2019billion} is a popular tool of similarity search, which is published by Facebook. It implements many product quantization-based methods to accelerate the embedding retrieval. Besides, Faiss integrates a GPU k-selection algorithm, which can utilize GPU more efficiently.

\subsection{Future Directions}
    More research has paid attention to the memory and energy cost of recommender systems for environmental purposes. However, one serious problem is the lack of a framework to measure and predict the energy consumption for recommender systems specifically, like SyNERGY \cite{rodrigues2018synergy}. Such an estimation framework will help researchers to study energy-efficient recommender systems. As for model compression techniques, NAS is a promising direction to shrink model sizes. However, most existing works aim to improve accuracy. How to get the trade-off between the size and accuracy of recommendation models is an interesting problem for NAS. For the aspect of acceleration, the design of collaboration between hardware and software may be a future direction. Edge recommendation is a great example.

\section{Accountability \& Auditability}
\label{sec:account}
Accountability for recommendation refers to what extent users can trust recommender systems  and who is responsible for the devastating effects brought by the recommender systems. Because recommender systems  play the role of information messenger, it is vital to equip recommender systems with accountability to be trustworthy. For example, during a crisis such as the COVID-19 pandemic, recommender systems  encouraged the spread of many fake news and conspiracy theories \cite{papadamou2022just}, which caused distrust in recommender systems and had harmful effects on society. Besides, due to the lack of transparency and explainability of deep recommendation models, not only general users but also professional experts are unable to control recommender systems absolutely\cite{jannach2016user,bellogin2021improving}. Recently, the emergence of auditabilty, which refers to the methodology of evaluating recommender systems, has helped build the accountable recommender systems from a new perspective. To further discuss the accountability and auditability of recommender systems, we first introduce the definition of accountability and the taxonomy of auditability. Then, we present some relevant surveys and tools. At the final of this section, we discuss some future directions to inspire the readers who concern about this topic.

\subsection{Concepts and Taxonomy}
In this subsection, we will introduce the concepts of accountability and the taxonomy of auditability for the recommendation.

    \subsubsection{Accountability} Accountability has various concepts among different applications of artificial intelligence. Loi et al. \cite{loi2021towards} indicate that the general concept of accountability at least includes three dimensions: responsibility, answerability, and sanctionability. From the view of recommendation, we interpret the three dimensions: (1) \textit{Responsibility}. If a user accepts an uncomfortable or illegal recommendation, accountability requires recommender systems to know which part of the system should be blamed. (2) \textit{Answerability}. If an recommender system is accountable, it can reveal the reasons when recommender system has a bad effect. (3) \textit{Sanctionability}. Sanctionability refers that recommender systems should punish and mend the parts that cause harmful impacts. According to these dimensions and AI regulations \cite{2021europe} published by European Commission, we summarize four roles for an accountable recommender systems: (1) \textbf{Content Governors}. Content governors are responsible for examining the facticity and noxiousness of "items" in a recommender system. When a malignant event is reported, they should give an explanation from the aspect of content and decide to remove which item, so they undertake the answerability and sanctionability. (2) \textbf{Model Designers}. Model designers build the recommendation models for service. On the one hand, they can design explainable models for answerability. On the other hand, they should make the models reproducible, which benefits the dimension of sanctionability. (3) \textbf{System Deployers}. System deployers not only need to deploy recommendation models online but also check the possible trustworthy problems brought by recommender systems to avoid detriments in advance, so they take on the task of Responsibility. (4) \textbf{Third-party Auditors}. Third-party auditors are vital to guarantee accountability for general AI systems \cite{raji2020closing}, not excluding recommender systems. They play the role of Responsibility for pointing out existing and potential problems in recommender systems. Besides, they will also reveal the reasons which refer to the role of answerability.
    
    \subsubsection{Auditability} Algorithm audits indicate a class of methods to analyze the existence and reasons for harmful problems for AI systems. It can help implement an accountable recommender systems. The methods of recommendation algorithm audits can be categorized into two classes: external and internal audits.
    
    \begin{itemize}
        \item \textbf{External Audits}. External audits regard recommendation models as a black box, and utilize input and output data from recommender systems to evaluate the algorithm \cite{le2022algorithmic}. Based on the concepts, we know two roles for accountable recommender systems that are relevant to external audits: Content Governors and Third-party auditors. Content governors aim to identify and remove the items that contain harmful contents regardless of the algorithm itself \cite{buzzi2011children,araujo2017characterizing}. However, their works are always complex and burdensome, so that it is impossible to eliminate problematic issues of recommender systems only by content governors. Recently, the methods by third-party auditors became popular. Many works focus on recommendations of YouTube, one of the most popular video platforms, to examine malicious problems, such as inappropriate videos for children \cite{papadamou2020disturbed}, user radicalization \cite{ribeiro2020auditing} and pseudoscientific misinformation \cite{papadamou2022just}. These three works conduct three procedures for audits similarly. Firstly, they collect publicly available data from YouTube. Then, they classify normal and bad videos (such as radicalized videos) by manual annotations or well-trained classifiers. At last, they analyze the annotated data to probe problems. In conclusion, external audits can avoid the problem of subjectivity because no system developers and deployers join. However, it cannot be conducted before a recommender system is deployed.
        
        \item \textbf{Internal Audits}. Internal audits examine the problems with access to training data by the other two roles for accountable recommender systems, i.e., Model Designers and System Deployers. One of the most useful audits means for model designers is to enhance explainability for recommendation models \cite{doshi2017accountability}, which can output reasons for a bad case. Another efficient way is to achieve reproducibility of recommendation models \cite{bellogin2021improving} because a reproducible environment gives auditors more chances to evaluate recommender systems with different strategies. To achieve reproducible recommendation models, many researchers propose recommendation benchmarks, such as Recbole \cite{zhao2021recbole,zhao2022recbole}, FuxiCTR \cite{zhu2020fuxictr}, DaisyRec \cite{sun2020we,sun2022daisyrec} and ELLIOT \cite{anelli2021elliot}. As for System deployers, they audit the system thoroughly based on the designed models before deployment. For example, Wilson et al. \cite{wilson2021building} propose a five-step (scoping, mapping, artifact collection, testing, and reflection) audit method to explore fairness problems in job recommendation systems. In detail, they analyze the source code and the data to explore some questions relevant to fairness. According to the methods mentioned above, we find that internal audits can minimize the probability of harmful impact before deployment, but it may cause the problem of subjectivity because designers and auditors are the same groups of people.
        
    \end{itemize}

\subsection{Surveys and Tools}
In this subsection, we will summarize the surveys and tools relevant to accountability and auditability.

    \subsubsection{Surveys} A survey \cite{wieringa2020account} on algorithmic accountability summarizes the theory of accountability and organizes existing works according to five necessary parts of an accountable system \cite{bovens2007analysing}. As for auditability, one recent work \cite{bandy2021problematic} surveys algorithm audits from the aspects of behavior, domain, organization, and audit methods and focuses on the four problematic behaviors (discrimination, distortion, exploitation, and misjudgment) that auditors should pay attention to. Another survey \cite{le2022algorithmic} focuses on external audits. It first formulates the external audit process and then organizes existing works based on two proposed canonical audit forms. However, there is no specific survey about accountability and auditability for recommendations.
    
    \subsubsection{Tools} As mentioned above, annotation is one of the most important steps for external audits. Many researchers choose manual annotation for high accuracy, but it is a hard job. A crowdsourced platform is a good tool for this task, such as Amazon Mechanical Turk (AMT)\footnote[1]{https://www.mturk.com/}. Besides, there are many benchmarking code library that can give convenience to auditors for reproducibility, e.g. Recbole \footnote[2]{https://github.com/RUCAIBox/RecBole} \cite{zhao2021recbole,zhao2022recbole}, FuxiCTR \footnote[3]{https://github.com/xue-pai/FuxiCTR} \cite{zhu2020fuxictr}, DaisyRec \footnote[4]{https://github.com/AmazingDD/daisyRec} \cite{sun2020we,sun2022daisyrec} and ELLIOT \footnote[5]{https://github.com/sisinflab/elliot} \cite{anelli2021elliot}.

\subsection{Future Directions}
As we know, the accountability of recommender systems is related to many aspects, such as explainability, fairness, and so on. However, most of the existing studies  only focus on one aspect, which may lead to inadequate accountability. Therefore, the combination of many aspects for accountable recommender systems should be further considered. As for auditability, many works focus on external audits, but few on internal audits. To minimize the risk before deploying recommendation models, we can explore automated internal audits in the future. Furthermore, the collaboration between external and internal audits can be a promising direction because it can benefit from both merits.

\section{Interactions among Different Dimensions}
\label{sec:relation}
The ideal trustworthy recommender systems would possess all of six features and advantages. However, in real-world context, it is challenging to consider the modeling of multiple features simultaneously, as these features may have many varying levels of interdependence, and even conflict in some aspects.

Despite that a number of studies have investigated the interactions between dimensions of trustworthy AI ~\cite{xu2021robust,fan2021jointly,liu2021trustworthy}, research on trustworthy recommender systems is still limited. Fortunately, some researchers have recognized the importance of potential interactions among different dimensions, and attempted to explore and utilize them. In this section, we focus on the interactions between dimensions with extensive and close ties to other dimensions.

\noindent \textbf{Interactions with Robustness.} 
Since the robustness of a system is an intrinsic characteristic to ensures its normal operation, it undoubtedly maintains a high degree of close crossover and connection with other dimensions in a recommendation system. 
Previous research on trustworthy AI \cite{liu2021trustworthy} shows that the robustness of such systems is positively correlated to their explainability \cite{etmann2019connection,noack2021empirical}, while partly conflicts with their privacy \cite{song2019privacy} and fairness dimensions \cite{xu2021robust}. 
These characteristics are particularly evident in adversarial attacks and robust training. For recommender systems, the issue is comparable. Therefore, how to use positive dimensions to promote robustness, and maintain the balance between robustness and potentially conflicting dimensions, while maintaining system's  robustness against adversarial attacks without affecting other dimensions, is a non-trivial issue.

Most recently, Bilge et al. investigate the robustness of four recommendation algorithms based on collaborative filtering with privacy enhancement to determine, which improvement is more appropriate for the interaction between the two sides. Recent research on Trustworthy AI \cite{lee2021machine,small2022robust} has shifted the focus of researchers to the interaction between robustness and other dimensions of recommendation systems within the context of machine learning. In \cite{zhang2020gcn}, Zhang et al. design a robust model to combat the attacks and ensure the fairness of the recommender system. In \cite{zheng2021disentangling}, Zheng et al. develop an additive causal model for disentangling user interest and conformity for recommendation with causal embedding. In the meantime, this method ensures the robustness and explainability of the recommendation.

\noindent \textbf{Interactions with Fairness.} 
With the gradual expansion of the research on fairness in terms of trustworthy recommender systems and people's recognition of the importance of fairness, researchers began to take fairness as one of the goals when designing recommender systems, resulting in a number of studies on the interactions between fairness and other dimensions. 
One of the main directions is the interaction between fairness and explainability. In \cite{chen2020bias}, Chen et al. provide a survey of the research on fairness and analyzes the explainability of the model at the same time. In \cite{fu2020fairness}, Fu et al. propose a fairness-aware explainable recommendation model. In \cite{ge2022explainable}, Ge et al. provide research on explainable fairness in recommendation. In addition, there are additional interaction studies with fairness, such as the interaction between robustness and fairness \cite{zhang2020gcn} mentioned in the previous subsection.

\noindent \textbf{Interactions with Explainability. }
As mentioned in the previous section, the primary focus of interaction research in this regard is the interaction between explainability and fairness. In addition, there are several additional interaction studies. In \cite{anelli2019make}, Anelli et al. analyze the robustness of their proposed explainable model.
In ~\cite{fan2021jointly}, Fan et al. study the interactions between adversarial vulnerability and explainability, and take advantage of explainability to enhance adversarial attacks. 
In \cite{ghazimatin2020prince}, Ghazimatin et al. provide a new counterfactual explanation mechanism for recommendation, which also solved the privacy exposure problem. 
\section{Future Directions}
\label{sec:future}
In this survey, we detail the existing works for trustworthy recommendations from six vital dimensions. However, there are also some other potential directions to be explored for the supplement to the definition of trustworthy recommender systems (TRec). In this section, we will illustrate some promising directions for further research on this topic.

\noindent \textbf{Interactions among different dimensions}.
A trustworthy recommender system should possess six mentioned dimensions simultaneously, but most present researches only focus on one of them. Though few works have paid attention to achieving two dimensions, such as robust-fairness \cite{zhang2020gcn}, explainability-privacy \cite{ghazimatin2020prince}, etc., no one tries to go forward to three or more. Therefore, how to reach more requests of trustworthy dimensions is still an urgent problem for the community of recommender systems. Besides, the conflicts between different dimensions should not be ignored. For example, some recommendation works \cite{wang2018reinforcement,singh2020model} add an additional module for explainability, which may bring the risk of violating the aspect of environmental well-being. Such conflicts may ruin the efforts for trustworthiness, so how to resolve the conflicts and get a trade-off is an important direction for TRec. In a word, both positive and negative interactions among different dimensions should be a spotlight in future directions.

\noindent \textbf{Other Dimensions to achieve trustworthy recommender systems}. 
Though we have issued six essential dimensions of TRec, there also exist some other dimensions worthy of being noted. For instance, security is a necessary dimension in many scenes, such as medication recommendation \cite{tan20224sdrug} and industrial recommendation \cite{ezenwa2022toward}. In these scenes, the recommender systems will affect human decisions directly, and any improper decision can cause uncountable losses to life and property. Therefore, the characteristic of security is needed. Another aspect that TRec should possess is controllability. When a recommender system causes a devastating effect, accountability can only give out whether and who should be blamed. By comparison, controllability can help stop harmful recommendations and minimize the horrible effects. The two dimensions mentioned above are still far from the full content of TRec, and more other dimensions of TRec should be explored in the future.

\noindent \textbf{Technology Ecosystem for trustworthy recommender systems}. 
With the increasing demands of TRec, many researchers have devoted themselves to this field. However, the lack of a technology ecosystem causes huge inconvenience to the developments and experiments. A TRec technology ecosystem should contain datasets, metrics, toolkits, etc., but none of these parts have been well developed. For example, a few recent works aim to build up a standard dataset, such as KuaiRec \cite{gao2022kuairec} for non-discrimination, but no related work for environmental well-being and accountability specifically. Besides, there is no toolkit for evaluating various dimensions of the recommender system, which is one of the most important reasons why few research focuses on the interaction between various dimensions. Therefore, an integrated technology ecosystem is a vital procedure for achieving trustworthy recommender systems.

\section{Conclusion}
\label{sec:con}

In this survey, we provide a comprehensive overview of trustworthy recommender systems (\textbf{TRec}) from a computational perspective. More specifically, we elaborate on  six of the most critical dimensions for the trustworthiness of recommender systems: safety \& robustness, non-discrimination \& fairness, explainability,  privacy, environmental well-being, and accountability \& auditability.  
For each dimension, we provide the basic concepts and taxonomy for  readers to have a better understanding  of this topic, as well as  summarize the representative methods in achieving trustworthy recommender systems. 
In addition, we introduce widely-used applications in  real-world systems for achieving trustworthy recommender systems from multiple dimensions. 
Surveys and tools are also provided for readers' further exploration in this demanding topic.
Finally, we also analyze the potential interactions among different dimensions and possible future research directions for trustworthy recommender systems.

\bibliographystyle{ACM-Reference-Format}
\bibliography{main}

\end{document}